\documentclass[printer]{aa}  

\usepackage{txfonts}
\usepackage{bm}
\usepackage{graphicx}
\usepackage{subfig}
\usepackage[]{hyperref}
\usepackage{xcolor}
\usepackage{array}
\usepackage{bbm}
\usepackage{tikz}
\renewcommand*{\vec}[1]{\bm{#1}}
\newcommand*{\mat}[1]{\mathbf{#1}}

\newcommand{\ch}[1]{{\textcolor{black}{#1}}}

\newcolumntype{C}[1]{>{\centering\let\newline\\\arraybackslash\hspace{0pt}}m{#1}}

\DeclareMathOperator{\e}{e}

\renewcommand*{\vec}[1]{\bm{#1}}

\def\H{\mathcal{H}}
\def\K{\mathcal{K}}

\begin{document} 
        
\title{ Testing whether a signal is strictly periodic  }

        \subtitle{Application to disentangling planets and stellar activity in radial velocities}
        
   \author{Nathan C. Hara 
        \inst{\ref{i:geneve}}\thanks{CHEOPS fellow} 
        Jean-Baptiste Delisle\inst{\ref{i:geneve}}, Nicolas Unger\inst{\ref{i:geneve}}, Xavier Dumusque\inst{\ref{i:geneve}}}

\institute{
                Observatoire Astronomique de l’Université de Genève, 51 Chemin de Pegasi, 1290 Versoix, Switzerland\label{i:geneve}  \email{nathan.hara@unige.ch}
}

        
        
        \abstract
{Searches for periodicity in time series are often done with models of periodic signals, whose statistical significance is assessed via false alarm probabilities or Bayes factors. However, a statistically significant periodic model might not originate from a strictly periodic source. In astronomy in particular, one expects transient signals that show periodicity for a certain amount of time before vanishing. This situation is encountered for instance in the search for planets in radial velocity data. While planetary signals are expected to have a stable phase, amplitude and frequency --- except when strong planet-planet interactions are present --- signals induced by stellar activity will typically not exhibit the same stability. In the present article, we explore the use of  periodic functions multiplied by time windows to diagnose whether an apparently periodic signal is truly so. We suggest diagnostics to check whether a signal is consistently present in the time series, and has a stable phase, amplitude and period. The tests are expressed both in a periodogram and Bayesian framework.   
Our methods are applied to the Solar HARPS-N data as well as HD 215152, HD 69830 and HD 13808.  We find that (i) the HARPS-N Solar data exhibits signals at the Solar rotation period and its first harmonic ($\sim$ 13.4 days). The frequency and phase of the 13.4 days signal appear constant within  the estimation uncertainties, but its amplitude presents significant variations which can be mapped to activity levels. (ii) as previously reported, we find four, three and two planets orbiting HD 215152, HD 69830 and HD 13808.
}        
        
        
        \maketitle
        %
	

	\section{Introduction}
	\label{sec:intro}
	
	In many astronomical fields, one searches for periodicity in a time series of measurements. Due to observational constraints, the samples often are unevenly spaced, possibly with large gaps in the sampling. The search for periodicity is then typically done with periodograms, which can be defined in several ways. The different definitions can be viewed as a metric (most often the $\chi^2$ difference) comparing two models: a base model, and a model that includes this base model plus a periodic signal at a given frequency, for a grid of frequencies.  Typically, periodic components are sinusoids, and the base model includes white, Gaussian noise~\citep{lomb1976,scargle}. It can also include a constant~\citep{ferrazmello1981,cumming1999,reegen2007,zechmeister2009},  general linear models~\cite{baluev2008}, non-sinusoidal functions~\cite{baluev2013_vonmises, baluev2015}, several periodic components~\cite{baluev2013, baluev2013_freqdecomposer} or correlated noise~\citep{delisle2020a}. 
	Once the periodogram is computed, detections are claimed if the false alarm probability (FAP) of the highest peak is below a certain threshold. 
	
	The search for periodic signals can also be performed in a Bayesian framework. \ch{Periodograms can be defined as comparisons of marginal likelihoods of the two competing models, with or without a given frequency \citep{mortier2015, feng2017}. The analysis can also be done directly from the posterior distributions of orbital elements. In that case,}  to determine how many periodic components can be confidently detected, one computes the ratio of Bayesian evidences -- or Bayes factors~\citep{kassraftery1995} -- of models with $n$ and $n+1$ components. If the Bayes factor is above a certain threshold (usually 150), the addition of a periodic component is validated\ch{~\citep[e. g.][]{gregory2007, gregory2007a, tuomi2012, faria2016}}. Alternatively, one can use a detection criterion based on Bayesian model averaging~\citep{hara2021a}.

	A low FAP or high Bayes factor give a measure of the confidence in a detection of a periodic component within a certain model of the data. If the model is inappropriate, there might be significant signals that do not correspond to strictly periodic signals. 
	Certain noises are particularly likely to yield significant peaks: the ones that present a periodicity that is localised in time, similarly to wavelets. This situation is encountered in particular in the search for planets in radial velocity (RV) data. This type of data consists of a time series of velocity projected along the line of sight (the radial velocity) of a given star, measured thanks to the Doppler effect. If a planet orbits the star, then one expects periodic variation of the velocity. However, the surface of a star is not uniform: spots and faculae break the flux balance between the approaching and receding limb of the star, furthermore, they inhibit convective blue shift. As a consequence, during their lifetime, stellar features introduce radial velocity signals that might mimic planetary ones, in particular at the stellar rotation period or its harmonics~\citep[e.g.][]{saar1997, boisse2011, meunier2010b, dumusque2014}, but also possibly at periods apparently unrelated to the stellar rotation \citep[][]{nava2020}.
	
	One approach to this problem consists in trying to improve the data model  and the detection metrics so that a significant detection has a meaning as close as possible to the detection of the signal of interest (here, a planet). In RV data analysis, stellar activity is typically modelled with a Gaussian process. 
	This approach is taken in~\cite{haywood2014, rajpaul2015, jones2017, gilbertson2020} and~\cite{hara2021a}. The stellar activity models are however imperfect, and another approach consists in being more agnostic to the form of stellar activity. For instance,~\cite{gregory2016} suggests to compute the Bayesian evidence of a an apodized model, that is a periodic planetary signal model multiplied by a function of the form $\e^{-\frac{(t-t_0)^2}{\tau^2}}$, where $\tau$ and $t_0$ are free parameters. Denoting by $T_\mathrm{obs}$ the total timespan of observation, a signal is claimed to be of planetary origin if it is significant, and if the posterior probability of the event  $\tau \geqslant T_\mathrm{obs}$ is greater than a given threshold. The rationale is that a planetary signal is a purely periodic signal, and should therefore be identical from the beginning to the end of the dataset. In a periodogram setting, \cite{schuster1898, mortier2017} adopt a similar approach. They suggest to compute the periodogram adding one point at a time and  determine whether the significance of a peak at a certain period grows steadily with the number of points. 
	
	
	

	One can envision other diagnostics to determine if an apparently periodic signal presents signs of variability. In the present work, we present tests to assess the time-scale of a signal, and to determine if its period, phase and amplitude are constant. We suggest tests based on the periodogram and Bayesian formalisms. 

	
	
	The article is organised as follows. In section~\ref{sec:methods}, we define the diagnostics mentioned above.
	 In section~\ref{sec:applications}, we show examples of applications to the Solar HARPS-N data~\citep{dumusque2020}, HD 215152~\citep{delisle2018},  HD 69830~\citep{lovis2006} and HD 13808 \citep{ahrer2021}. 
	 We finally present our conclusions in section~\ref{sec:conclusion}.

	\section{Methods}
	\label{sec:methods}
	\subsection{Signal time-scales}
	\label{sec:detec}
	
	\subsubsection{Apodized sinusoids periodograms}
	\label{sec:def}
	
	Let us consider a time series $\vec y = (y(t_k))_{k=1..N}$. Following the method of~\cite{baluev2008} to define periodograms, we consider two alternative models of the time series $\vec y$: a linear base model $\mu_\H$ of $p$ parameters, and a model $\mu_\K$ that includes $\H$ plus an apodized sinusoidal component. The base model, \ch{as a vector with $N$ components}, is defined as
	\begin{align}
	\vec \mu_\H(\vec \theta_\H) = \mat \varphi_\H \vec \theta_\H 
	\label{eq:mh}
	\end{align}
 where \ch{$\mat \varphi_\H$ is a $N\times p$ matrix and $\vec \theta_\H$ is the vector of $p$ parameters to be fitted}. \ch{For instance, the matrix $\mat \phi_\H$ can be simply taken as a column vector with all $N$ entries equal to one, thus modelling an offset. If one assumes that sinusoidal signal of frequency $\omega$ is in the data by default, $\mat \phi_\H$ can be defined as a $N\times 3$ matrix with line $i$ equal to  $[1, \cos \omega t_i, \sin \omega t_i]$.  } The alternative model is defined as
	\begin{align}
    \vec \mu_\K(\vec \theta_\H, \vec \theta) = \varphi_\H \vec \theta_\H + \vec \mu(t,  \vec \theta )
	\label{eq:mk} 
	\end{align}
	where $\vec \mu(\vec t,\theta)$ is an apodized periodic function. \ch{ We compare the hypotheses $\H$: $\vec y = \vec \mu_\H(\vec \theta_\H) + \vec \epsilon$   for some $\vec \theta_\H$ and $\K$: $\vec y = \vec \mu_\K(\vec \theta_\K) + \vec \epsilon$  for some $\vec \theta_\K$, where}
	$\epsilon$ is a Gaussian \ch{noise model (or random variable)} of covariance matrix $\mat V$.
	
	In the present work, $\mu$ is specified to 
	\begin{align}
	\mu(t, \omega, \tau,t_0, A,B ) = w(\tau, t_0) (A \cos \omega t + B \sin \omega t ).
	\label{eq:mu1}
	\end{align}
	where  $w(\tau, t_0) $ is the apodization function. We will here use Gaussian and box-shaped functions, that is
	\begin{align}
	w_G(\tau, t_0) &:= \e^{-\frac{(t-t_0)^2}{2\tau^2}}   \label{eq:gauss}\\
	w_B(\tau, t_0) &:= \mathbbm{1}_{\left[t_0 - \frac{\tau}{2}, t_0 + \frac{\tau}{2}\right]} (\tau, t_0) \label{eq:box}
	\end{align}
	where $ \mathbbm{1}_X(x)$ is a function that is zero everywhere except  when $x \in X$. 
	but  other choices are possible. One can envision further generalisations. Indeed, the periodic signal need not be sinusoidal. It can be for instance a Keplerian one~\citep[][]{baluev2015}. Also, the base model $\H$ can be generalised to non linear models. 
	

	\ch{Considering a model $\mu_i$, $i= \H$ or $\K$ with parameters $\vec \theta_i$, and column vector residuals $\vec r = \vec y  -  \vec \mu_i(\vec \theta_i)$, the $\chi^2$ is defined as $ \vec r^T \mat V^{-1} \vec r $ where the suffix $T$ denotes the matrix transpose. 
	We denote by $\chi^2_\H$ the $\chi^2$ of the residuals after fitting the parameters $\vec \theta_\H$ of model $\eqref{eq:mh}$ and}, for a set of fixed $\omega, t_0,\tau$, by  $\chi^2_\K(\omega, t_0,\tau)$ the $\chi^2$ of the residuals after fitting the parameters $\vec\theta_\H, A$ and $B$ of the model $\eqref{eq:mk}$, on which the model depends linearly. We now define a periodogram type counterpart to~\cite{gregory2016}. 
	\begin{align}
	z(\omega, t_0,\tau) &= \chi^2_\H - \chi^2_\K(\omega, t_0,\tau), \label{eq:z}
	\end{align}
	which represents the improvement of the fit by adding the $\mu$ component compared to the base model $\H$. 
    In the following sections, we will use the following representation.
	For a given $\tau$ we represent the quantity 
	\begin{align}
	z'(\omega,\tau) &= \max\limits_{t_0} z(\omega, t_0,\tau) .
	\label{eq:zm}
	\end{align}
	The values of $z'(\omega,\tau)$ are overplotted for different values of $\tau$. In Fig.~\ref{fig:solardata_perios}), we represent the values of $z'$ for four values of $\tau$ for the Solar HARPS-N data (see Section \ref{sec:sun} for more details). We call such a figure an ASP for apodized sinusoids periodogram.

	Our definition of Eq.~\eqref{eq:mu1} and~\eqref{eq:gauss} is similar to a continuous wavelet transform~\citep[][]{grossmannmorlet1984} applied to irregularly sampled time-series~\citep{foster1996}. Our choice of Eq.~\eqref{eq:gauss} is inspired by the Gabor wavelet, which minimizes the product of time and Fourier domain variance~\citep{daugman1998}. However, in the wavelet transform context, the time-scale $\tau$ and frequency $\omega$ are dependent on each other. Indeed, a family of wavelet $(\psi_{a,b})_{a,b}$ is defined as the translation and dilation of a mother wavelet $\psi$, $\psi_{a,b}(x) =  \psi((x-b)/a)/\sqrt{a}$. In our case, the lifetime $\tau$ of a quasi-periodic feature is not unequivocally linked to its frequency $\omega$. As a consequence, even though having three free parameters instead of two introduces extra computational load, we leave $\omega, \tau$ and $t_0$ as independent parameters. 
	
	\subsubsection{Grid}
	\label{sec:grid}
	
	Regular periodograms usually are  computed on a grid of equispaced frequencies $\omega_k = k \Delta \omega$. The choice of $\Delta \omega$ depends on the typical width of the periodogram peaks, which is approximately equal to $ 2\pi/T_\mathrm{obs}$, where $T_\mathrm{obs}$ is the total timespan of the observations. Then $\Delta \omega$ is chosen as  $ 2\pi/T_\mathrm{obs}*1/N_{\omega}$, where $N_{\omega}$ is called the oversampling factor. 
	
	For the ASPs, we need to construct a grid in three dimensions: $\omega, t_0$ and $T$. The naive approach, consisting in having a fixed grid for each parameter and trying all the possible combinations, is unnecessarily computationally expensive. Indeed, for a given time-scale $T$ such that $T < T_\mathrm{obs}$, the typical width of the peaks in $\omega$ is $\approx 2\pi/T > 2\pi/T_\mathrm{obs}$. 
	To set the grid, the approach adopted here is to define the following quantities
	\begin{itemize}
		\item A maximum frequency $\omega_\mathrm{max}$
		\item Two oversampling factors, $N_\omega$ and $N_t$.  
		\item A fixed number $\alpha < 1$  and a number of time-scales $N_\tau$ such that the grid of $\tau$ is defined as $\tau_k = \alpha^{k-1} T_\mathrm{obs}$, $k=1..N_\tau$.
	\end{itemize} 
	We  define two reference grids for $\omega$ and $t_0$ as, $\omega^{\mathrm{ref}} = (\omega_k)_{k=1..n_\omega}$ and  $t_0^{\mathrm{ref}}= ({t_0}_k)_{k=1..n}$, where $\omega_k = k \Delta \omega$, with $\Delta \omega = 2\pi/T_\mathrm{max}/N_{\omega}$ and ${t_0}_k = k\Delta t_0$ with $\Delta t_0 = \tau_\mathrm{min}/N_{t}$. 
	
	For each $\tau_k$, the waveletogram is computed on a sub-grid of $\omega^{\mathrm{ref}} $ and $t_0^{\mathrm{ref}}$, denoted by $\omega^{\tau} $ and $t_0^{\tau}$. These are defined as $\omega^{\tau}_k =  \omega^{\mathrm{ref}}_{k s_\omega^\tau} $ and ${t_0^{\tau}}_k =  {t_0^{\mathrm{ref}}}_{k s_t^\tau} $, where $ s_\omega^\tau = \lfloor \tau_\mathrm{max}/\tau \rfloor$ and $ s_t^\tau = \lfloor \tau/\tau_\mathrm{min} \rfloor$. The rationale behind these numbers is that, for a given $\tau$, the grid spacing in $\omega$ is approximately $2\pi/\tau/N_\omega$ and the grid spacing in $t_0$ is $\tau/N_t$. Secondly, since the omega grid is not recomputed for every time-scale $\tau$, one has to compute only once the $\cos \omega t$ and $\sin \omega t$ terms.
	Once this method is specified, one has to choose specific values for  $N_\omega$, $N_t$, $\omega_\mathrm{max}$, $\alpha$ and $n_T$.  The \ch{list of symbols and, when relevant, the default} values we choose are reported in table~\ref{tab:grid}.
	
	\begin{table}
		\caption{\ch{Symbols.}}
		\begin{tabular}{p{1.3cm}|p{4.3cm}|p{1.9cm}}
				\hline  \hline
		\multicolumn{3}{c}{\ch{Data}} \\
		\hline  \hline
		Symbol & \multicolumn{2}{p{6.2cm}}{Definition} \\
		$\vec t$ & \multicolumn{2}{p{6.2cm}}{Epochs of measurements, the $N$ data points are taken at times $\vec t = (t_i)_{i=1..N}$} \\
		$\vec y$ & \multicolumn{2}{p{6.2cm}}{Colummn vector of the data, $\vec y = (y(t_i))_{i=1..N}$} \\
		\multicolumn{3}{c}{} \\
				\hline  \hline
		\multicolumn{3}{c}{\ch{Model parameters}} \\ \hline \hline
		Parameter & \multicolumn{2}{p{6.2cm}}{Definition} \\
		\hline  
		$\mat V$ & \multicolumn{2}{p{6.4cm}}{Assumed covariance matrix of the noise.} \\
		$\vec \mu_\H$ & \multicolumn{2}{p{6.4cm}}{Model of the null hypothesis, column vector with $N$ components, $\vec \mu_\H = \varphi_\H \vec \theta_\H $ where $\mat \varphi_\H$ is a $N\times p$ matrix and $\vec \theta_\H$ are free parameters.} \\
		$\vec \mu_\K$ & \multicolumn{2}{p{6.4cm}}{Signal models. As defined in Eq.~\eqref{eq:mk}, $\vec \mu_\K = \varphi_\H \vec \theta_\H + \vec \mu(\vec t, \omega,\tau,t_0, A,B)$. The parameters $\vec \theta_H, A,B$ are fitted onto the data for a grid of parameters $\omega,\tau,t_0$. } \\
		$\mu$ & \multicolumn{2}{p{6.4cm}}{Function of time $t$ and parameters $ \omega$, $\tau$, $t_0$, $A$, $B$  defined in Eq.~\eqref{eq:mu1} as $\mu(t, \omega,\tau,t_0, A,B ) = w(\tau, t_0) (A \cos \omega t + B \sin \omega t )$. } \\
		$w$ & \multicolumn{2}{p{6.2cm}}{Apodisation window (for instance $\e^{-(t-t_0)^2/(2\tau^2)}$) } \\
		$\tau$ & \multicolumn{2}{p{6.2cm}}{Time-scale of the apodisation window} \\
		$t_0$ & \multicolumn{2}{p{6.2cm}}{Center of the apodisation window} \\
		\multicolumn{3}{c}{} \\
				\hline  \hline
		\multicolumn{3}{c}{Grid parameters} \\ \hline \hline
			Parameter & Definition & Default value \\ \hline 
			$N_\omega$ & Oversampling in frequency & 8 \\
			$N_t$ & Oversampling in time & 5 \\
			$ \alpha$ &  Downscaling of the time-scale& $1/3$\\
			$N_\tau$ & Number of time-scales & 5 \\
		\end{tabular}
		\label{tab:grid}
	\end{table}

	\subsubsection{Statistics}
	\label{sec:stat}
	
	In~\cite{gregory2016}, one computes the Bayesian evidence (or marginal likelihoods) of models containing $k$ periodic components multiplied by an apodization factor $ \e^{-\frac{(t-t_0)^2}{2\tau^2}} $. A planet detection is claimed if a signal is statistically significant (the ratio of evidence of the model including to the evidence without it is greater than a certain threshold), and the posterior distribution of $\tau$ favours values greater than the total time-span of the observations. 
	
	We here propose an alternative way to estimate the timescale on which the signal remains coherent, that can be computed very quickly from the periodogram. Denoting by $t_{\tau,\omega}$ the value of $t_0$ maximising the value of the periodogram~\eqref{eq:z} for a given $\omega$ and $\tau$,  we compute the the distribution of $D_z = z(\omega, t_{\tau,\omega}, \tau) - z(\omega, t_{\tau',\omega}, \tau')$ with the hypothesis that model $\K(\omega, t_{\tau,\omega}, \tau, A^\star, B^\star )$ is correct,  where the fitted cosine and sine amplitude $A^\star, B^\star$ are obtained by fitting model $\K$ to the data.  $D_z$ can easily be expressed as a generalised $\chi^2$ distribution, its mean and variance is given by an analytical expression, given in Appendix~\ref{app:dz}, eq.~\eqref{eq:dze} and eq.~\eqref{eq:dzv}. This method does not take into account that $\omega$ is chosen as a maximum of the periodogram and could be refined in future work. 
	
	Secondly, one can define a false alarm probability associated to~\eqref{eq:z}, like a regular periodogram. \ch{In the remainder of the article, we will not use the false alarm probability (FAP) for ASPs, however we will use a FAP for regular periodograms \citep[][]{baluev2008,delisle2020a}}. For the sake of conciseness, the method is described in Appendix~\ref{app:fap}.
	
	\ch{In general, assuming that the signal is in the form of the model $\K$ (eq.~\eqref{eq:mk}) will be an incorrect assumption. In particular, if aliasing cannot be neglected and the data contain several signal sources, not accounting for them might result in an inaccurate uncertainty on the time-scale $\tau$ of the signal of interest. We suggest to include in the base model all the periodic signals found to be significant (we give an example of this procedure in Section~\ref{sec:applications}). Alternatively, at the price of a higher computational cost, one can search for several signals simultaneously, as shown in the next section.}
	
	\subsubsection{Bayesian approach}
	\label{sec:bayesian}
	
	
	In \cite{gregory2016}, a detection is claimed if the posterior probability of the event $\tau > T_\mathrm{obs}$ is greater than a certain threshold. 
	However, such a criterion can be misleading. For instance, there might be datasets such that there are a few isolated measurements with a large time gap to the bulk of the measurements. In that case, most of the samples could correspond to $\tau < T_\mathrm{obs}$ \ch{while there is a} possibility that the signal is consistent with a fully periodic signal. Additionally to the posterior distribution of $\tau$, we suggest to consider \ch{a quantity $f$ that quantifies the fraction of information captured by a certain window. We compute the posterior distribution of}
	\begin{align}
	    f = \frac{\vec w(\tau, t_0)^T \mat V^{-1} \vec w(\tau, t_0)}{\vec 1^T\mat V^{-1} \vec 1}
	    \label{eq:f}
	\end{align}
	conditioned on the frequency of the planet being within $\pm \Delta \omega$ of a given frequency of interest. In Eq.~\eqref{eq:f} $\vec w = (w(t_i))_{i=1..N}$ is the apodization window (see Eq.~\eqref{eq:mu1}), $\mat V$ is the covariance matrix and $\vec 1$ is a vector of size $N$. The values of the vector $\vec w$ are between 0 and 1. If $f$ is close to 1, then the apodized signal is close to \ch{a strictly periodic} signal. On the contrary, a value of $f$ close to 0 implies that the $\vec w$ has concentrated information in time. \ch{Note that $f$ can be computed for any apodised signal, in particular the planet model can be circular or Keplerian, and for all the signals in the data.}
	
	In \cite{gregory2016}, the number of planets is determined by checking that the Bayes factor comparing the $n+1$ and $n$ planets model is greater than a certain threshold. The significance of  the planet can also be established with the false inclusion probability (FIP)~\citep{hara2021a}, defined as follows. Considering a certain parameter space $S$ (for instance a period and eccentricity interval), the true inclusion probability (TIP) is the probability to have planet with orbital elements in $S$ marginalised over the number of planets in the system. The FIP is defined as $1-$TIP. Instead of claiming a detection if the Bayes factor is above a certain threshold, we use the fact that the FIP of having a signal with period in $[\omega - \Delta \omega, \omega + \Delta \omega]$ --- marginalised on all parameters including $t_0$ and $\tau$ ---  is below a certain threshold. 
	

	\subsection{Amplitude and phase consistency}
	
	\label{sec:amphase}
	
	The second diagnostic we present in this work relies on computing the phase and amplitude of a signal at a given period with a moving time window. Let us consider, as in section~\ref{sec:def}, the linear models $\H$~eq.~\eqref{eq:mh} and  $\K$~eq.~\eqref{eq:mk}. For the function $\mu$, we adopt the definition of~\eqref{eq:mu1} where the window function $w$ is box shaped.  
	For each value of $\omega$, $t_0$ and $\tau$, one has estimates of parameters $A$ and $B$ (\eqref{eq:mu1}).
	The uncertainties on $A$ and $B$ are obtained from the diagonal elements of the covariance matrix of the fit. From the estimates of $A$ and $B$, we compute the local semi-amplitude $K$ and phase $\phi$
	\begin{align}
	K(\omega, t_0, \tau) &= \sqrt{A^2 + B^2} \label{eq:k}\\ 
	\phi(\omega, t_0, \tau) &= \mathrm{atan2}(-B,A), \label{eq:phi}
	\end{align}
	as well as the uncertainties on $K$ and $\phi$ propagated from those on $A$ and $B$ with  simple Monte Carlo simulation. Alternatively, one can estimate the uncertainties on $A$, $B$, $K$ and $\phi$ with their posterior distributions, computed with a Monte-Carlo Markov chain algorithm, which is more computationally expensive but propagates the uncertainties on the orbital elements more accurately. 
	



	For a given frequency $\omega$ and time-scale $\tau$, the quantities $A$,$B$, $\phi$ and $K$ and their uncertainties (computed from the least square fit or posterior distributions) can be used to check the consistency of phase and amplitude of a signal at a given period. Assuming that the signal contains a pure sine at $\omega$, the values of $A$ and $B $ or $K $ and $\phi $ as a function of $t_0$ should be consistent with the hypothesis that $A$ and $B$ or $K$ and $\phi$ are constant. 
	
	Besides a visual inspection of $A$ and $B $ or $K $ and $\phi $ as a function of $t_0$, the hypothesis that the signal has a constant phase or amplitude can be checked with statistical tests. In particular, one can isolate values of $A$ and $B$ estimated in different windows. Let us denote by $(A_i)_{i=1..M}$ and $(B_i)_{i=1..M}$ estimates of $A$ and $B$ in disjoint windows.  If one neglects correlations that might happen between those estimates due to correlated noise,  under the hypothesis that phase and amplitude are constant, then the statistics
	\begin{align}
	\chi^2_A &=  \sum\limits_{i=1}^M \frac{(A_i - \mu_A)^2}{\sigma_{A_i}^2} \\
	\chi^2_B &= \sum\limits_{i=1}^M  \frac{(B_i - \mu_B)^2}{\sigma_{B_i}^2} 
	\label{eq:test}
	\end{align}
	where
	\begin{align}
	\mu_X = \sum \frac{X_i}{\sigma_i^2} \left/ \sum \frac{1}{\sigma_i^2} \right.
	\label{eq:testmu} 
	\end{align}
	with $X = A$ or $X=B$, should follow a $\chi^2$ distribution with $M-1$ degrees of freedom. We can measure the quantiles of $\chi^2_A$ and $\chi^2_B$: if they are below a certain threshold, then the hypothesis that they are constant through time can be rejected. A $\chi^2$ distribution with $M-1$ degrees of freedom has mean $M-1$ and standard deviation $\sqrt{2(M-1)}$.
	 For a quick diagnostic, we will use the quantity 
	\begin{align}
N\sigma_{\chi^2_i}  = \frac{\chi^2_i - (M-1)}{\sqrt{2(M-1)}} \;\;\;, \;\;\; i = A, B
\label{eq:nsigma}
\end{align}
which is the difference of the $\chi^2_i $ and its expected value divided by the standard deviation.

	

	In the test suggested in this section, the time window $w$ in eq.~\eqref{eq:mu1} is assumed to be a box-shaped one. One could envision a more general one. However, for a consistent analysis, since the noise is also multiplied by the window, the covariance matrix of the noise would have to be modified according to the chosen window. If this one is non-zero everywhere, then this comes down to doing the analysis of the data without any window, unless some numerical errors lead the covariance to be non invertible. In that case, using a window comes down to a rectangular window. As a consequence, we here only consider box-shaped time windows (see eq.~\eqref{eq:box}).

	We stress that to evaluate the phase and amplitude consistency of a signal, one must have an accurate estimate of the period of the signal to analyse. If this one is poorly estimated, then the phase will spuriously appear \ch{as a linear phase drift}. As a consequence, we advise to select the value of $\omega$ to be studied in depth from the regular periodogram containing all the data.

	\ch{As mentioned in Section \ref{sec:stat},} in general, assuming that the signal is in the form of the model $\K$ (eq.~\eqref{eq:mk}) will be an incorrect assumption, \ch{and aliasing can be a problem. For instance,}  if the signal contains several sources that are strictly periodic, due to aliasing, the phase and amplitude of the signal under consideration might appear to vary significantly even though the signal is periodic. 
As a consequence, the fact that the quantities defined in eq.~\eqref{eq:test} deviate from the mean is not always interpretable as testing the hypothesis that the signal at frequency $\omega$ is constant in phase and amplitude. One of the assumptions in $\K$ is the noise model, however this one might not be known and needs to be adjusted to the data. We suggest to include in the base model all the periodic signals found to be significant, this is precised in Section~\ref{sec:applications}.

	Besides testing for the consistency of the phase and amplitude of a signal, it might be interesting to study its relative phase and amplitude with another signal. For instance, one might want to see the evolution of the phase difference between the RV and $\log R'_{HK}$ as a function of time to study the stellar activity. An example of such an analysis is given in Section \ref{sec:sun}.

	\begin{figure*}
	\noindent
	\centering
	\hspace{-1,5cm}
	\begin{tikzpicture}
	\path (0.1,0) node[above right]{\includegraphics[scale=0.62]{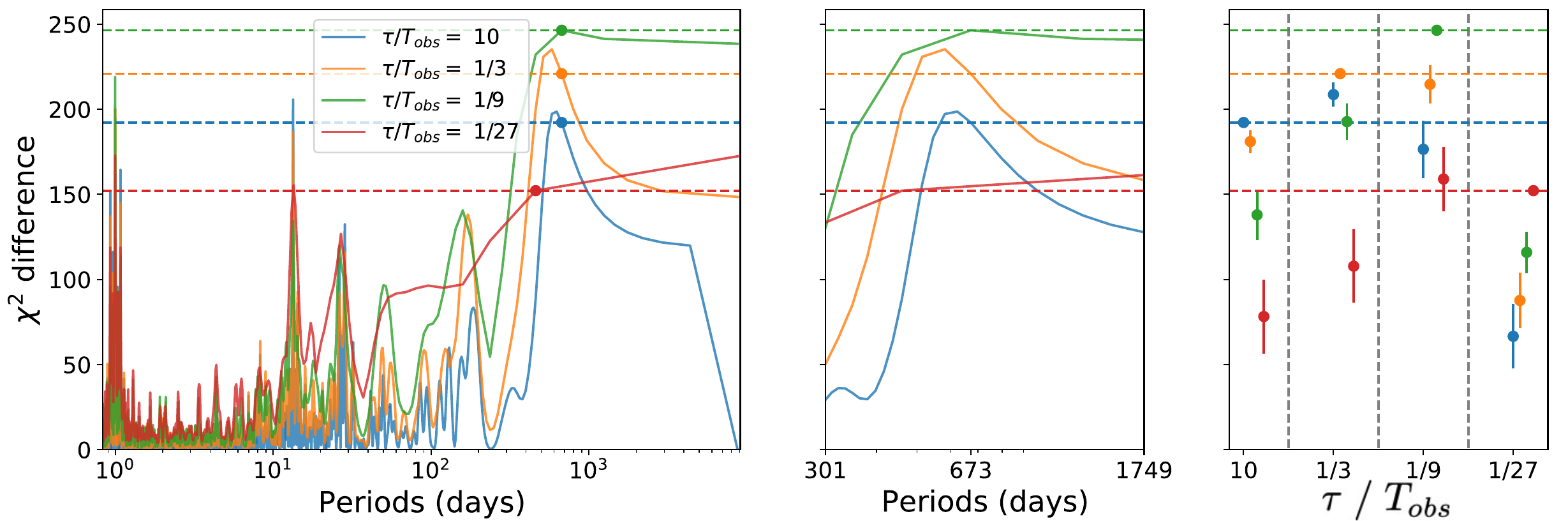}}; %
	\path (0.,4.9) node[above right]{\large(a)};
	\begin{scope}[yshift=-5.7cm]
	\path (0,0) node[above right]{\includegraphics[scale=0.62]{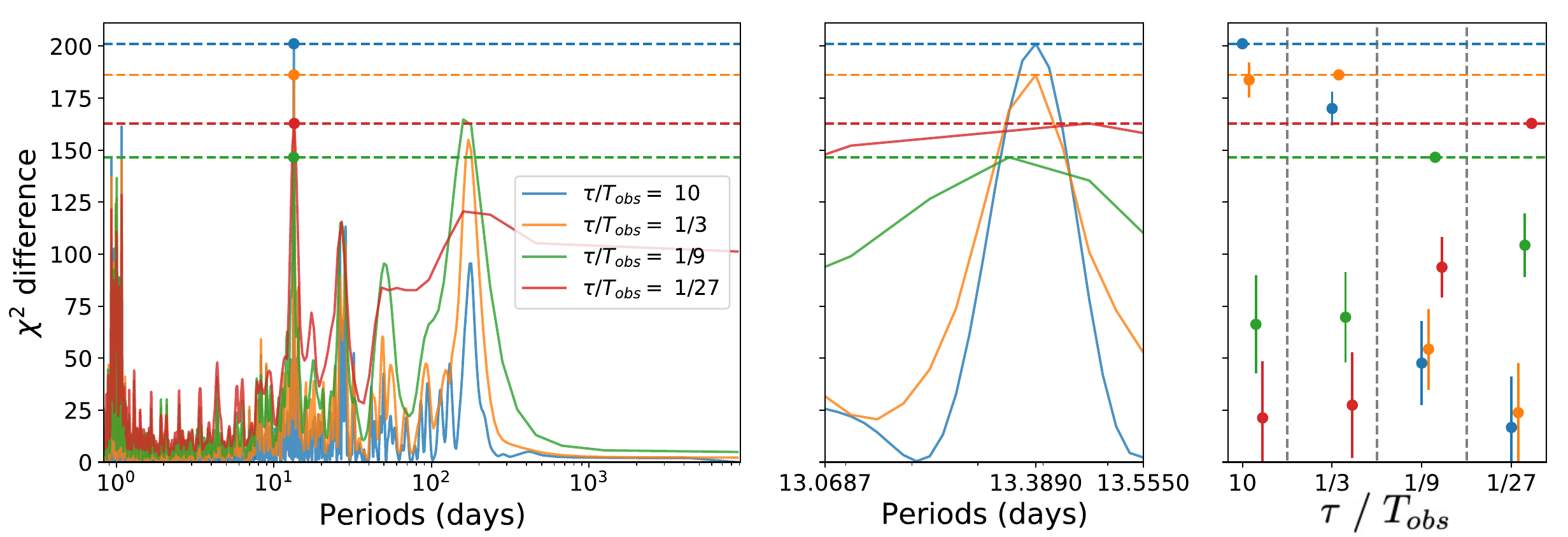}};
	\path (0.,4.9) node[above right]{\large(b)};
	\end{scope}
	\begin{scope}[yshift=-11.4cm]
	\path (0,0) node[above right]{\includegraphics[scale=0.62]{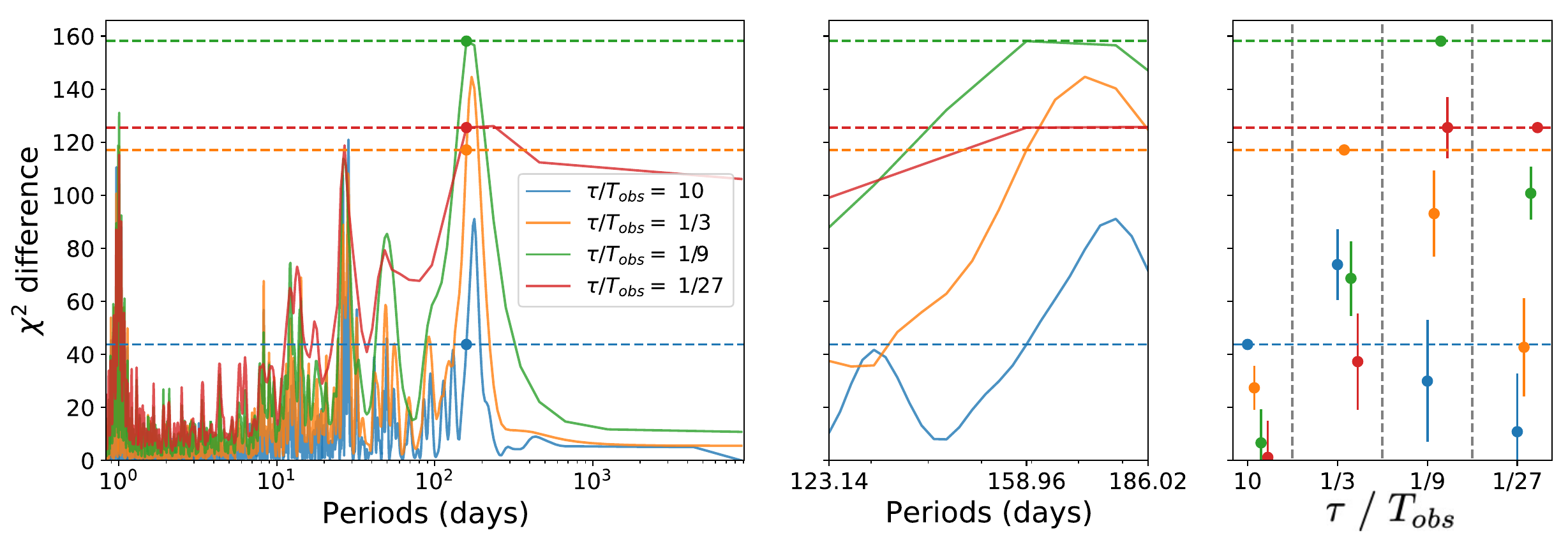}};
	\path (0,4.9) node[above right]{\large(c)};
	\end{scope}
	\begin{scope}[yshift=-17.1cm]
	\path (0.1,0) node[above right]{\includegraphics[scale=0.62]{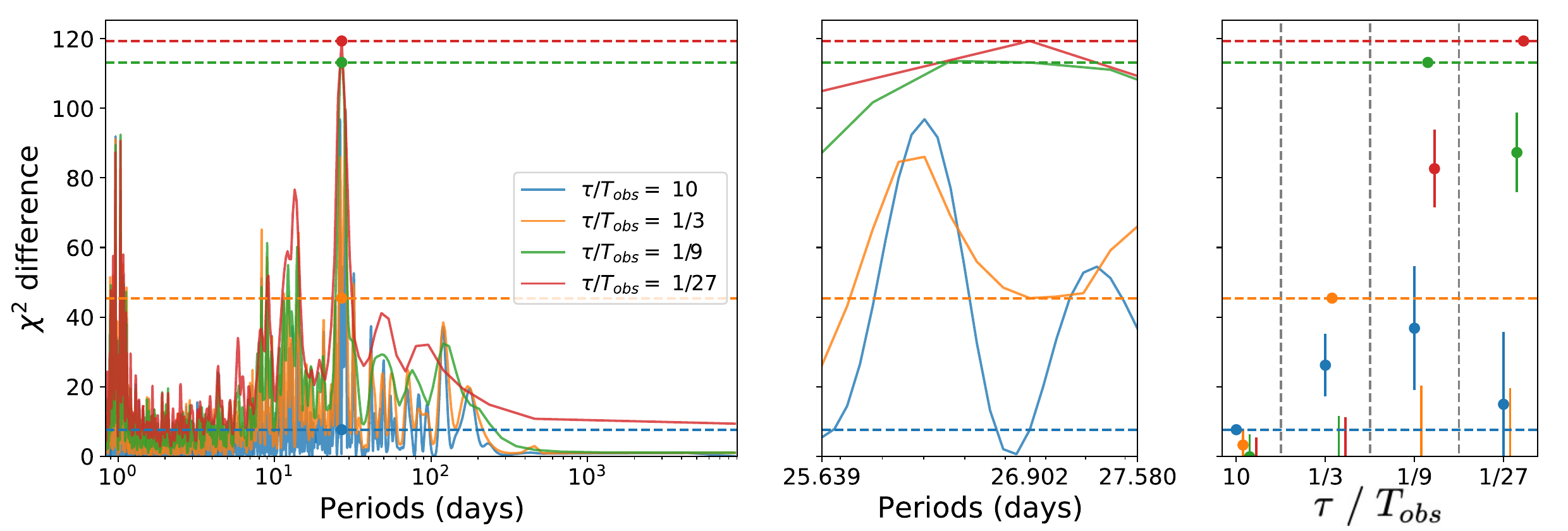}};
	\path (0,4.9) node[above right]{\large(d)};
	\end{scope}
	\end{tikzpicture}
	\caption{Each row correspond to \ch{Apodised sine periodograms} (ASPs) computed iteratively on the Solar data. Left column: ASPs corresponding to equation~\eqref{eq:zm} for different values of $\tau$.  Middle column: zoom on the maximum peak of the ASP. Right: statistical test on the time-scale described in Section \ref{sec:stat}. \ch{On the $x$ axis, we represent the ratio of $\tau$ to the total observation time assumed to be true. For each $\tau_i, i=1..4$ in abscissa, the position of the markers corresponding to $\tau_j, j=1..4$ is represented as the value of the periodogram peak for $\tau_i$  minus the expectancy of $D_z = z(\omega, t_{\tau,\omega}, \tau_i) - z(\omega, t_{\tau_j,\omega}, \tau_j)$ and the error bar is the square root of the variance of $D_z$. Informally, the markers represent what the periodogram values at $\tau_j$ should be if $\tau_i$ in abscissa was the true timescale. } }
	\label{fig:solardata_perios} 
\end{figure*}
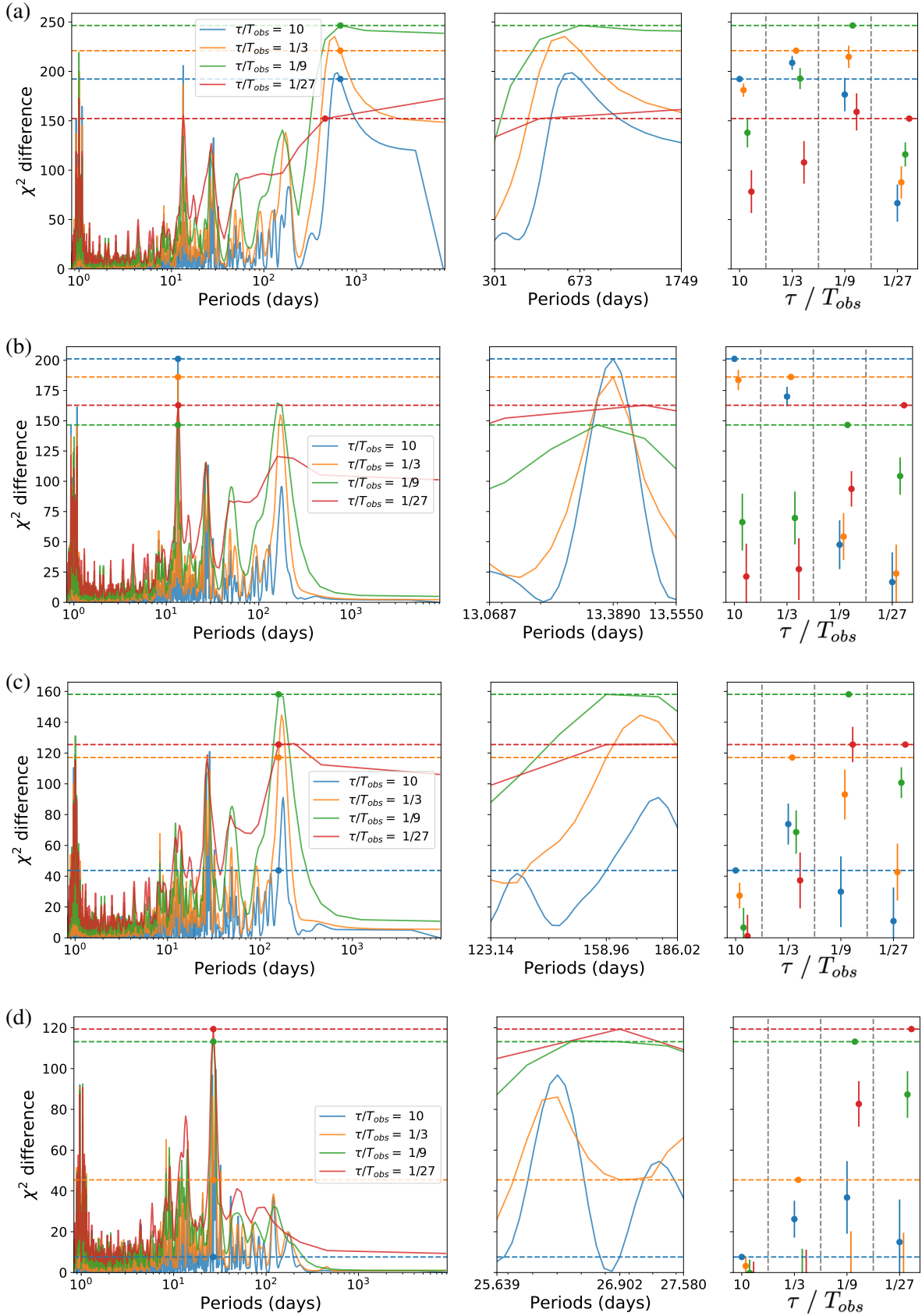

	\subsection{Period consistency }
	\label{sec:period}
	
	Assuming a sinusoidal shape of the signal, the last parameter of an apparently periodic signal whose consistency in time can be evaluated is the period. A convenient representation is, for a given time-scale $\tau$, to represent as a colormap $z(\omega, t_0, \tau)$ as defined in eq.~\eqref{eq:z}. For a given quasi-periodic signal, a local change of the $\chi^2$ value might be due to a clustering or spacing of the observation time. A convenient representation to evaluate jointly period and amplitude consistency is to represent the fitted semi-amplitude $K :=\sqrt{A^2 + B^2}$ (see Eq.~\eqref{eq:mu1} as a function of $\omega$ and $t_0$. To give meaningful diagnostics, it might be suitable to mask the values of $K$ with an uncertainty above a certain threshold. 
	
	We suggest another test. We consider a frequency range $[\omega_l,\omega_r]$ containing the frequency of a candidate periodic signal $\omega_0$. For a given time-scale $\tau$ we plot 
	\begin{align}
	\omega^\star (t_0) := \arg \max\limits_{ \omega \in [\omega_l,\omega_r]} z(\omega, t_0, \tau)
	\label{eq:periodmax}
	\end{align}
	as a function of $t_0$, where $w_B$ is defined by eq.~\eqref{eq:mu1} and~\eqref{eq:box}. 
	
	Like in section~\ref{sec:amphase}, one can test the hypothesis that approximately independent measurements of $\omega^\star$ stem from the same distribution.  
	One can envision further generalisation where one looks for shape variations of periodic signals. However, in the cases considered here, the signal to noise ratio would not allow to conclude on this aspect. As a consequence, we do not consider it here. 
	
	We note that estimating the local phase, amplitude and frequency of a signal is a problem also encountered in signal demodulation in the context of telecommunications~\citep[e. g.][]{madhow2008}. Phase consistency tests have been suggested in the context of discrete events, more specifically $\gamma$ ray pulsars, to detect phase shifts relative to an expected position \citep{jager1994}.
	

	\section{Applications}
	\label{sec:applications}
	\subsection{Solar data}
	\label{sec:sun}
		\begin{figure}[!h]
		\noindent
		\centering
		\hspace{-1,5cm}
		\begin{tikzpicture}		
		\path (0,0) node[above right]{\includegraphics[width=0.8\linewidth]{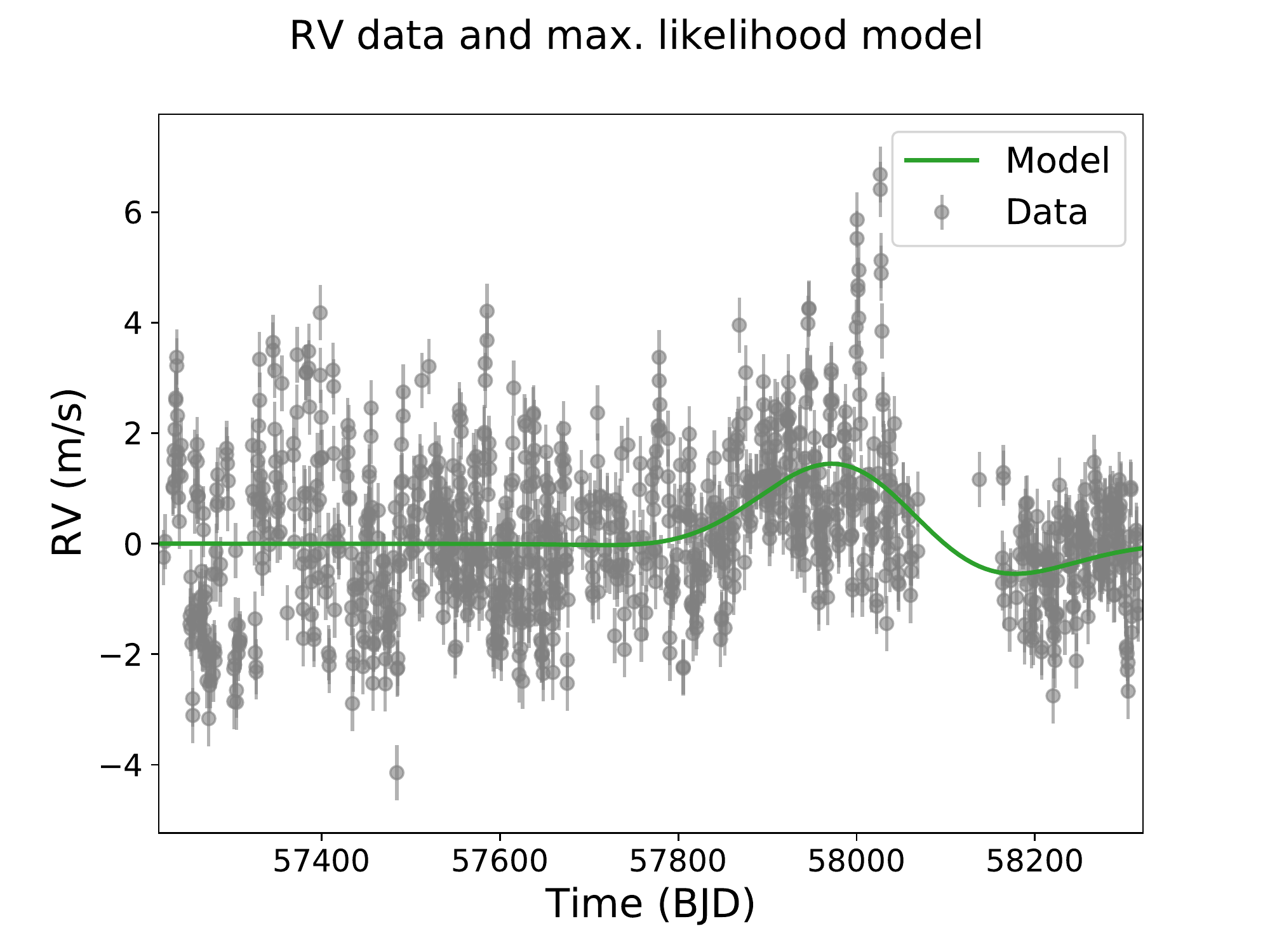}};
		\path (0.15,4.5) node[above right]{\large(a)};
		\begin{scope}[yshift=-5.5cm]
		\path (0,0) node[above right]{\includegraphics[width=0.8\linewidth]{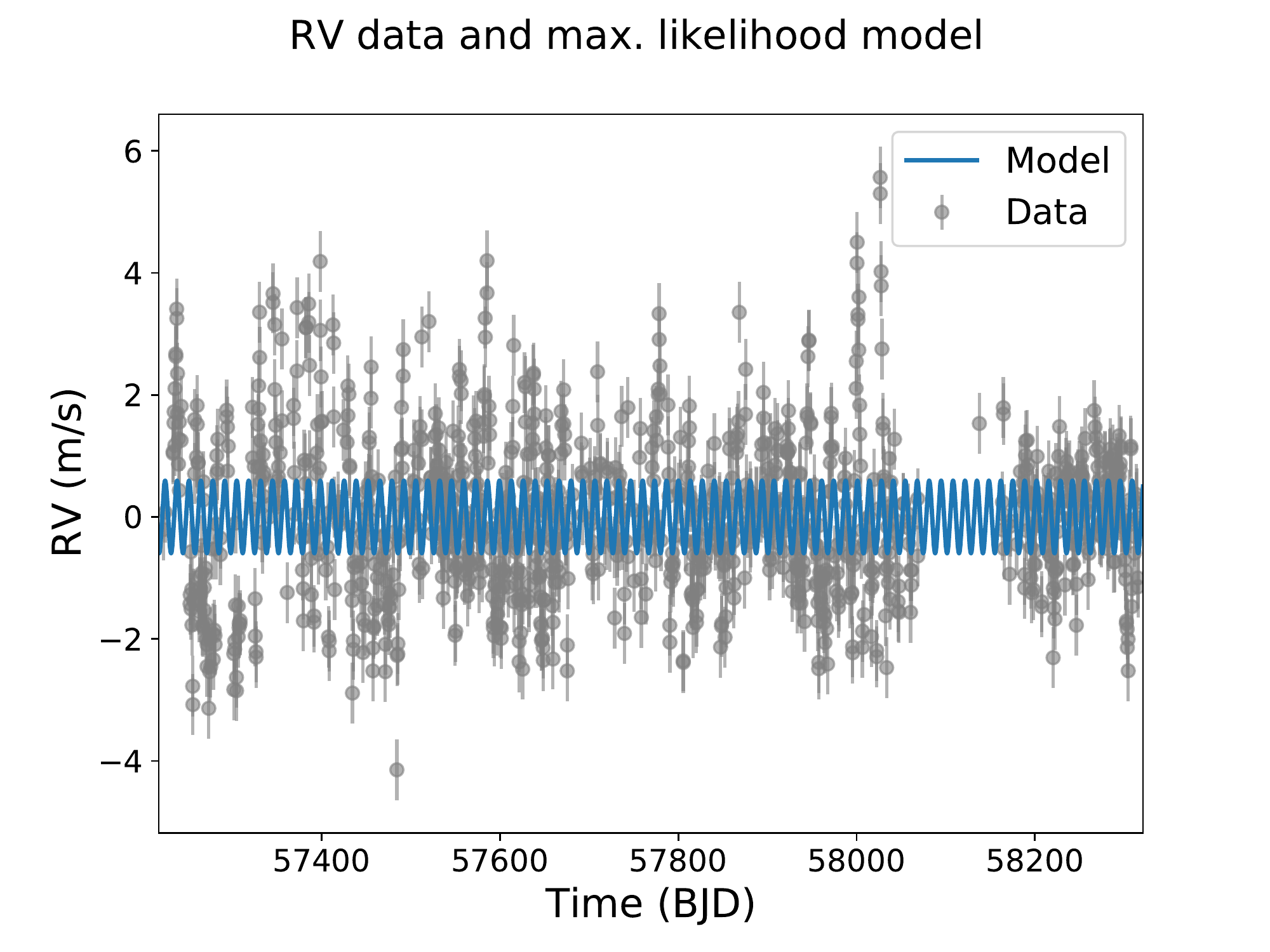}};
		\path (0.15,4.5) node[above right]{\large(b)};
		\end{scope}
		\begin{scope}[yshift=-11cm]
		\path (0,0) node[above right]{\includegraphics[width=0.8\linewidth]{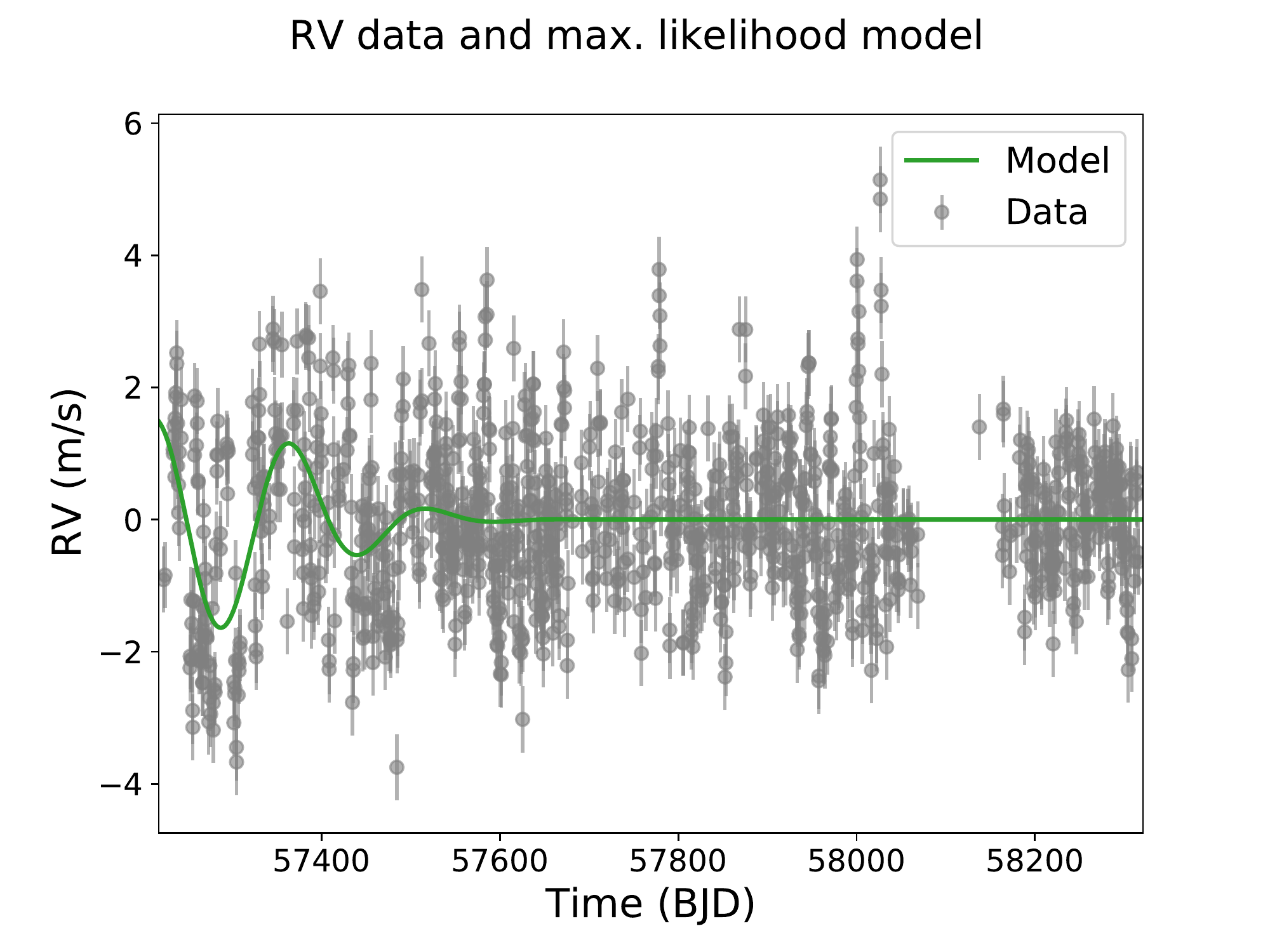}};
		\path (0.15,4.5) node[above right]{\large(c)};
		\end{scope}
		\begin{scope}[yshift=-16.5cm]
		\path (0,0) node[above right]{\includegraphics[width=0.8\linewidth]{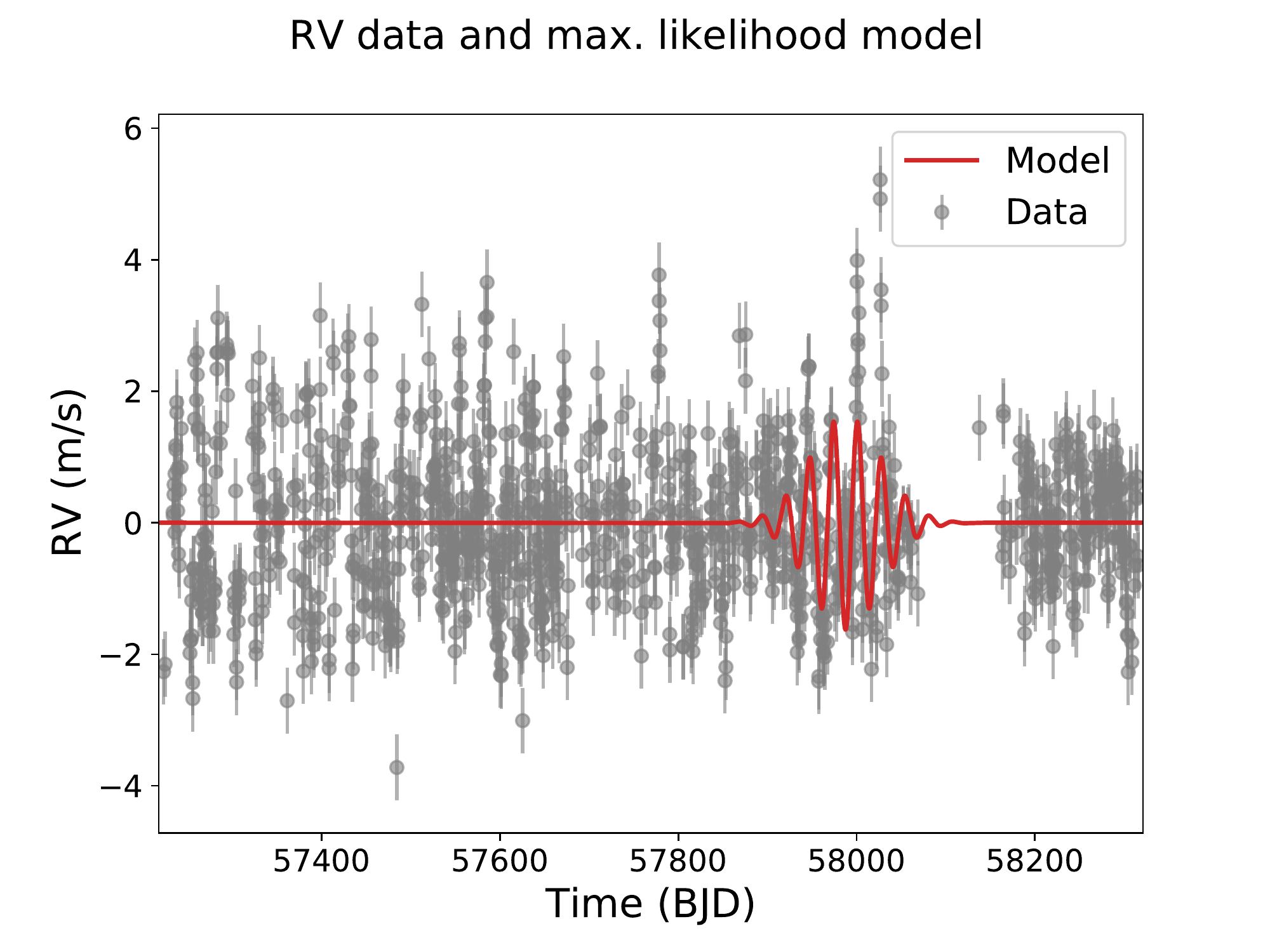}};
		\path (0.15,4.5) node[above right]{\large(d)};
		\end{scope}
		\end{tikzpicture}
		\caption{Models corresponding of the maximum of the ASPs of Fig. \ref{fig:solardata_perios} }
		\label{fig:solardata_models}
	\end{figure}
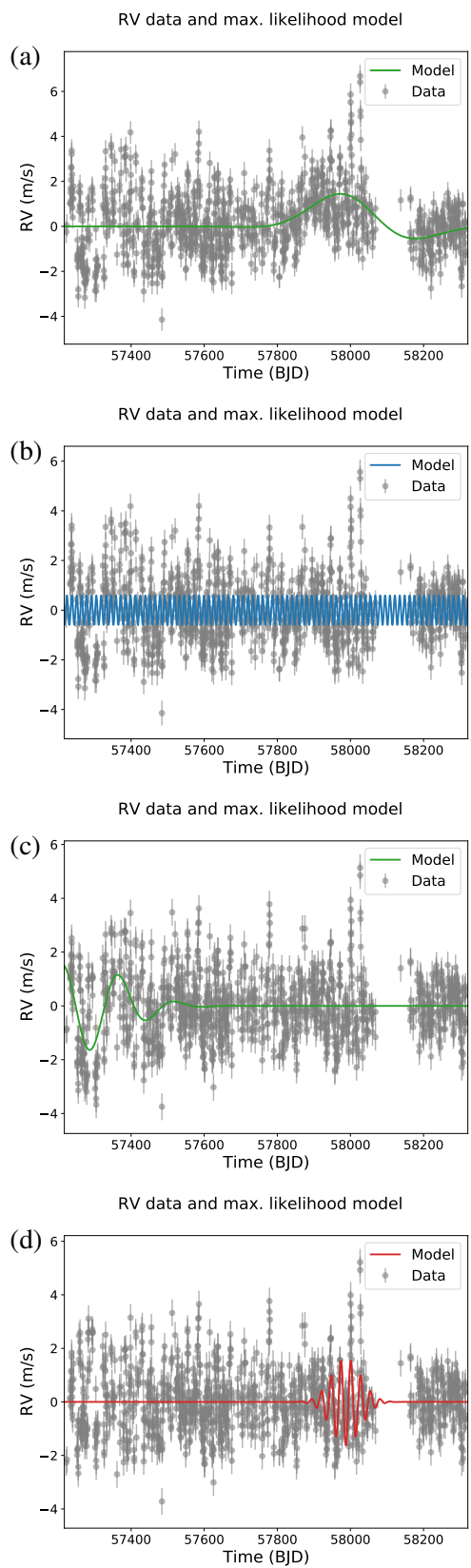

\begin{figure*}
	\noindent
	\centering
	\begin{tikzpicture}
	\path (0,0) node[above right]{\includegraphics[scale=0.4]{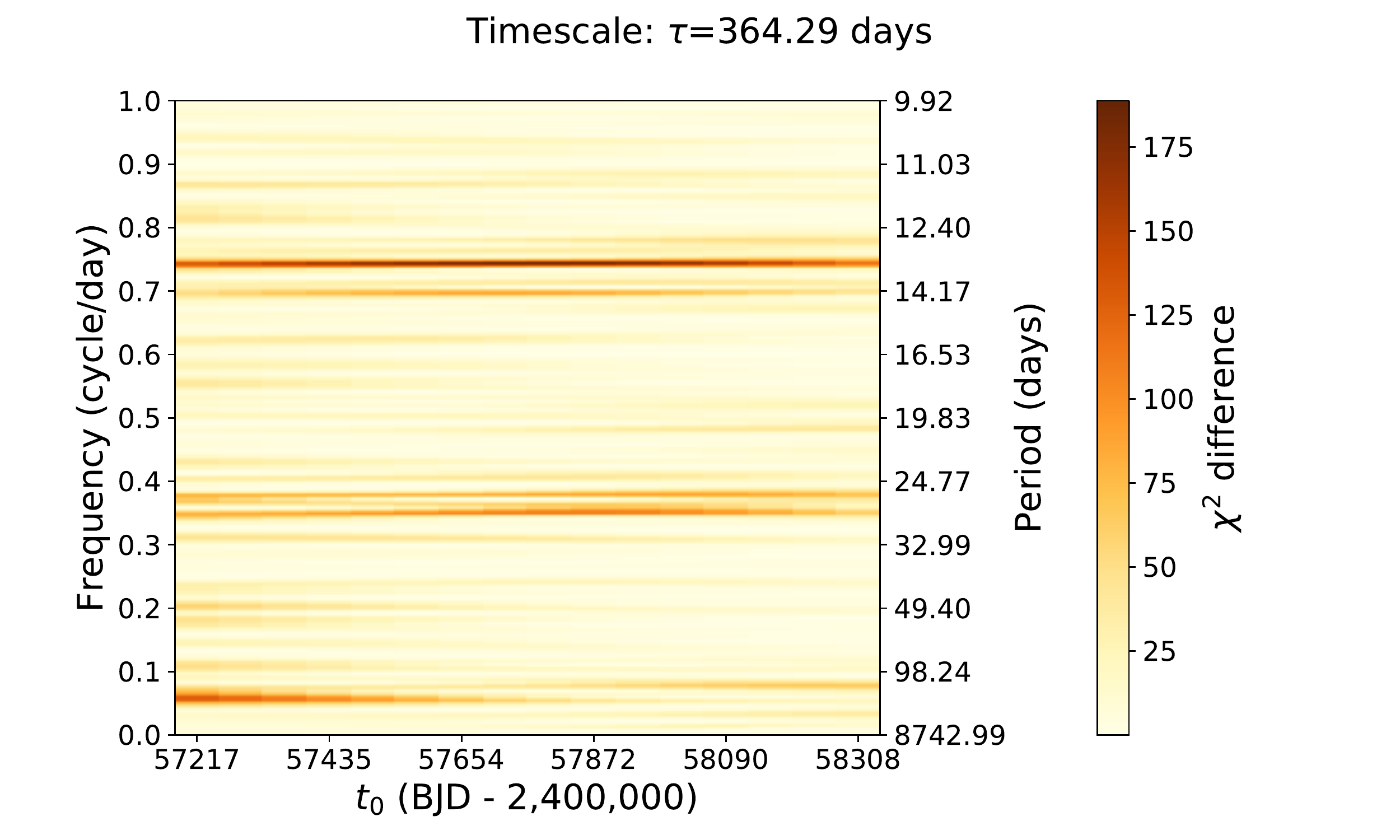}};
	\path (1.2,6) node[above right]{\large(a)};
	\path (9.5,0) node[above right]{\includegraphics[scale=0.4]{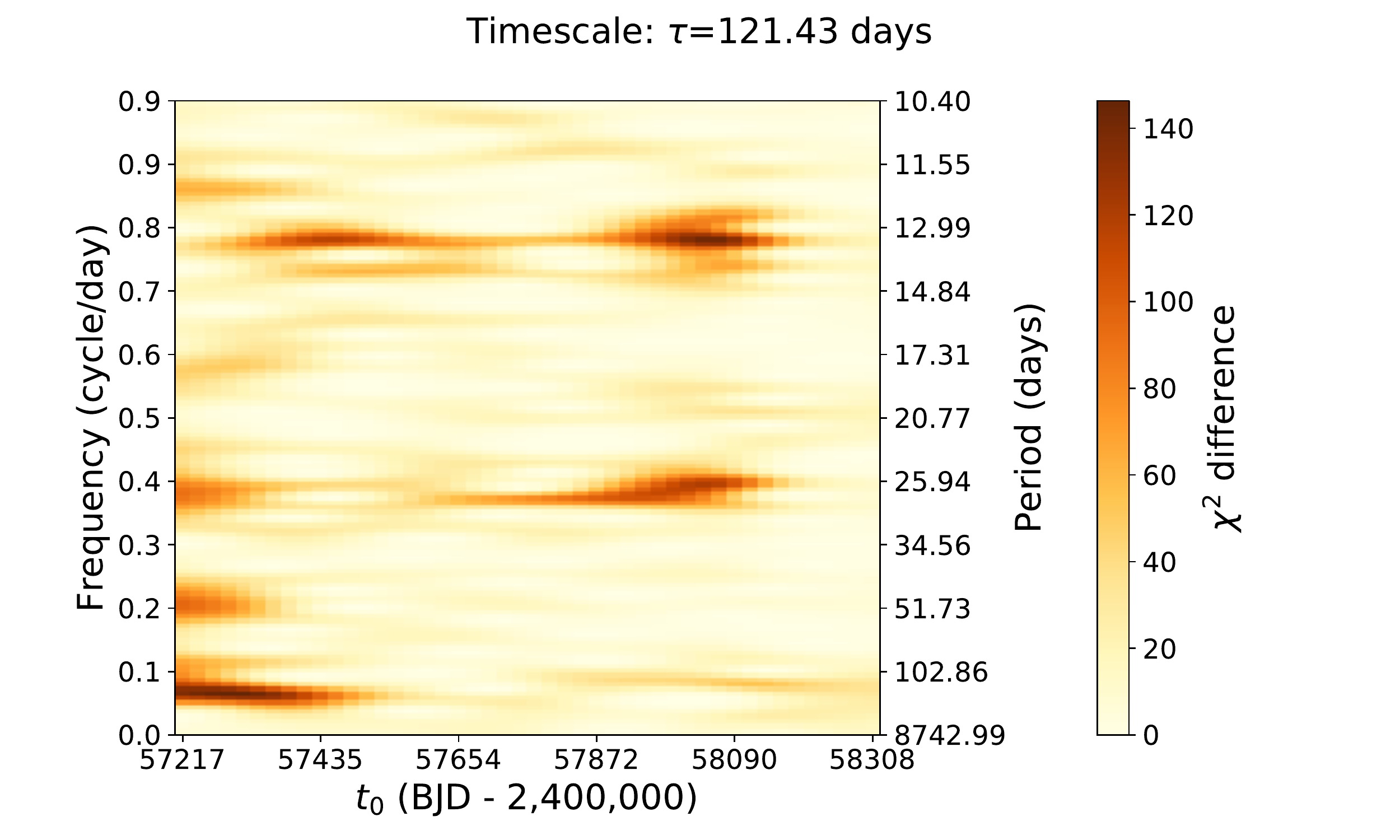}};
	\path (10,6) node[above right]{\large(b)};
	
	\begin{scope}[yshift=-6.2cm]
	\path (0.33,0) node[above right]{\includegraphics[scale=0.34]{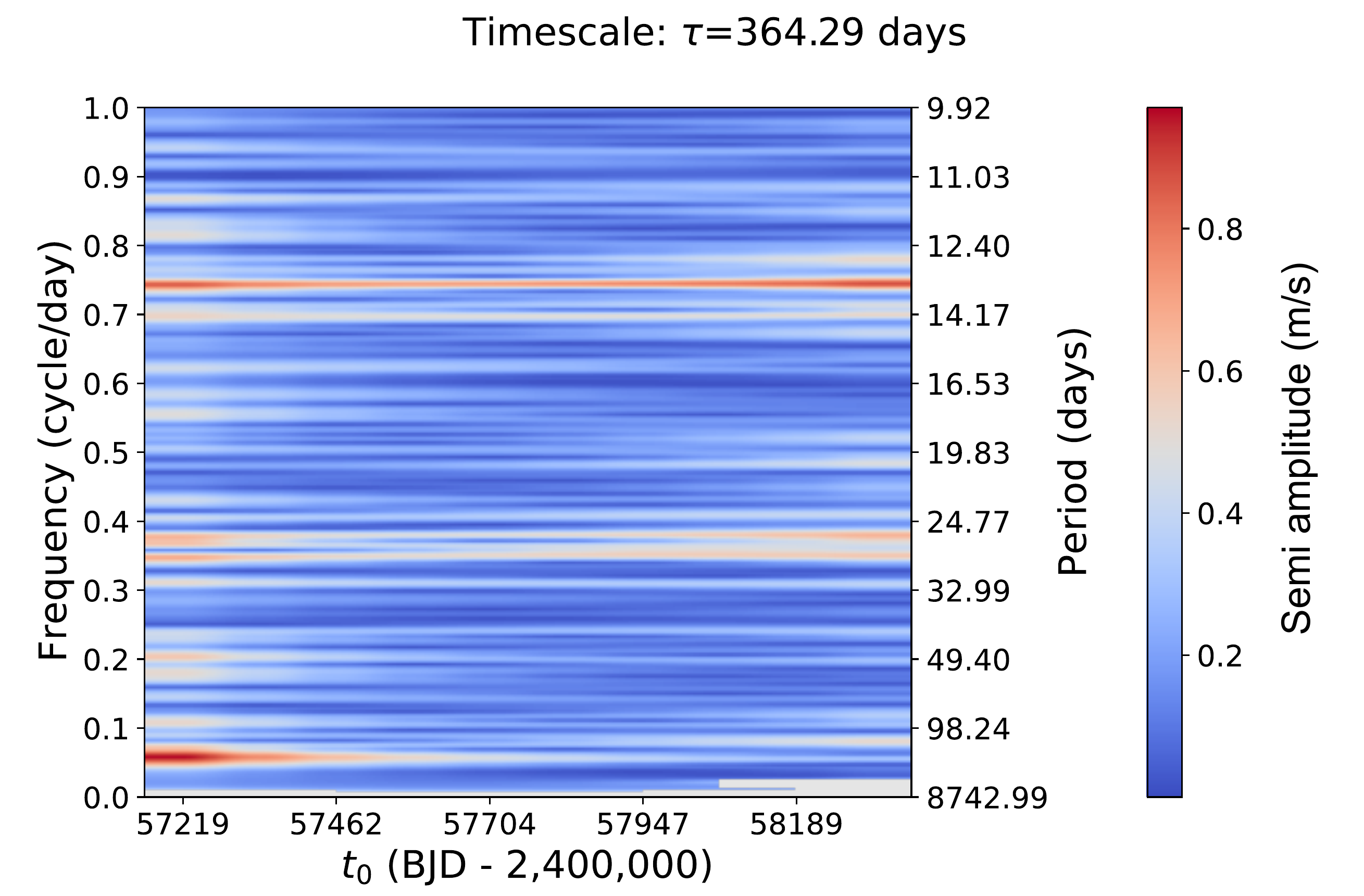}};
	\path (1.2,6) node[above right]{\large(c)};
	
	\path (9.83,0) node[above right]{\includegraphics[scale=0.34]{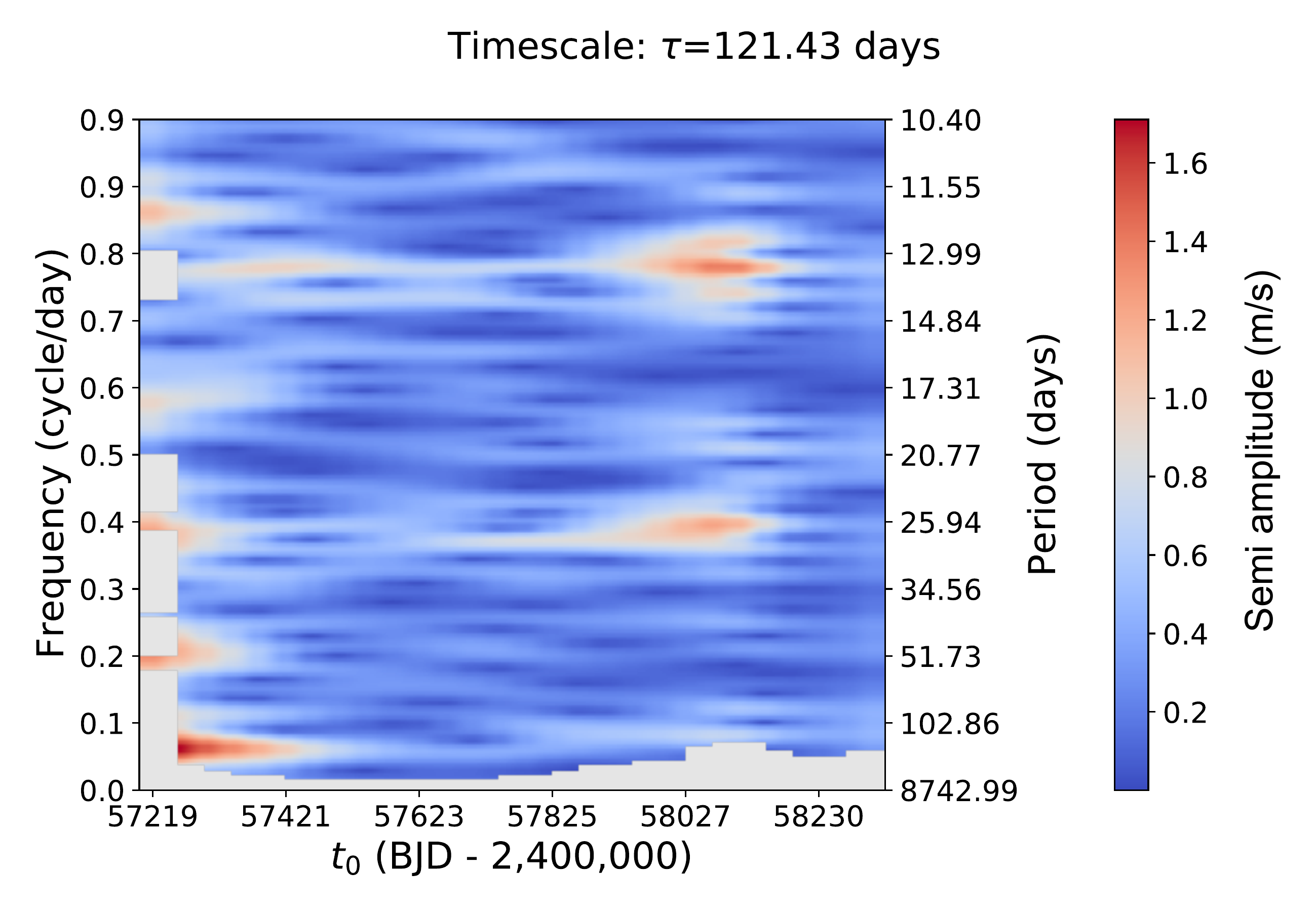}};
	\path (10,6) node[above right]{\large(d)};
	\end{scope}

	\begin{scope}[yshift=-13cm]
	\path (0,0) node[above right]{\includegraphics[scale=0.46]{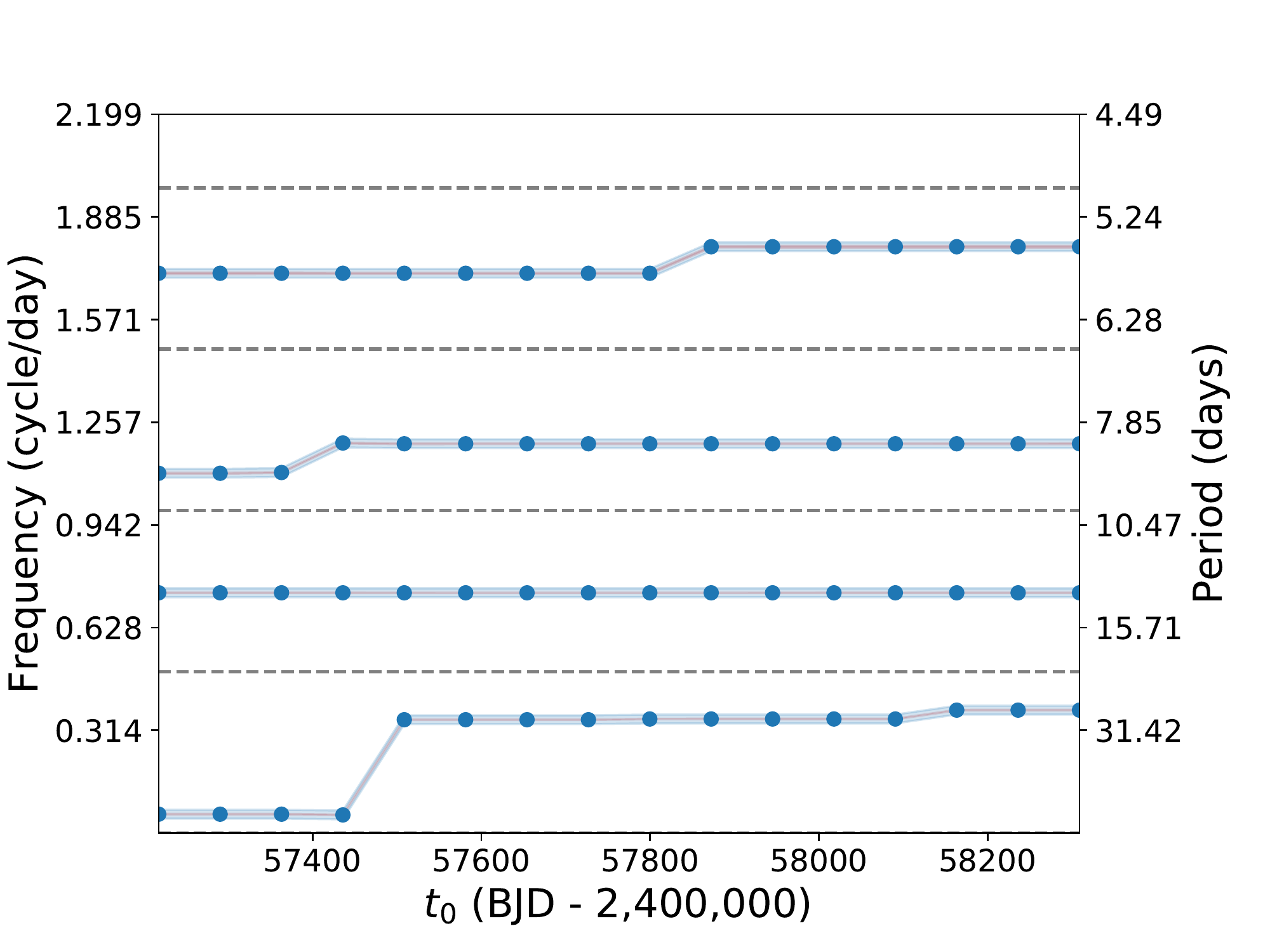}};
	\path (1.2,6.6) node[above right]{\large(e)};
	
	\path (9.5,0) node[above right]{\includegraphics[scale=0.46]{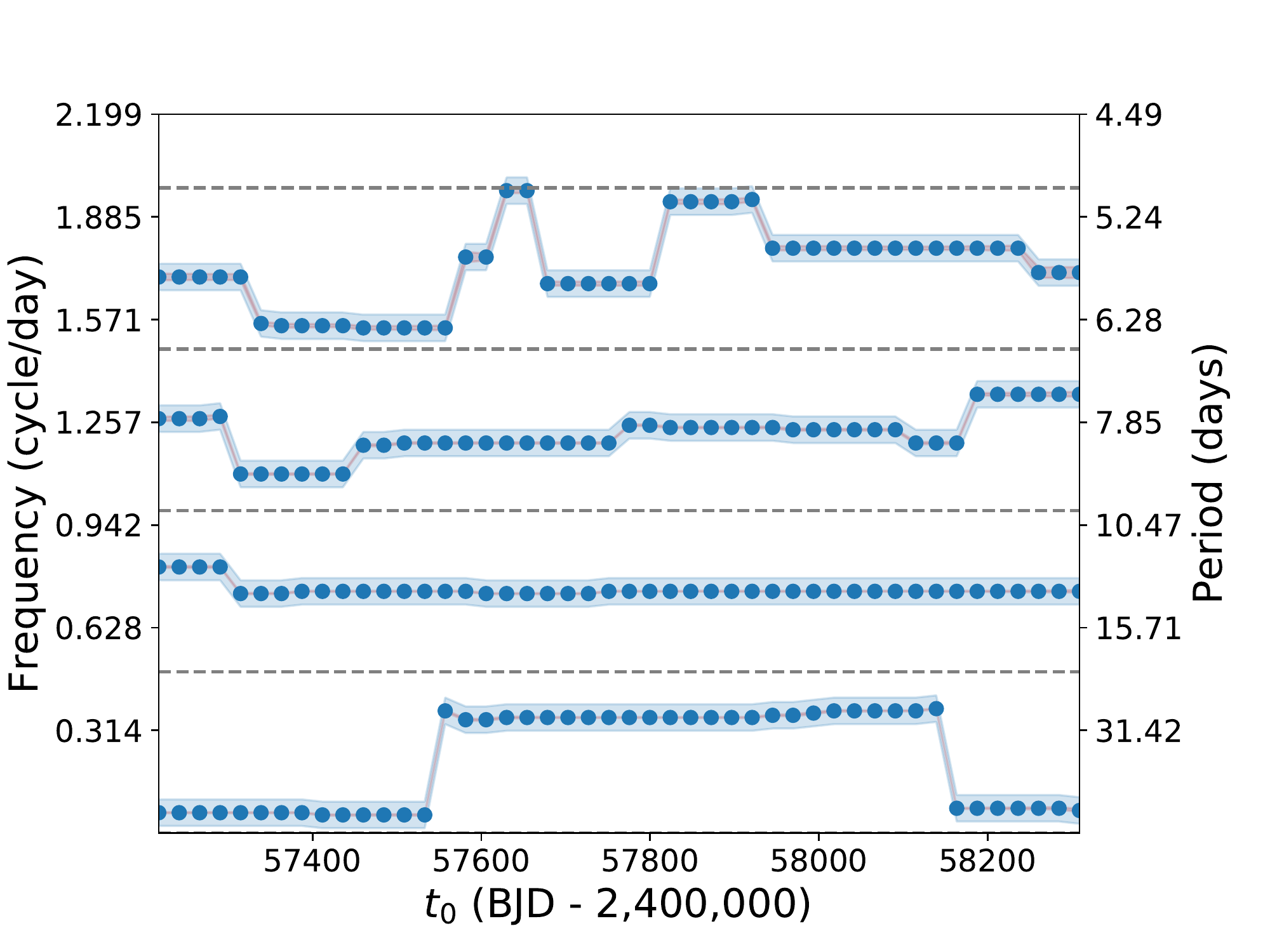}};
	\path (10,6.6) node[above right]{\large(f)};
	\end{scope}
	\end{tikzpicture}
	\caption{(a) and (b): value of the $\chi^2$ defined in Eq.~\eqref{eq:z} for the Solar HARPS-N RV data as a function of $t_0$ for $\tau=365$ (=$T_\mathrm{obs}/3$) days (a)) and $\tau=121$  (=$T_\mathrm{obs}/9$) days  (b)). (c) and (d) value of the semi amplitude as a function of $t_0$ for $\tau=365$ (=$T_\mathrm{obs}/3$) days (c) and $\tau=121$  (=$T_\mathrm{obs}/9$) days  (b). Grey areas correspond to 1 $\sigma$ uncertainties greater than 30 cm/s. (e) and (f) Blue points represent the position of the mode of the ASP between frequencies represented in grey as a function of $t_0$, as defined in Section~\ref{sec:period}. Computed for   $\tau=365$ (=$T_\mathrm{obs}/3$) days (c) and $\tau=121$  (=$T_\mathrm{obs}/9$) days  (d).  }
	\label{fig:solardata_period}
\end{figure*}
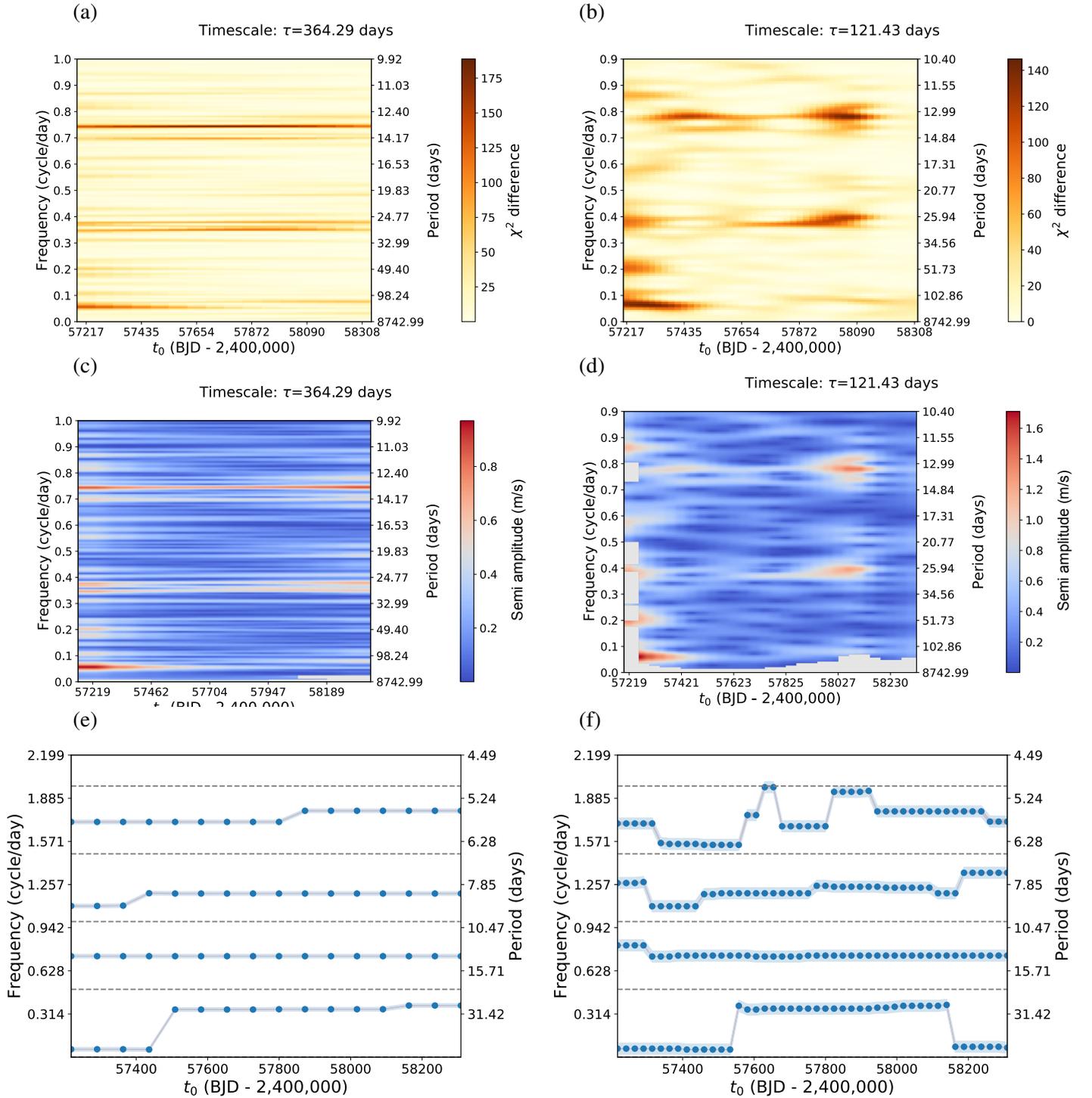

\begin{figure*}
	\noindent
	\centering
	\begin{tikzpicture}
	\path (0,0) node[above right]{\includegraphics[scale=0.46]{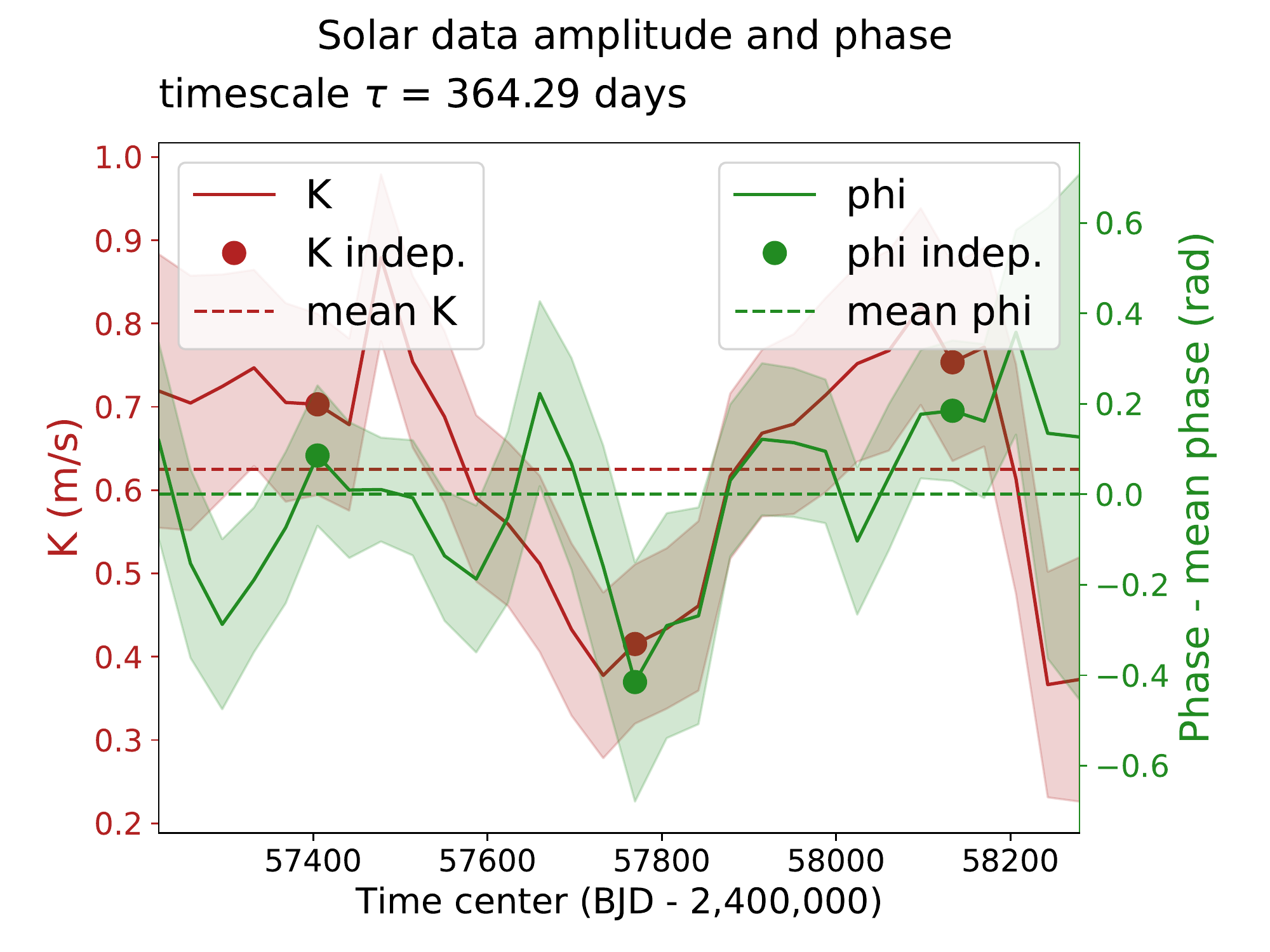}};
	\path (1.2,6.6) node[above right]{\large(a)};
	\path (9.5,0) node[above right]{\includegraphics[scale=0.46]{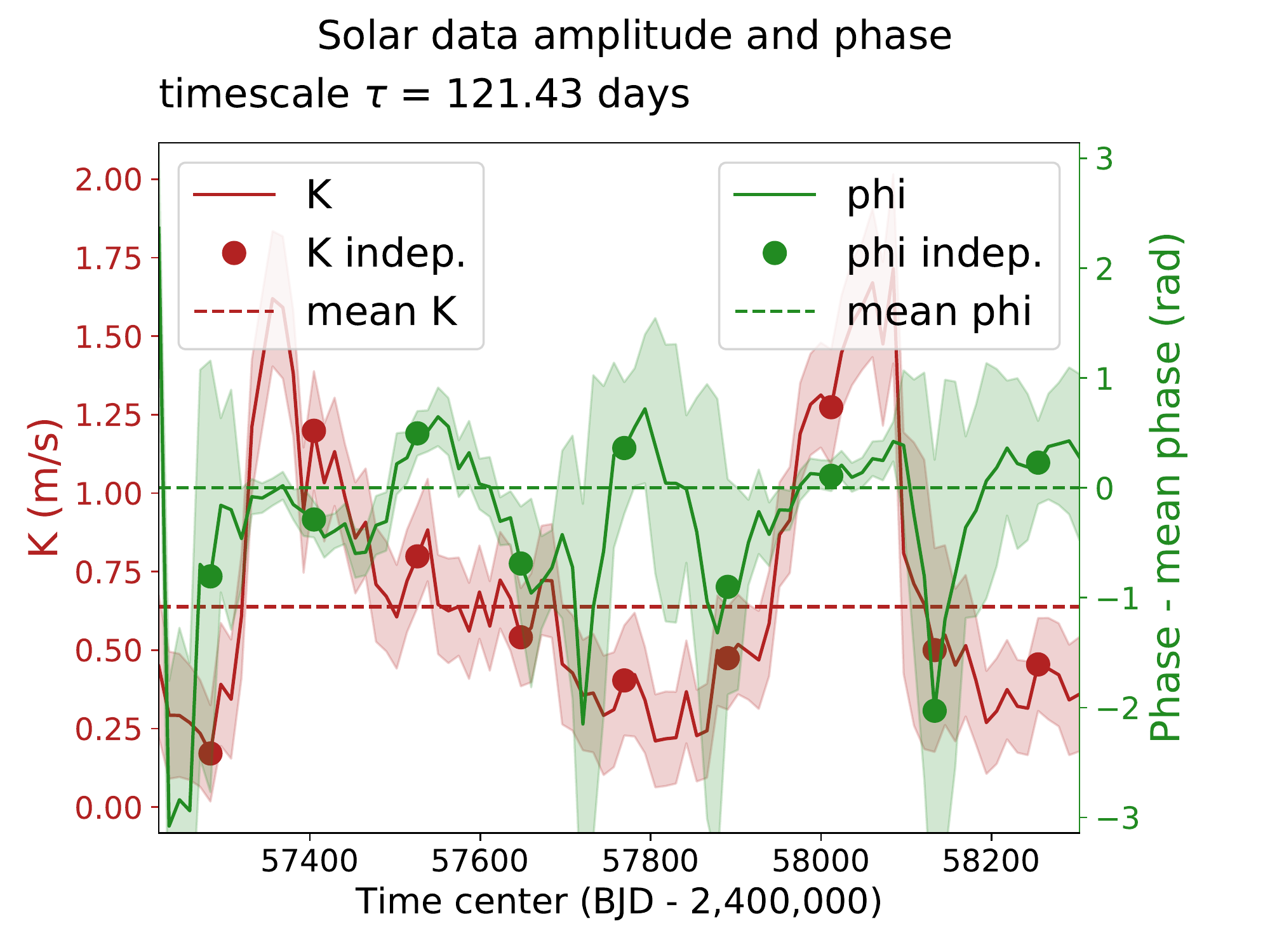}};
	\path (10,6.6) node[above right]{\large(b)};
	
	\begin{scope}[yshift=-7cm]
	\path (0,0) node[above right]{\includegraphics[scale=0.46]{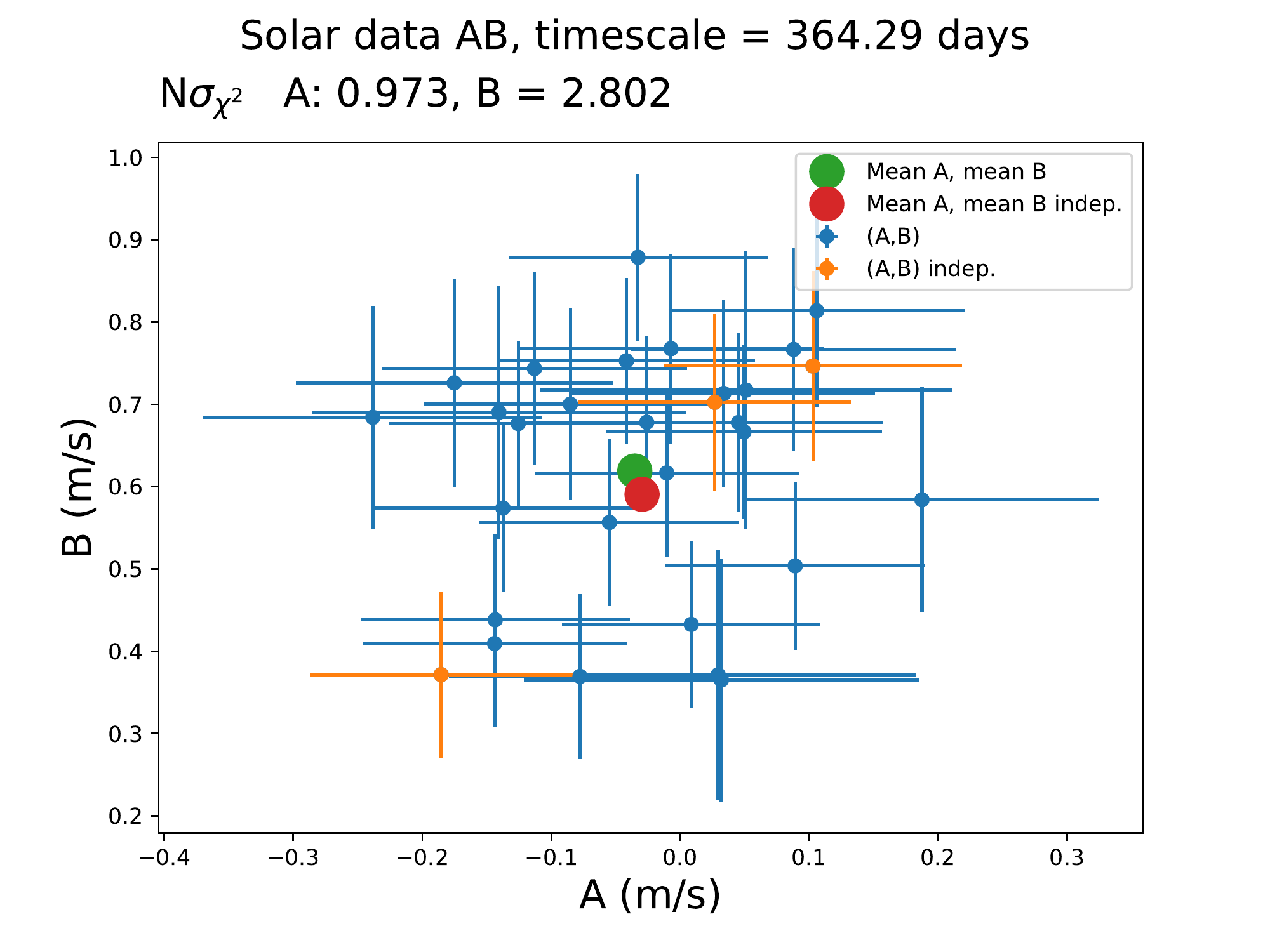}};
	\path (1.2,6.6) node[above right]{\large(c)};
	
	\path (9.5,0) node[above right]{\includegraphics[scale=0.46]{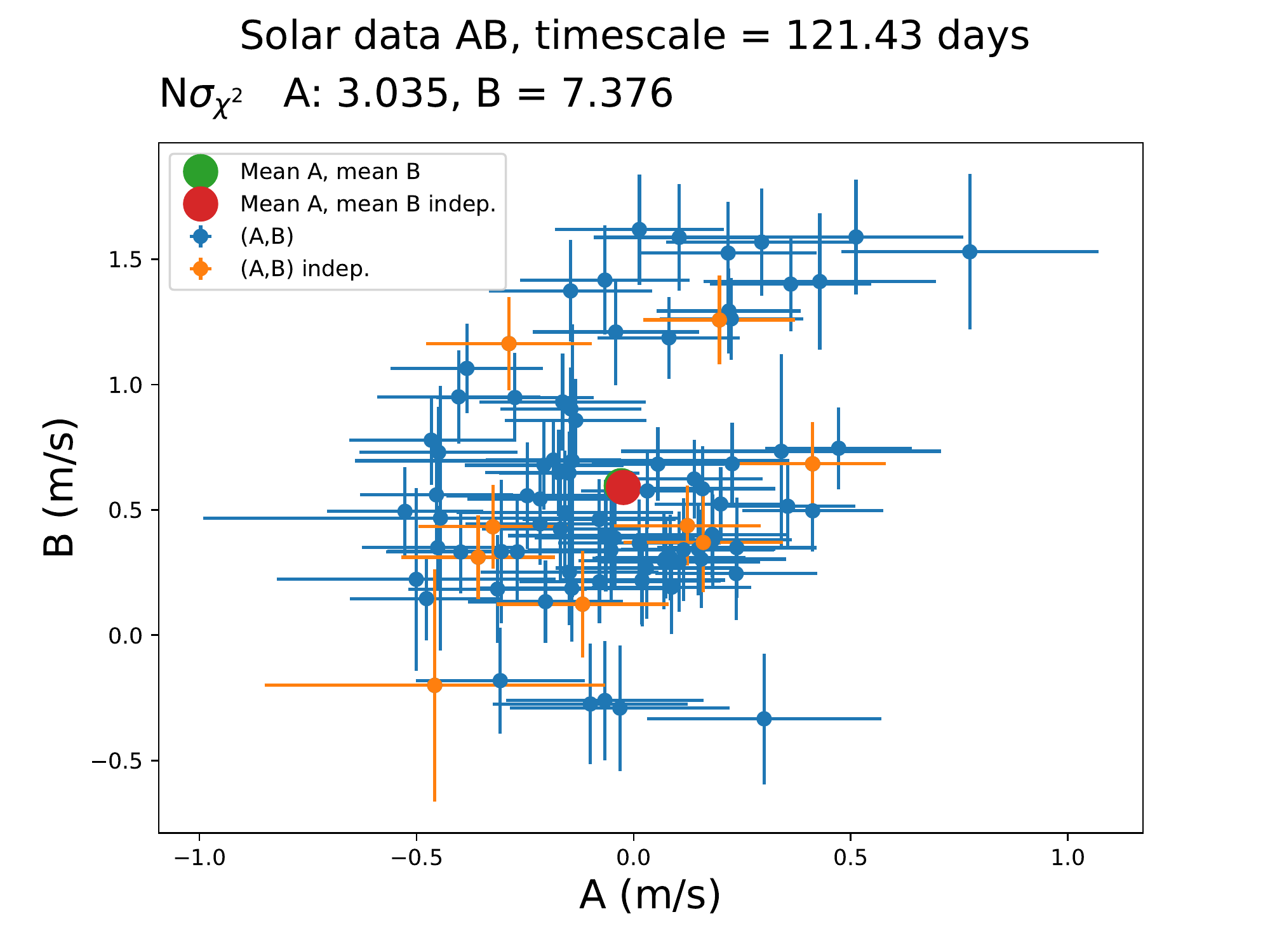}};
	\path (10,6.6) node[above right]{\large(d)};
	\end{scope}
	\end{tikzpicture}
	\caption{Phase and amplitude \ch{of the 13.39 days signal} as a function of the time center of the window $t_0$ for the Solar HARPS-N data, as described in Section~\ref{sec:amphase} (solid lines, red and green respectively). Markers correspond to statistically independent estimates. Red and green dashed lines represent the average phase and amplitudes. The figures are all computed for a time-scale $\tau =539$ days corresponding to $\tau/T_\mathrm{obs}$  = 1/9. Plots (a), (b), (c), (d) are computed by adding iteratively the signals corresponding to the highest peaks to the base model.  }
	\label{fig:solardata_amphase}
\end{figure*}
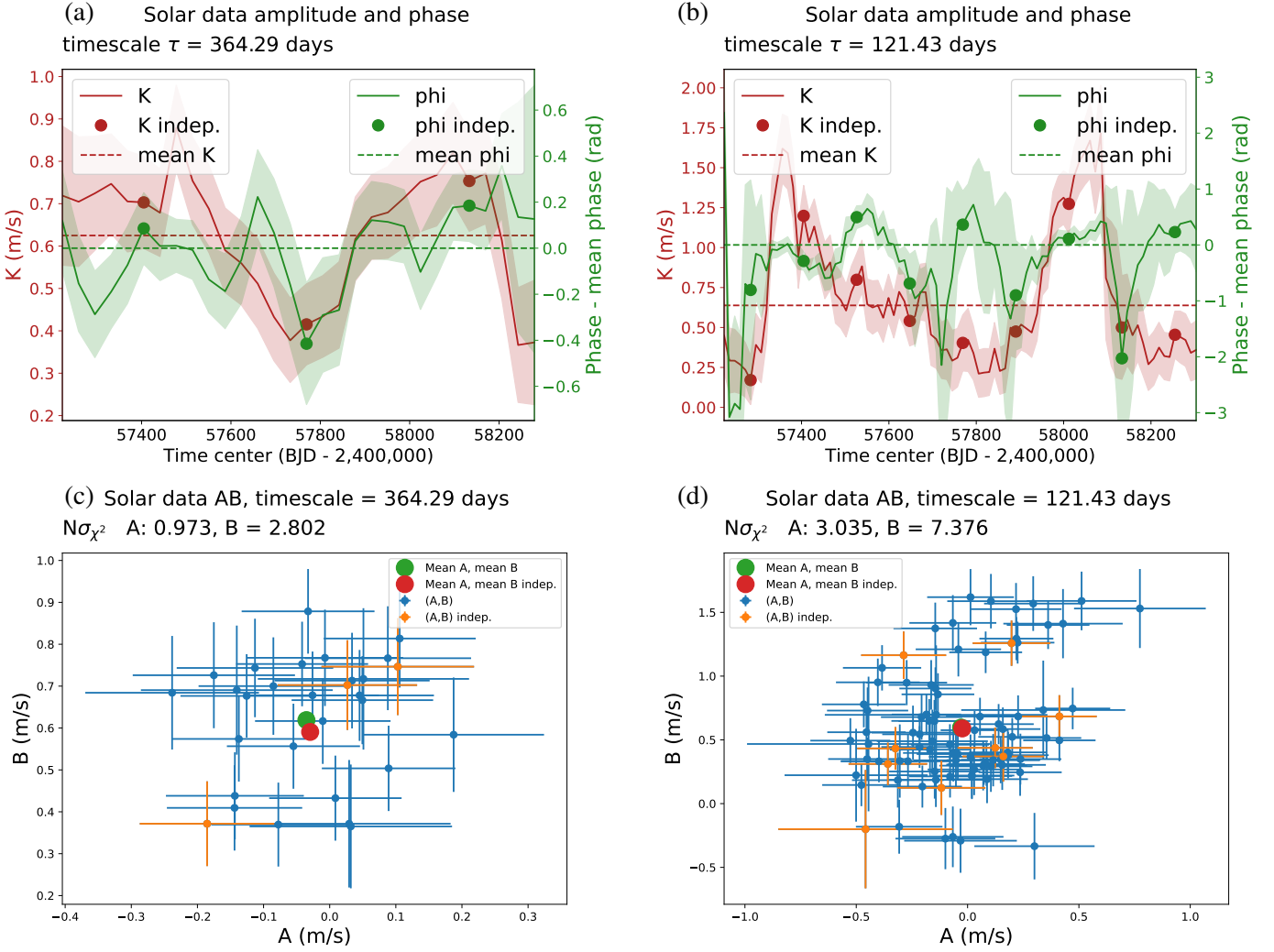

\begin{figure}
\centering
	\includegraphics[width=\linewidth]{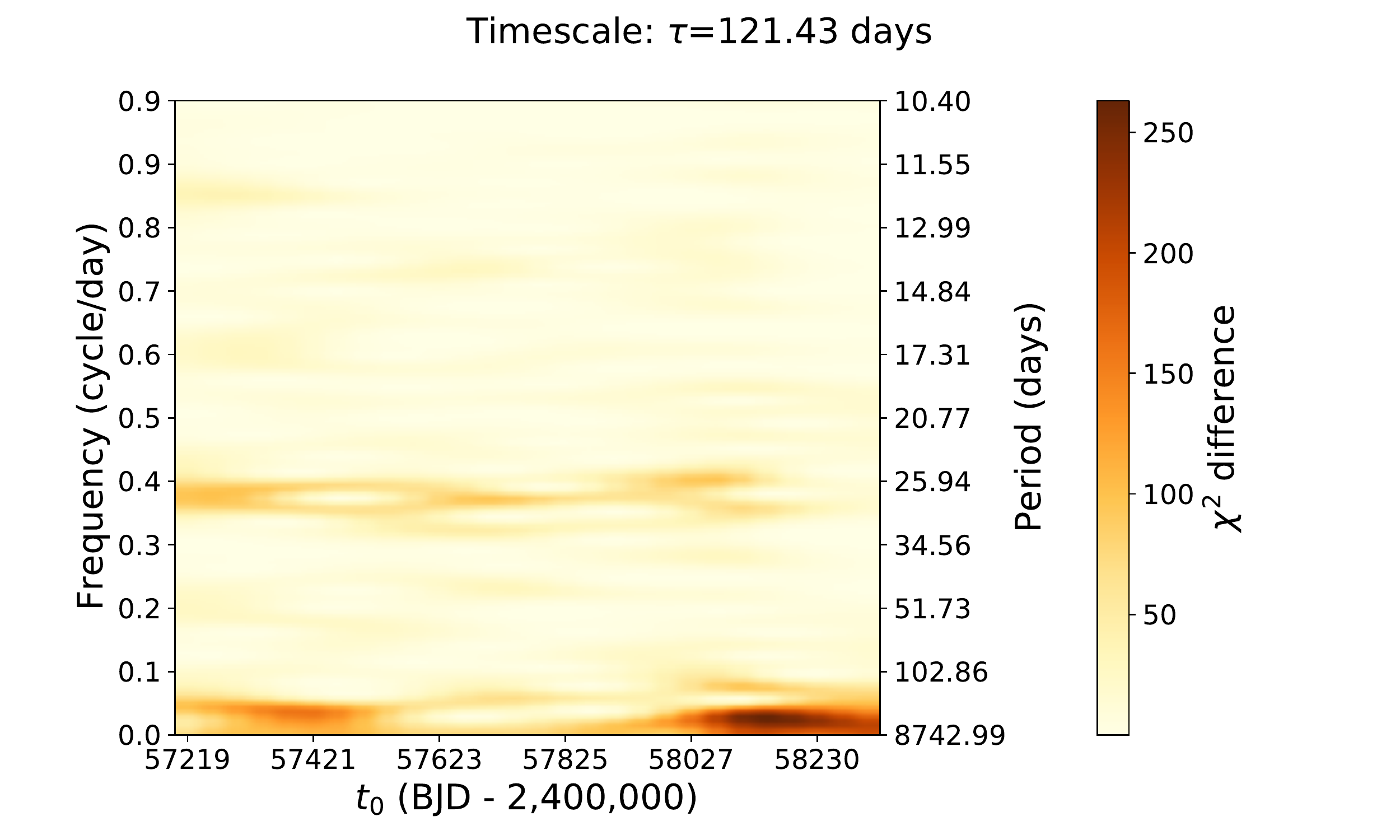}
	\caption{value of the $\chi^2$ defined in Eq.~\eqref{eq:z} for the Solar HARPS-N $\log R'_{HK}$ as a function of $t_0$ for $\tau 121$ (=$T_\mathrm{obs}/9$) days}
	\label{fig:chi2rhk}
\end{figure}	
\begin{figure}
\centering
	\includegraphics[width=\linewidth]{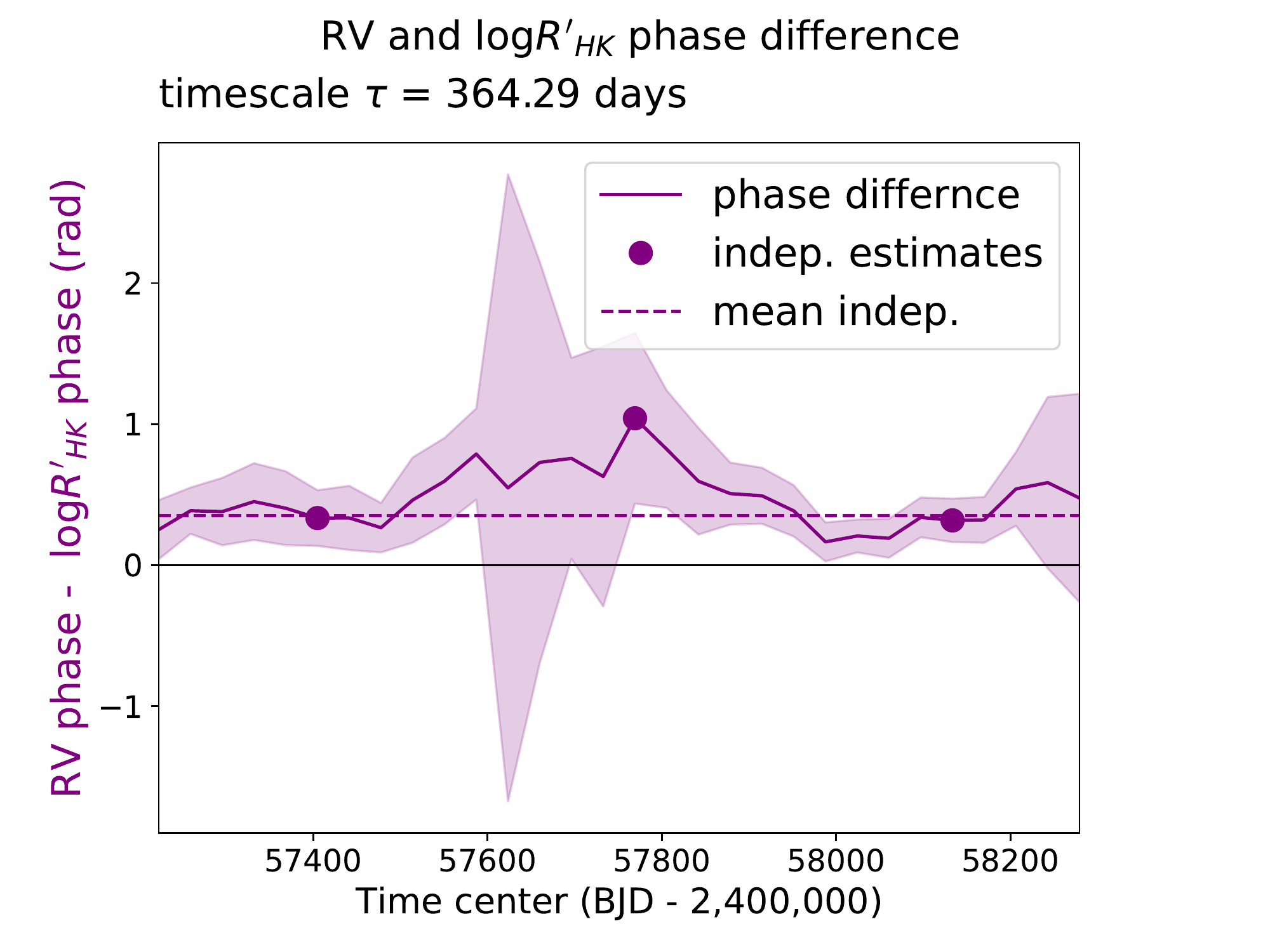}
	\caption{Difference between the RV phase and $\log R'_{HK}$ phase as a function of time for $\tau$ = $T_{\mathrm{obs}}/3$}
	\label{fig:phaserhk}
\end{figure}

\subsubsection{Data analysis}

In this section, we apply our methods to the HARPS-N Solar data~\citep{dumusque2020}. This dataset contains the radial velocity of the Sun measured with the HARPS-N solar telescope \citep[][]{Dumusque2015}. \ch{This dataset spans three years, from BJD 2457222.1788 to 2458315.9982 and contains 34550 measurements with a median interval between two samples of 5 min 38 s and a median nominal error of 0.22 m/s. }
The data is  publicly available in the DACE platform\footnote{Data can be downloaded from \url{https://dace.unige.ch/sun/?}}. 
We bin the data by half day, which means that data acquired between midnight and noon, and noon and midnight (local time at La Palma observatory) are averaged, weighted by their nominal uncertainties. We consider a Gaussian, uncorrelated noise model, with a 1 m/s jitter added in quadrature to the nominal uncertainties. 

We first perform an iterative search of signals using the ASPs defined in Section~\ref{sec:def}. We consider four time-scales, denoting by $T_\mathrm{obs}$ the total timespan of the observations, we compute Eq.~\eqref{eq:z} for $\tau/T_\mathrm{obs}=10$, 1/3,  1/9 and 1/27.
 The grid in $t_0$ and $\omega$ is defined as in Section~\ref{sec:grid}. We compute the ASP and then add to the base model the model corresponding to the maximum peak.

 Fig~\ref{fig:solardata_perios} (a), left panel shows the first ASP, which presents a maximum at  period $\omega^{(0)}=9000$ days for $\tau^{(0)} = T_\mathrm{obs}/9$ and $t_0^{(0)}$. We then add to the base model $w(\tau^{(0)}, t_0^{(0)})\cos \omega^{(0)} t$ and $w(\tau^{(0)}, t_0^{(0)})\sin \omega^{(0)} t$ and compute the ASP, obtaining Fig.~\ref{fig:solardata_perios} (b). The same operation is repeated, yielding Fig.~\ref{fig:solardata_perios} (c) and (d). In Fig.~\ref{fig:solardata_models} (a,b,c,d), we represent the models that correspond to the highest peaks of Fig.~\ref{fig:solardata_perios} (a,b,c,d, respectively). 
 
  The middle panel of Fig.~\ref{fig:solardata_perios} presents a zoom on the maximum peak of the successive ASP.
Horizontal dashed lines correspond to the values of $z$ at the frequency where the maximum is attained. For instance in (b), the maximum value is attained at period = 13.39 days. The blue, orange, red and green horizontal lines represent the level of the periodogram peaks with  $\tau/T_\mathrm{obs}=10$, 1/3,  1/9 and 1/27 at $P^{(1)} =$13.39 days. \ch{We found empirically that a downsizing factor of 1/3 for $\tau$ allows a reasonable compromise between resolution and grid size.}  Finally, the right panel represents the statistical test presented in Section~\ref{sec:stat}.The four time-scale corresponds to a value of $\tau$, reported in abscissa. Let us consider Fig.~\ref{fig:solardata_perios} (b). For each time-scale $\tau$, we assume that the data contains a signal at frequency $P^{(1)}$ and time-scale $\tau$. The points with error bars correspond to the expected value of the periodogram peak and its standard deviation assuming $\tau$ is the correct time-scale. For instance in Fig.~\ref{fig:solardata_perios} (b), assuming $\tau = T_\mathrm{obs}/3  = 365$ days would lead to expect a $z$ value with  $\tau = T_\mathrm{obs}$ over four sigmas away than the observed one. It then appears that the \ch{13.39} days signal is consistently detected on the whole dataset. In comparison, signals at 9000 days, 158 days and 26.9 days  (Fig.~\ref{fig:solardata_perios} (a), (c) and  (d)) seem to be localized in time. The peaks at 26.9 and 13.39 days are compatible with the first and second harmonic of the Solar rotation period. Low frequency structures are expected to come from correlated patterns stemming either from the star or the instrument.

The 13.39 days signal seems the most coherent of the four signals found. To further estimate if it is purely periodic we apply the methods of Section~\ref{sec:amphase} and~\ref{sec:period}. As suggested in Section~\ref{sec:period}, we study the consistency of the signal frequency by representing the values of  $z(\omega, t_0, \tau)$ (see Eq.~\eqref{eq:z}) and the amplitude $K(\omega, t_0, \tau)$ for a given $\tau$ as a colormap. In Fig.~\ref{fig:solardata_period} we represent $z$ ((a) and (b)) and $K$ ((c) and (d)) for $\tau = T_\mathrm{obs}/3 = 365 $  days ((a) and (c)) and $\tau = T_\mathrm{obs}/9 = 121$ days ((b) and (d)). The base model is identical to the one used to compute Fig.~\ref{fig:solardata_perios} (b). It appears that the 13 days signal is present throughout the dataset. Notice that its apparent waning towards the edge of the data set is due to the fact that when the $t_0$ reaches the beginning or end of the time series, the window includes twice as less points as in the middle of the time series. Signals at $\sim\!\!160$ and $\sim\!\!25$ days also seem to be present but with more variability, which is consistent  with the analysis of Fig.~\ref{fig:solardata_perios}.  In Fig.~\ref{fig:solardata_period} , we represent the quantity defined in Eq.~\eqref{eq:periodmax}, that is the local maximum of $z$ between certain frequencies as a function of $t_0$ for fixed $\tau$ ($\tau = T_\mathrm{obs}/3 = 365 $ days for (c) and $\tau = T_\mathrm{obs}/9 = 121 $ days for (d). We choose a collection of frequencies $\omega_k = k\Delta \omega$ which are represented in gray in Fig.~\ref{fig:solardata_period} (c) and (d). The maxima of $z$ between 10 and 20 days consistently occurs at 13.39 days.

The 13 days signal seems to be consistently present in the data and to have a steady period. We further examine whether its phase $\phi$  and semi amplitude $K$ as defined in Eq.~\eqref{eq:phi} and Eq.~\eqref{eq:k} are constants of time. To avoid being polluted by other signals, we include in the base model a 26.9 days signal as well as a 9\textsubscript{th} order polynomial, filtering out low frequency signals.
In Fig.~\ref{fig:solardata_amphase} (a) and (b) we represent the evolution of  $\phi$  and  $K$ in red and green respectively as a function of $t_0$ for fixed $\tau$ ($\tau = T_\mathrm{obs}/3 = 365$ days for (a) and $\tau = T_\mathrm{obs}/9 = 121 $ days for (b), the uncertainties are represented with color shaded areas. Red and green dashed lines represent the mean values of  $\phi$  and  $K$. Dots corresponds to measurements done on disjoint \ch{box}-shaped windows, which are thus approximately statistically independent, \ch{as shown in Appendix \ref{app:independence}}.   In Fig.~\ref{fig:solardata_amphase} (c) and (d) we represent the values of $A$ and $B$ as defined in Eq.~\eqref{eq:mu1} with their uncertainties in blue ($\tau = T_\mathrm{obs}/3 = 365$ days  for (c) and $\tau = T_\mathrm{obs}/9 = 121 $ days for (d). Points in orange correspond to the measurement times marked with dots in Fig.~\ref{fig:solardata_amphase} (a) and (b), that yield approximately independent estimates of $A$ and $B$.  At the shortest time-scale, it appears that $N\sigma_\chi^2$ as defined in Eq.~\eqref{eq:nsigma}  is greater than 5 (see Fig.~\ref{fig:solardata_amphase} (d)). In Fig.~\ref{fig:solardata_amphase} (b), the variation of amplitude does seem to significantly vary over time with two peaks around BJD 2457400 and BJD 2458000, which is consistent with the behaviour seen in Fig.~\ref{fig:solardata_period} (b). The statistical tests depend on the assumed of the noise. To obtain more robust statistics, we perform the same analysis but by adjusting an extra jitter term at each trial $t_0$, and propagate the uncertainty on $K$, $\phi$, $A$ and $B$. The results are very similar, and qualitatively unchanged. 
As a conclusion, it appears that the 13.39 day signal has a long time scale, a consistent period and phase, but its amplitude significantly varies with time.

\subsubsection{Discussion}
 From a phenomenological point of view, the presence of the 13.39-day signal, as well as the one at $\sim$25 days, can be fully understood by the presence of active regions rotating with the solar surface, as those two periods corresponds to the second and $P_{rot}$ term. By modifying locally the flux intensity of the solar surface \citep[e.g.][]{saar1997} and changing convection \citep[e.g.][]{meunier2013}, spot and faculae on the surface of a solar-like star will impact RV data, with a semi-periodic signal that can be decomposed as a Fourier series, thus as a sum of periodic signals at the rotation period of the Sun and its first harmonics \citep{boisse2011}. 
 
 It is often observed on solar-like stars that activity inject more power at $P_{rot}/2$ than its rotation period $P_{rot}$ \citep[e.g.][]{boisse2011}. 
 This can be explained by the fact that not only one but several active regions perturbs the RV measurement at the same time, but also because the Sun is seen equator on, thus active regions are seen only during half the rotation. \ch{ On a star like the Sun, the main contribution of the RV stellar activity signal comes from faculae \citep[e.g.][]{meunier2010b, dumusque2014, colliercameron2019, milbourne2019} which induce a strictly positive RV offset whatever their position on the solar disc, and the effect of an active region, in first approximation, can be viewed as a sinusoid truncated at zero  \citep[e.g.][]{meunier2010b, dumusque2014}. This truncation alone cannot explain the predominance of the $P_{rot}/2$ term, as the Fourier expansion of such a truncated sinusoid has a stronger $P_{rot}$ term than $P_{rot}/2$, where $P_{rot}$ denotes the stellar rotation period. If spots are at different longitudes, the RV signal would resemble a sum of truncated sine signals with different phases, and random phase differences do not affect preferentially a given harmonic. 
 However, a phase difference of two signals of 180$^\circ$ cancels out the first harmonic. As noted in \cite{borgniet2015},  there are generally two persistent active longitudes per hemisphere,
shifted by  180$^\circ$ \citep{berdyuginausoskin2003}, and this might explain the second harmonic. We add that even if there are several such configurations of two diametrically opposed active longitudes at different phases, the sum of their contributions has a vanishing component at $P_{rot}$. 
The presence of active regions on thxe surface, not in opposite phase but randomly distributed, may introduce signal at $P_{rot}$.  We finally note that the RV effect of active regions is not only due to convective blueshift inhibition, but also to their difference in brightness relative to the surroundings, introducing an asymmetry in approaching and receding limbs. This signal of photometric origin resembles a sine function with period $P_{rot}/2$ on $[kP_{rot}/2, (k+1)P_{rot}/2] $ for $k$ even and equal to 0 for $k$ odd \citep[e.g.][]{dumusque2014}. This truncated signal has a dominant power at $P_{rot}/2$, which might partly explain the strong $P_{rot}/2$ signal in Fig.~\ref{fig:solardata_perios} (b).}

 
 Regarding the change in amplitude \ch{and stability in phase}, it can be explained by the presence of a few very large active regions, that last for several months on the solar surface. The induced RV effect will be large, and the signal consistent over a long-period of time. The fact that the amplitude is large around BJD 2457400 can be explained by the Sun being active in 2015, with a lot of large active regions on its surface. The lower $\chi^2$ closer to the beginning of the time-series around BJD 2457217 is due to an edge effect. 
 Regarding the sudden increase in amplitude around BJD 2458000, it can be explained by the last large active region that appeared on the solar surface before the Sun reached the minimum of magnetic cycle 24 \citep[clearly seen in the Log(R'$_{HK}$) time series in Figure 6 of ][]{dumusque2020}.
 
 \ch{Overall, our tentative explanation of the strength of the 13.39 days signal, its phase stability and amplitude variation, is that stable magnetic regions diametrically opposed might have caused it. This speculative explanation requires further work to be properly accepted or rejected. We note that thankfully, the amplitude of the solar 13.39-day harmonic is more variable than its phase, making it less likely that such a signal in stellar data could be mistaken for a planet.
}

\subsubsection{Radial velocity and $\log R'_{HK}$}
We now compare the RV and $\log R'_{HK}$ time series. We note that there is no apparent signal at 13-14 days in the $\log R'_{HK}$ time-series. In Fig.~\ref{fig:chi2rhk} we show the $\chi^2$ map of the solar $\log R'_{HK}$, binned with the same pattern as the RV. The base model is a second order polynomial, and we add a jitter to the noise model equal to the standard deviation of the $\log R'_{HK}$ time series divided by 2. In Fig.~\ref{fig:chi2rhk}, there are low frequency signals at 200 - 400 days as well as signals close to the stellar rotation period, but none close to its first harmonic. As for the RV, we then performed an iterative search for signals (6 iterations), and did not find signals at 13 days. 

Furthermore, as suggested in Section \ref{sec:amphase}, we study the phase difference of RV and $\log R'_{HK}$. In both time series, there is a signal at the Solar rotation period. We fix $P$ = 26.25 days, and compute the difference $\Delta \phi$ RV phase - $\log R'_{HK}$ as a function of $t_0$ for  $\tau = 364$ (=$T_\mathrm{obs}/3$) days, where the phases are defined in Eq.~\eqref{eq:phi}. The uncertainty on $\Delta \phi$ is defined as the quadratic addition of the uncertainties on the RV and $\log R'_{HK}$ phases. These are computed by fitting a free jitter error term for each choice of $t_0$. In Fig.~\ref{fig:phaserhk}, we represent in purple the phase difference as a function of $t_0$, the shaded areas represent $\pm$ $1\sigma$ uncertainties, and the dots represent statistically independent phase measurements. The average difference of phase is 20 $\pm 6^\circ$, and this difference seems approximately constant throughout the observations. The RV 26 days signal is consistently in advance on the $\log R'_{HK}$ by 1.5 $\pm$ 0.5 days.


\begin{figure}
\centering
	\includegraphics[width=9.2cm]{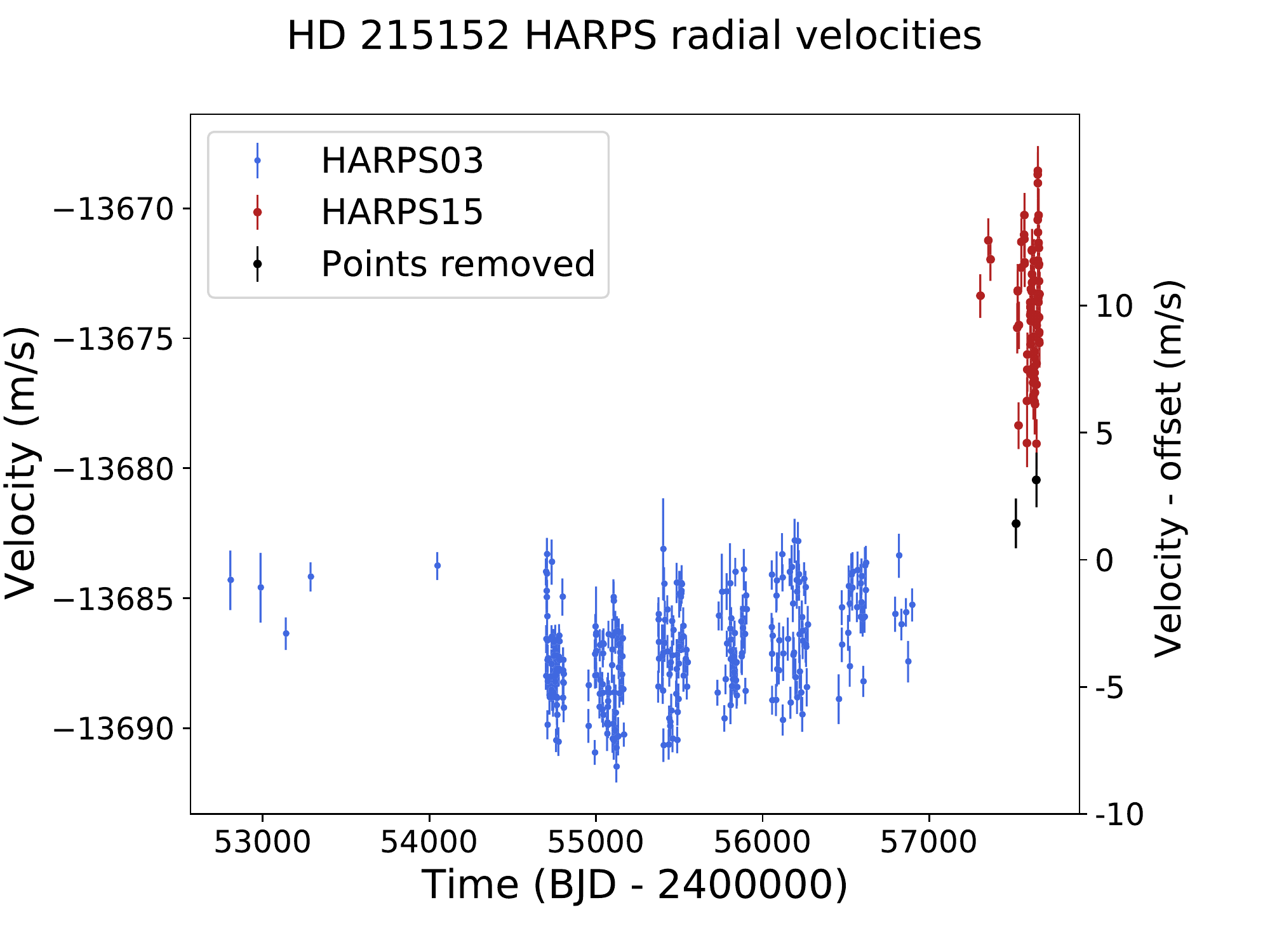}
	\caption{HD 215152 HARPS radial velocities. Data taken before and after the fiber update are represented in blue and red respectively. }
	\label{fig:hd215152}
\end{figure}

\begin{figure*}
	\centering
  \subfloat{
	\noindent
	\centering
	\hspace{-1cm}
	\begin{tikzpicture}
	\path (0,0) node[above right]{\includegraphics[scale=0.62]{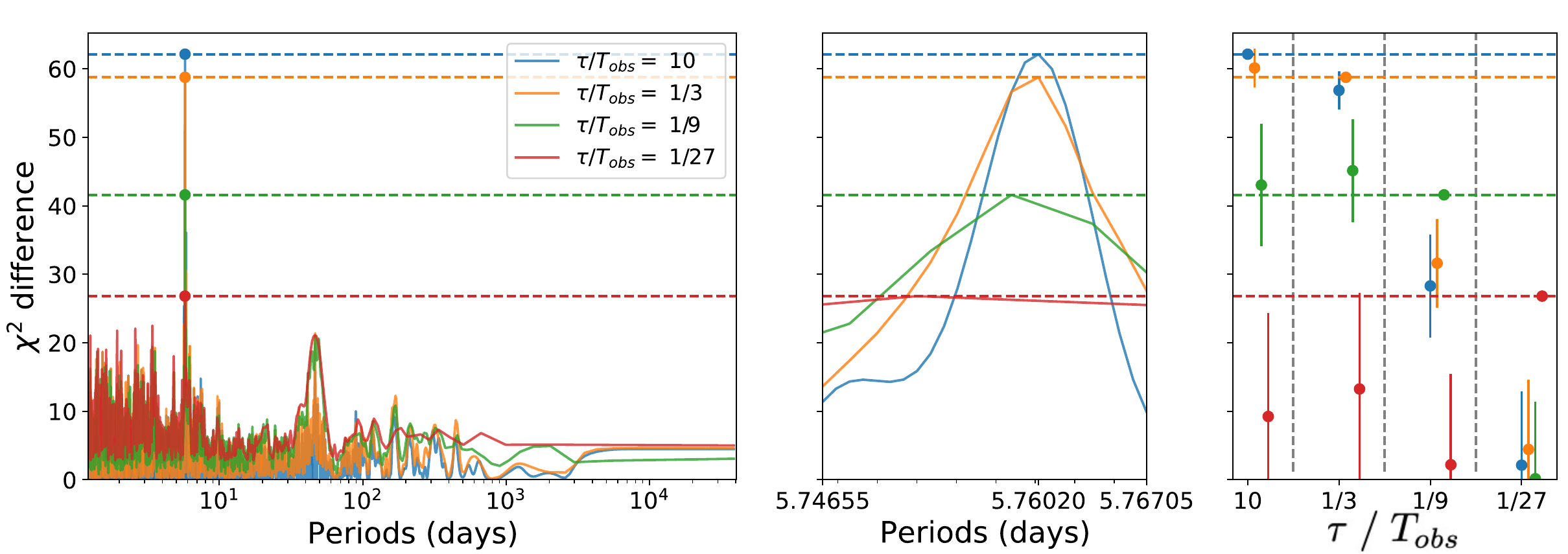}};
	\path (0,4.9) node[above right]{\large(a)};
	\begin{scope}[yshift=-5.5cm]
	\path (0,0) node[above right]{\includegraphics[scale=0.62]{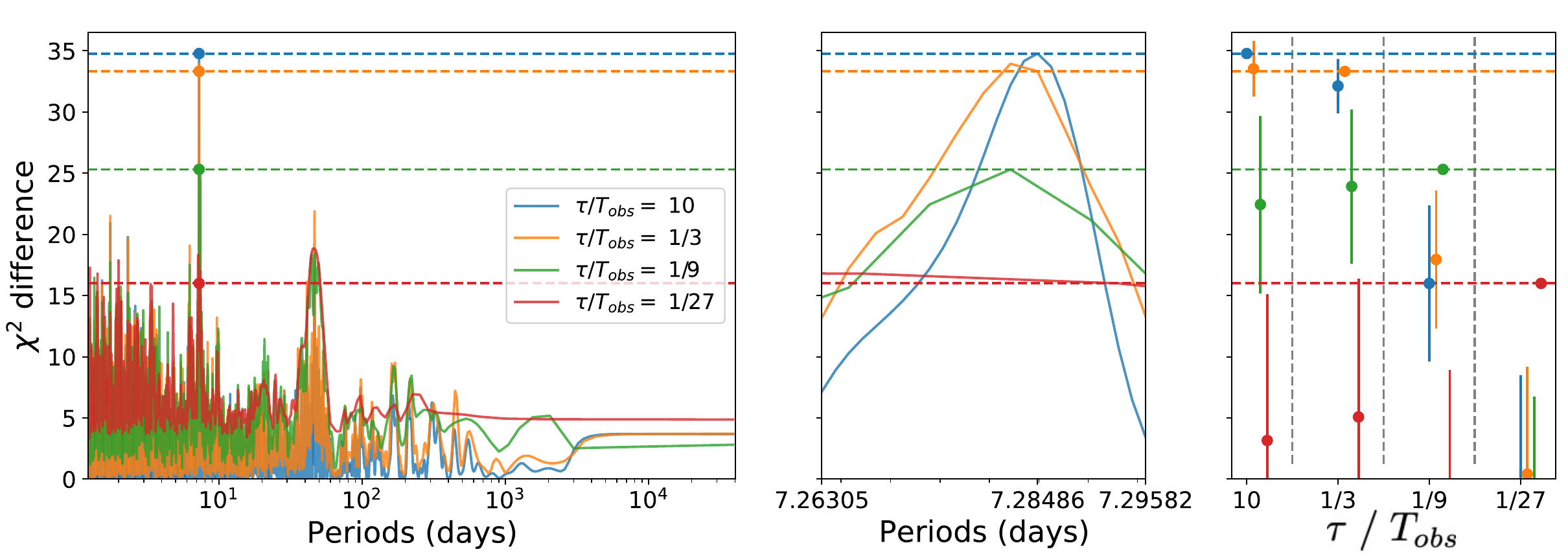}};
	\path (0,4.9) node[above right]{\large(b)};
	\end{scope}
	\begin{scope}[yshift=-11cm]
	\path (0.1,0) node[above right]{\includegraphics[scale=0.62]{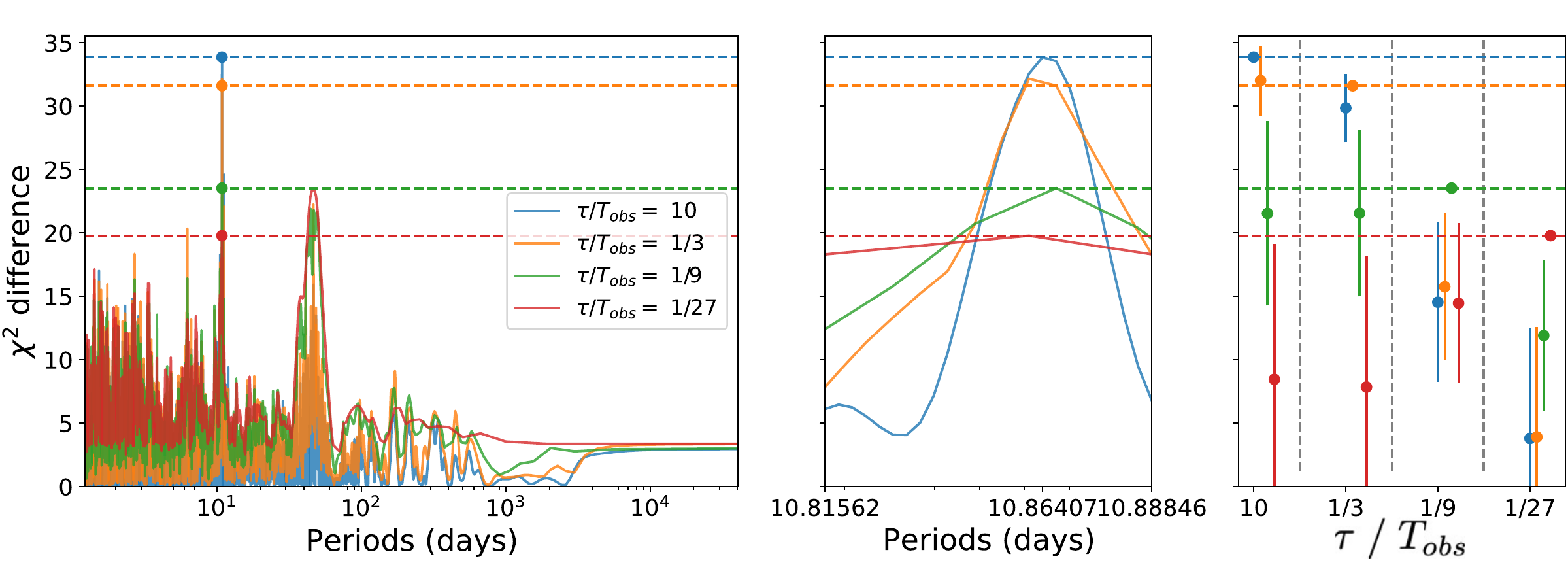}};
	\path (0.,4.9) node[above right]{\large(c)};
	\end{scope}
	\begin{scope}[yshift=-16.5cm]
	\path (0.1,0) node[above right]{\includegraphics[scale=0.625]{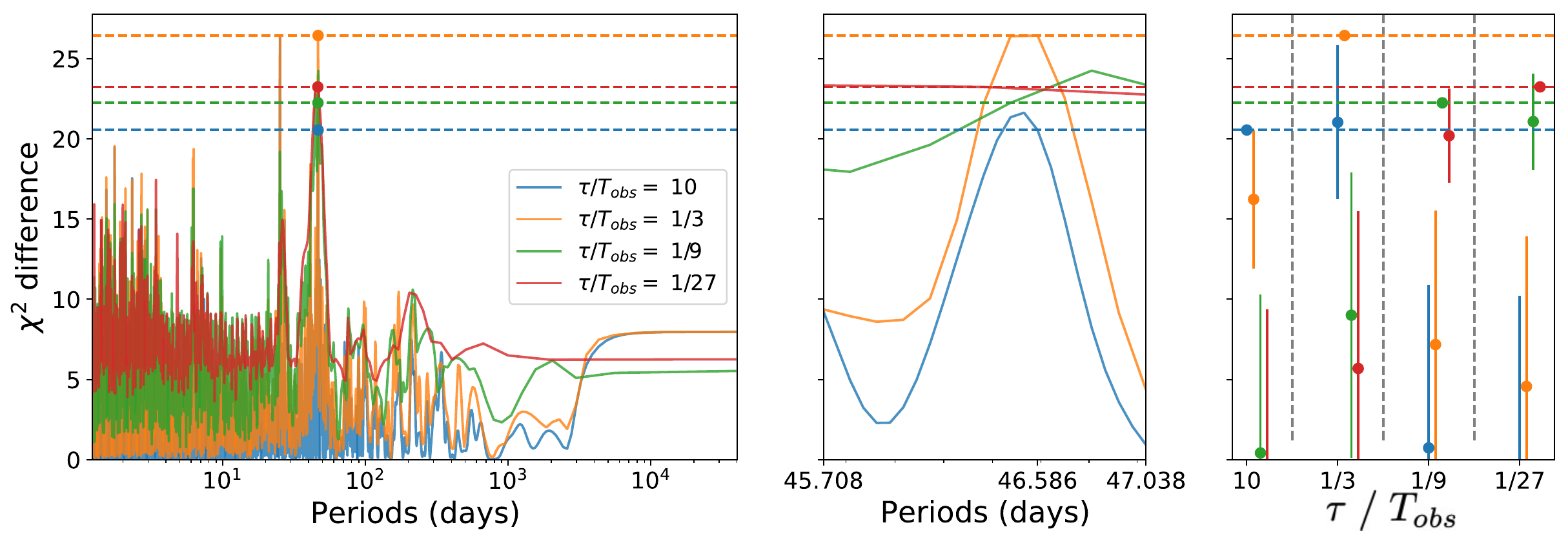}};
	\path (0,4.9) node[above right]{\large(d)};
	\end{scope}
	\end{tikzpicture}
	}
	\caption{Each row correspond to ASPs computed on the HD 215152 HARPS data. All are computed with a base model including linear activity model and offsets as described in Section~\ref{sec:hd215152}. They also include three of the four planets. Figures (a), (b), (c), (d) correspond respectively to leaving out planets at 5.75, 7.28, 10.86 and 25.20 days. 
		  Left column: ASPs corresponding to equation~\eqref{eq:zm} for different values of $\tau$ (blue, orange, green and red correspond to $\tau/T_\mathrm{obs}$ = 10, 1/3, 1/9 and 1/27).  Middle column: zoom on the maximum peak of the ASP. Right: statistical test on the time-scale. When subtracting the 46 day signal with $\tau = T_{\mathrm{obs}}/3$ (period at which the maximum of the ASP is attained in (d)), we obtain (e) (on the following page), where a coherent 25.2 d signal appears.   }
\label{fig:hd215152_apoperio}
\end{figure*}
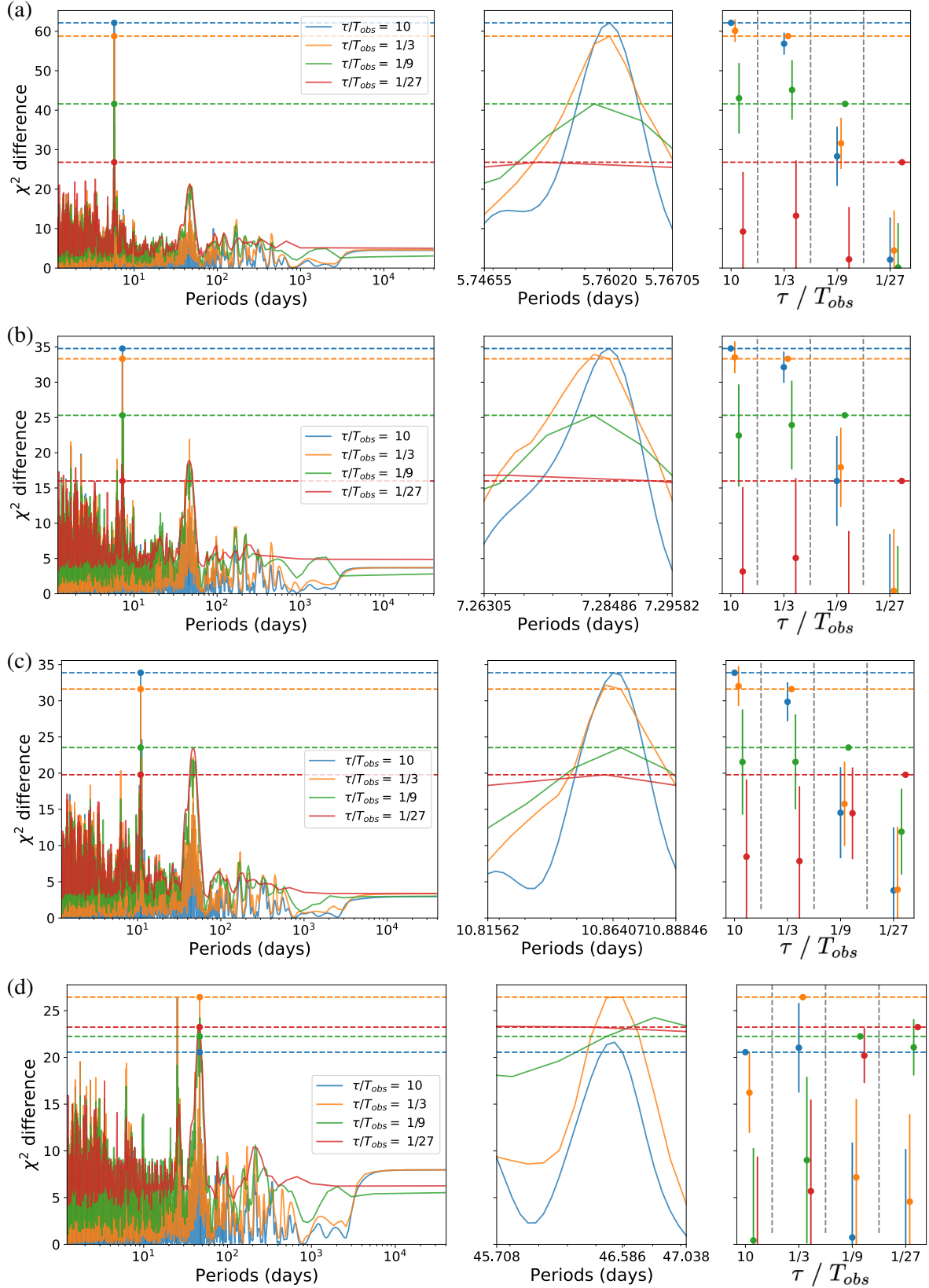
\begin{figure*}[h!]
	 \ContinuedFloat 
	 	\centering
	  \subfloat{  
	  	\begin{tikzpicture}
	\path (0,0) node[above right]{\includegraphics[scale=0.62]{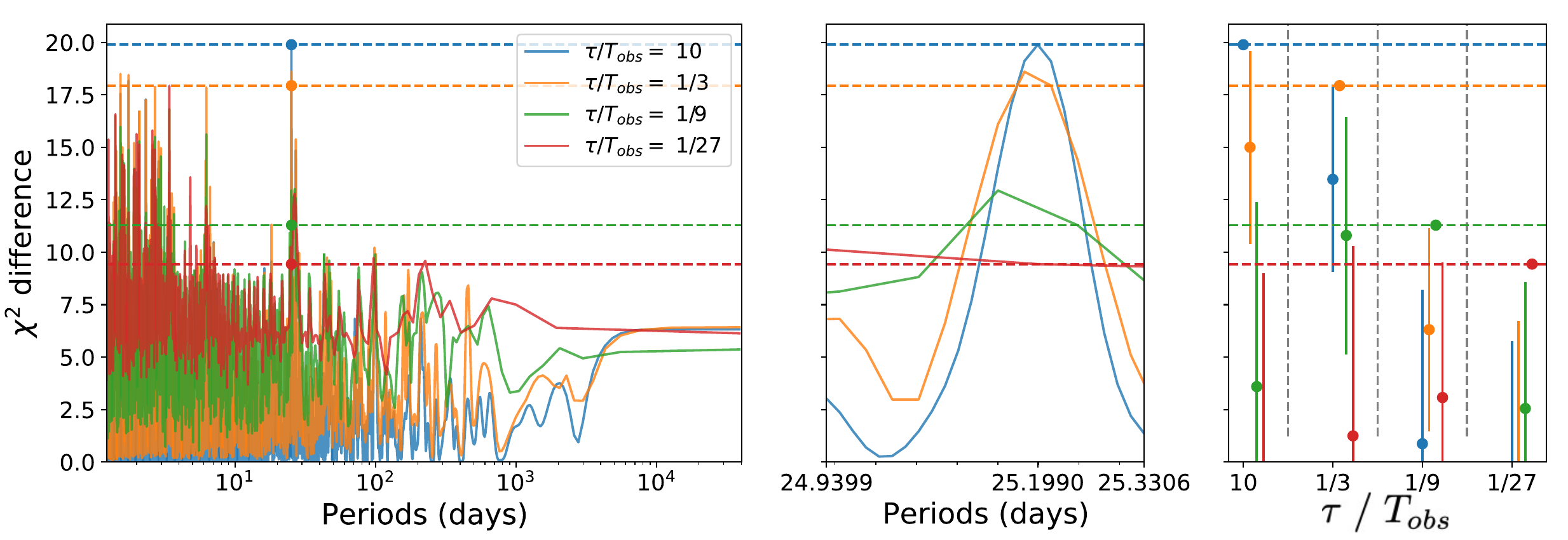}};
	\path (0.,4.9) node[above right]{\large(e)};
	 \end{tikzpicture}
	  }
\end{figure*}	  
\begin{figure*}[h!]
	\noindent
	\centering
	\begin{tikzpicture}
	\path (0,0) node[above right]{\includegraphics[scale=0.46]{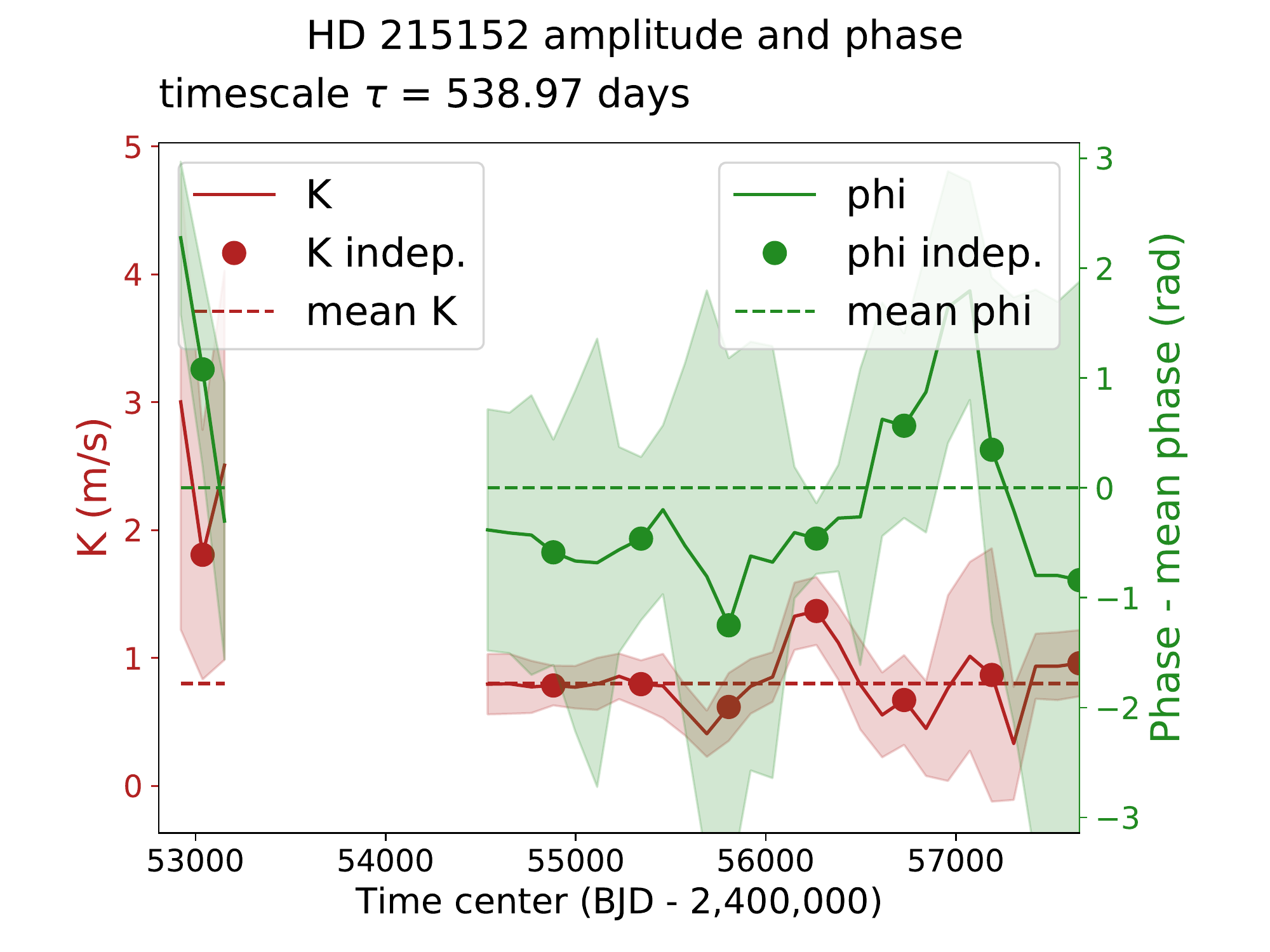}};
	\path (1.2,6.6) node[above right]{\large(a)};
	\path (9.5,0) node[above right]{\includegraphics[scale=0.46]{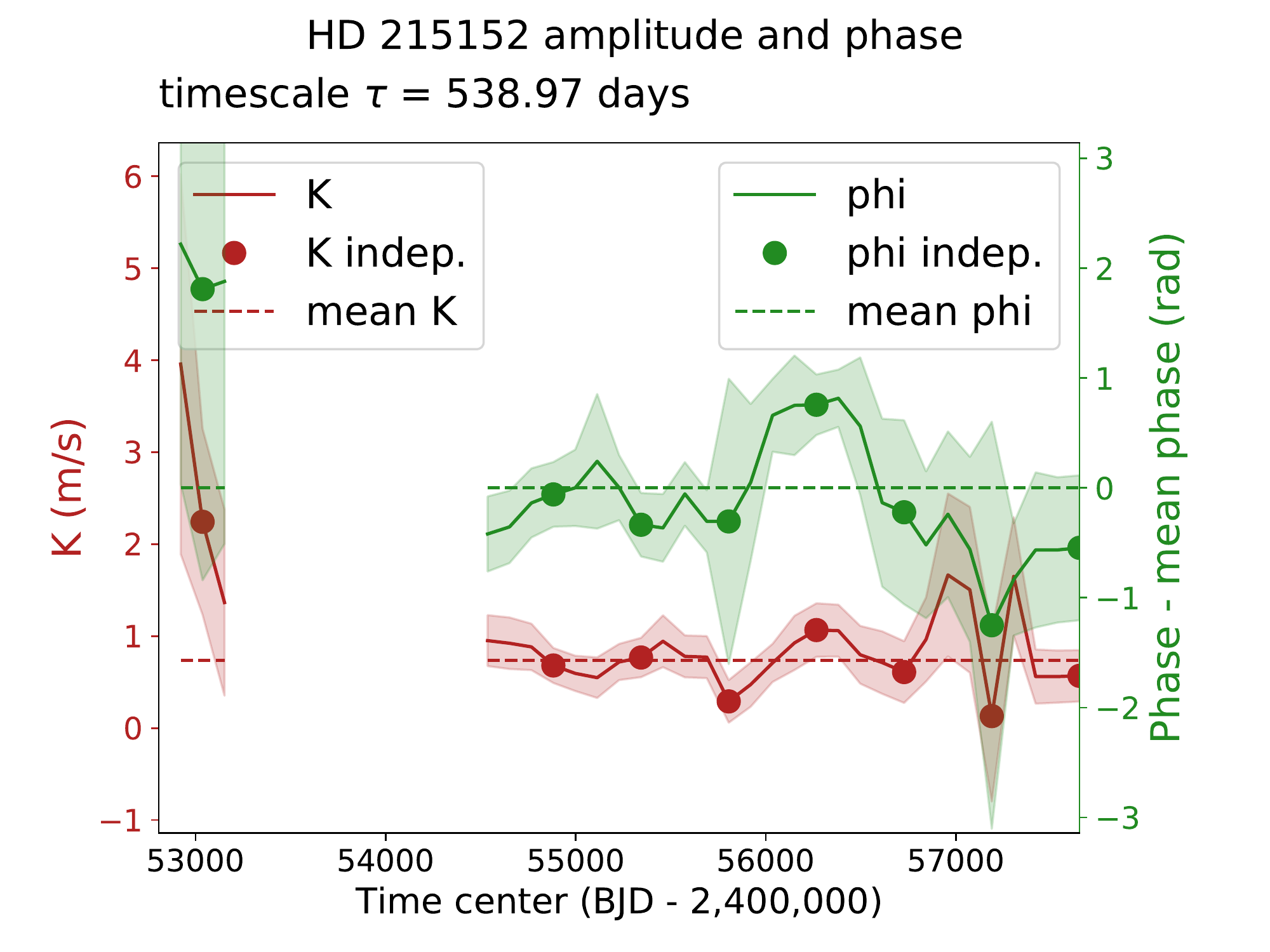}};
\path (10,6.6) node[above right]{\large(b)};
	
	\begin{scope}[yshift=-7cm]
	\path (0,0) node[above right]{\includegraphics[scale=0.46]{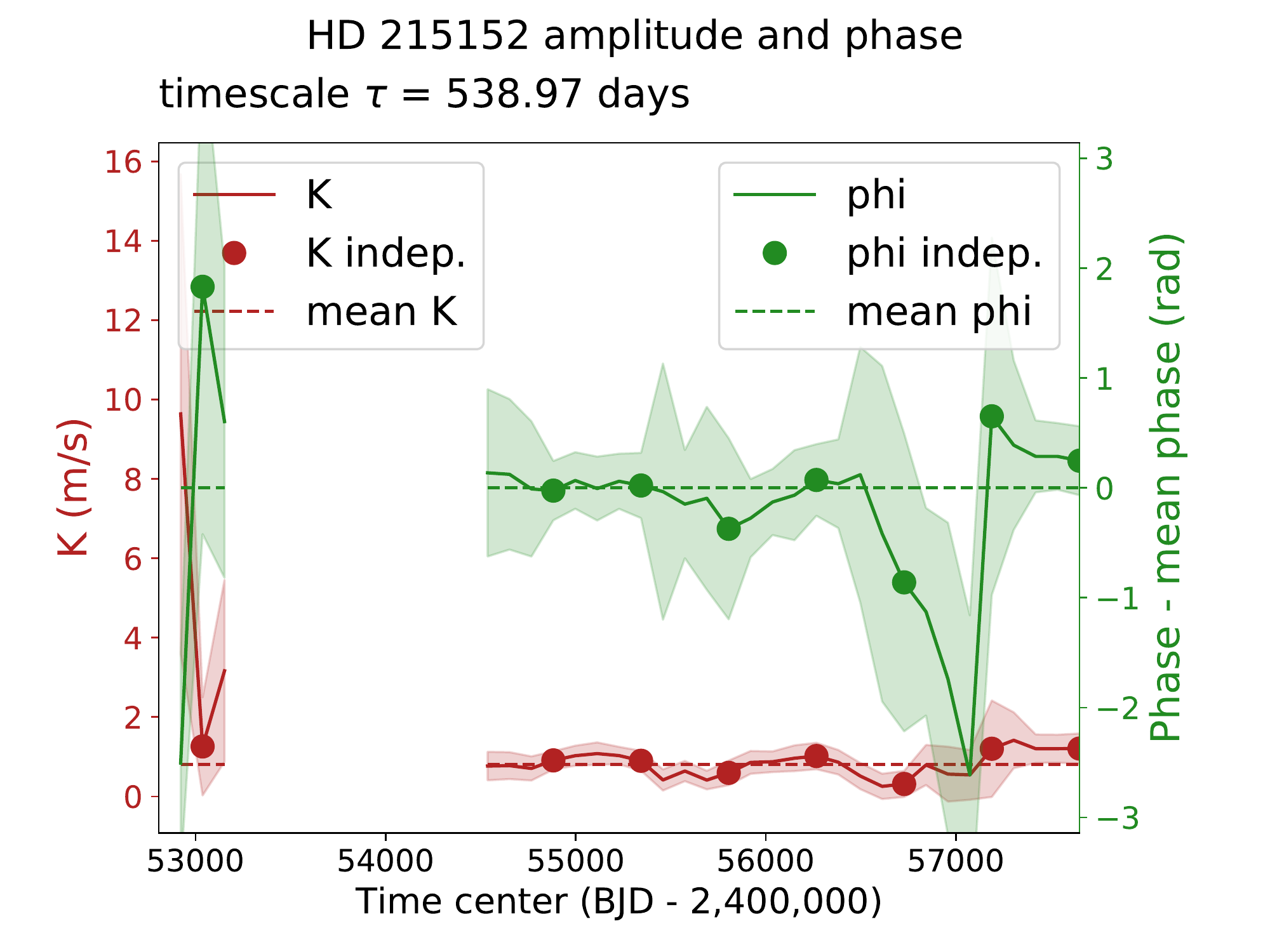}};
	\path (1.2,6.6) node[above right]{\large(c)};

	\path (9.5,0) node[above right]{\includegraphics[scale=0.46]{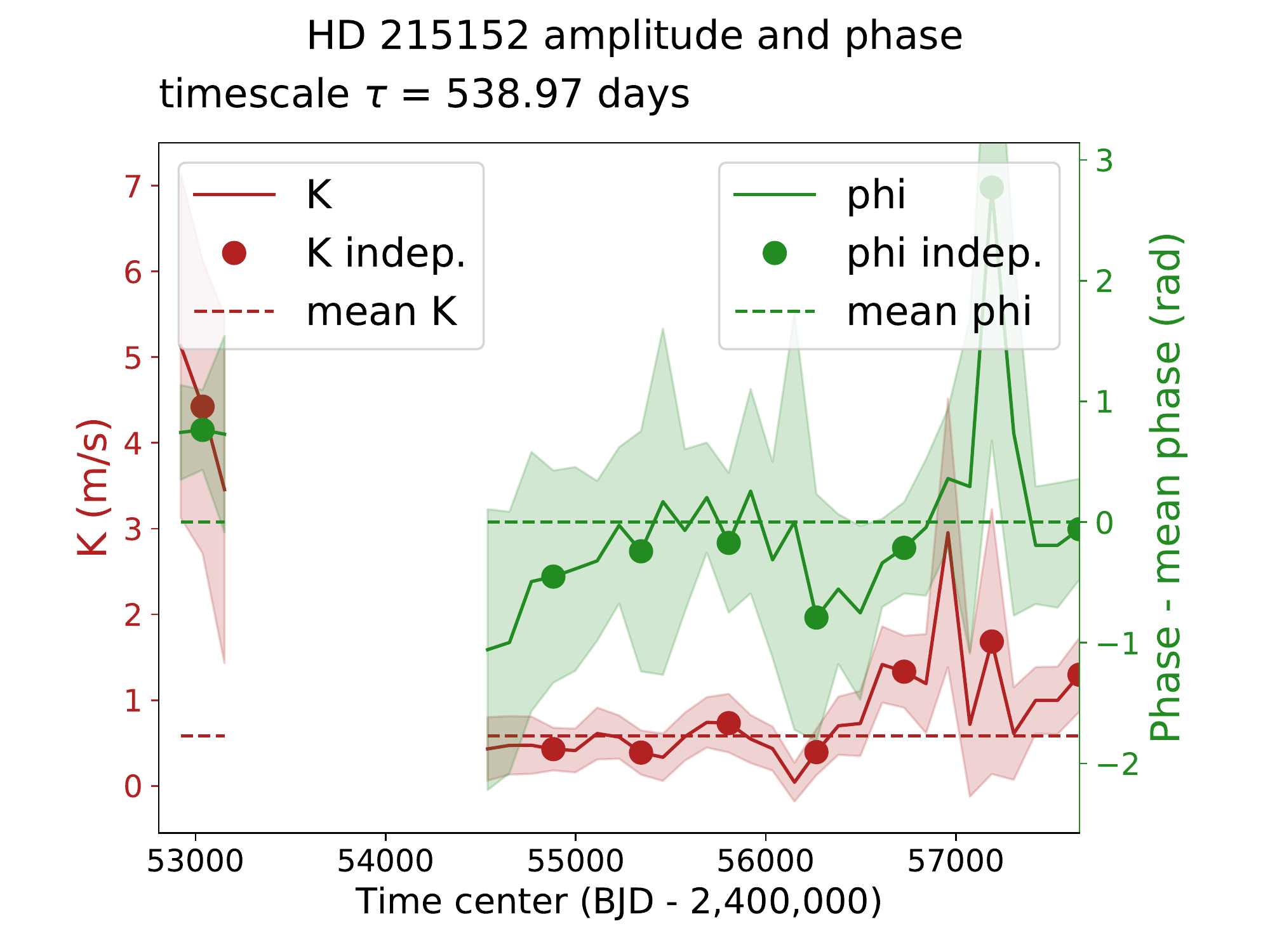}};
	\path (10,6.6) node[above right]{\large(d)};
	\end{scope}
	\end{tikzpicture}
	\caption{Phase and amplitude as a function of the time center of the window $t_0$, as described in Section~\ref{sec:amphase} (solid lines, red and green respectively). Markers correspond to statistically independent estimates. Red and green dashed lines represent the average phase and amplitudes. The figures are all computed for a time-scale $\tau =539$ days corresponding to $\tau/T_\mathrm{obs}$  = 1/9. Plots (a), (b), (c), (d) correspond to base models including offsets, linear activity indicators and all the claimed planets except 5, 7, 10 and 25 day respectively.   }
	\label{fig:hd215152_amphase}
\end{figure*}

\subsection{HD 215152}
\label{sec:hd215152}

The star HD 215152 has been observed with the HARPS spectrograph during a 13 years period from BJD 2452808 to 2457659. The data consists of 373 data points with mean error 0.73 m/s. The data are shown in Fig.~\ref{fig:hd215152}. \ch{HARPS has experienced an update of optical fibers in May 2015, which introduces a velocity offset.} Data taken before and after the fiber update, labelled HARPS03 and HARPS15, are represented in blue and red respectively. This system has been studied in~\cite{delisle2018}, \ch{who claims} the discovery of four super-Earth with orbital periods 5.76, 7.28, 10.86 and 25.20 days.  The analysis of~\cite{delisle2018} proceeds as follows. The data is modelled as a sum of Keplerian signals \ch{and} two offsets: one  corresponding to HARPS03 data and one  to HARPS15 data. The $\log R'_{HK}$~\citep{noyes1984} is smoothed with a low-pass filter, the resulting time series as well as the residuals (high-passed filter  $\log R'_{HK}$) are used as linear predictors. The data is analysed with a Gaussian noise model whose kernel is defined as the sum in quadrature of the nominal uncertainties,  a free jitter and a \ch{correlated component with a Gaussian autocovariance (or kernel). Denoting by $\Delta t$ the time difference between measurements, the autocovariance is $\sigma_R^2 \e^{- \Delta t^2/(2\tau_R^2)}$, where $\sigma_R^2 $ and $\tau_R$ are free parameters}. The significance of the signals is established by computing periodograms where $\sigma_J$, $\sigma_R$  and $\tau_R$ are fitted at all trial frequencies. The periodogram is computed as the difference of Bayesian information criterion~\citep{schwarz1978} of models $\H$ and $\K$. The rotation period of the star is estimated to be 43 days.

In this section, we present a re-analysis of the same data to assess the stability in amplitude, phase and period of the detected signals. We exclude from the datasets the two points represented in black in Fig.~\ref{fig:hd215152}, on the ground that they deviate from the median of the HARPS15 data (blue points) by over 4 median absolute deviation.

To simplify the discussion, we assume the same model as~\cite{delisle2018} and fix the values of the noise parameters to their posterior medians  $\sigma_J=0.6$ m/s, $\sigma_G=1.2$ m/s  and \ch{the timescale of the noise is $\tau_R = 2$ days (note that this is different from the apodisation time-scale $\tau$)}. In the base model~\eqref{eq:mh}, we also include the low-pass filtered $\log R'_{HK}$ as well as the two offsets. We then compute the ASP (eq.~\eqref{eq:z}) in four cases. In each one, we add to the base model six linear predictors $\cos \omega_1 t, \sin \omega_1 t, \cos \omega_2 t, \sin \omega_2 t,  \cos \omega_3 t, \sin \omega_3 t$ where $\omega_1, \omega_2$ and $\omega_3$ are the frequencies of three planets out of four.  This comes down to assuming that three planets are securely detected and the consistency of the fourth one is tested. The frequency grid goes from 0 to 0.8 cycles per day to avoid showing the aliases of the peaks. 

When leaving out \ch{the 5.76, 7.28, 10.86 days planets}  from the \ch{base} model, we obtain \ch{respectively} Fig.~\ref{fig:hd215152} (a,b,c). It appears clearly for planets at 5.76 \ch{days} and 10.86 \ch{days} that the maximum occurs for the longest coherence time (in blue). In the case of the 7.28 days signal, the maximum occurs at $\tau/T_\mathrm{obs} = 1/3$. However, as shown on the right panel, the $\tau/T_\mathrm{obs} =10$ hypothesis is not excluded. Note that the sampling of the HARPS data is very irregular, with a few samples between BJD 2453000 and  2454500 (see Fig.~\ref{fig:hd215152}). When leaving out  the 25 days planet, we obtain Fig.~\ref{fig:hd215152} (d). In that case, the highest peak corresponds to 46 days. If the $\tau = 10T_{\mathrm{obs}}$ were true, the periodogram peak corresponding to $\tau = T_{\mathrm{obs}}/3$ would be $\sim$ 2 $\sigma$ away from the highest peak (see right panel Fig.~\ref{fig:hd215152} (d). This is not surprising, as 46 days is close to the estimated rotation period of the star. When subtracting the 46 day signal with $\tau = T_{\mathrm{obs}}/3$, we obtain Fig.~\ref{fig:hd215152} (e). The maximum peak is attained at 25 days and appears to be consistent with   $\tau = 10 T_{\mathrm{obs}}$, and thus a planet. We note however that the the false alarm probability obtained with a regular periodogram of the 25 days planet is 1\%, so that it might be suitable to acquire more points for a more significant detection.


 We perform the test described in Section~\ref{sec:amphase} for each \ch{of the} four cases. Fig.~\ref{fig:hd215152_amphase} shows the resulting plot. It appears that the phase and amplitude of the signals are compatible with the hypothesis that they are constant in all cases.

\begin{figure*}
\centering
	\includegraphics[width=0.49\linewidth]{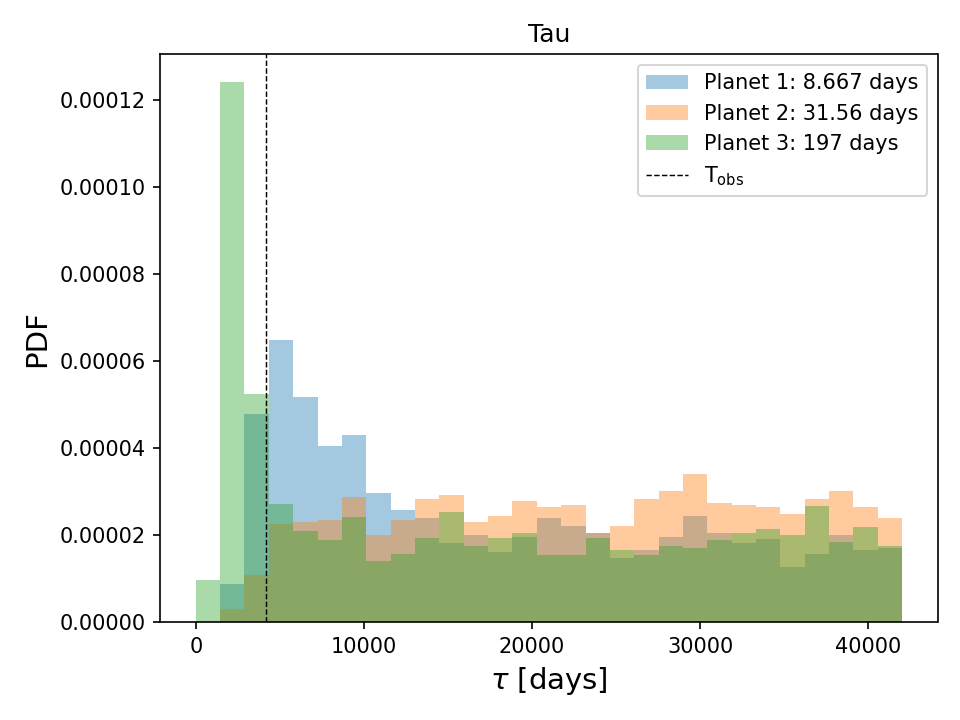}
	\includegraphics[width=0.49\linewidth]{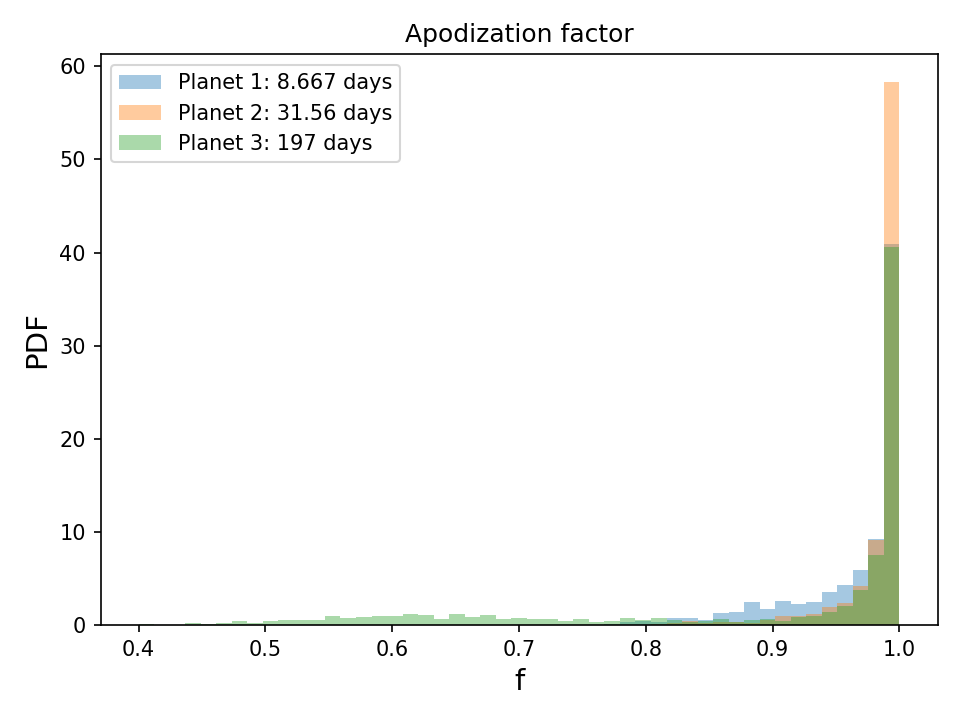}
	\caption{Left and right: posterior distribution of $\tau$  (defined in Eq. \eqref{eq:taukep}) and $f$ (defined in Eq. \eqref{eq:f}) conditioned on the period $P$ being such that $|1/P -1/P_i|<1/T_{\mathrm{obs}}$ for $i=1,2,3$ and $P_1 =8.66$ days, $P_2 = 31.56$ days and $P_3 = 197$ days. $i=1,2,3$ correspond respectively to the blue, orange and green histograms. The black dotted line corresponds to $\tau = T_{\mathrm{obs}}$  }
	\label{fig:hd69830}
\end{figure*}

	\subsection{HD 69830 }
\label{sec:hd69830}

In this section, we illustrate the Bayesian method proposed in Section~\ref{sec:bayesian} with the HARPS03 data of  HD 69830. This dataset contains 254 data points and spans on 11.5 years. The HD 69830 system is known to harbour three planets with minimum masses close to Neptune's~\citep{lovis2006} and periods of 8.667, 31.56 and 197 days. 

As suggested in Section~\ref{sec:bayesian}, we compute the joint posterior distribution of the orbital elements and the number of planets. The signal is represented as a sum of $k$ keplerian functions multiplied by an apodization factor and a noise \ch{model (a random variable)}
\begin{align}
    \vec y = \sum\limits_{i=1}^k  \e^{-\frac{(\vec t-t_0)^2}{2\tau^2}} \text{Kep}_i(P,e,K,\omega, M_0) + \vec \epsilon \label{eq:taukep}
\end{align}
where $\text{Kep} (P,e,K,\omega, M_0)$ is a Keplerian function \citep[e.g.][]{perryman2011} of period $P$, eccentricity $e$, semi-amplitude $K$, argument of periastron $\omega$ and initial  mean anomaly $M_0$.
We use a Gaussian noise model for $\vec \epsilon$ with a jitter term and an exponential decay, implemented with the \texttt{spleaf} software \citep[][]{delisle2020b}. The covariance matrix $\mat V$ has the form 
\begin{align}
    \mat V_{ij} = \delta_{ij} (\sigma_i^2 + \sigma_W^2) + \sigma_R^2\e^{-\frac{(t_i-t_j)^2}{2T^2}}
    \label{eq:rnoise}
 \end{align}
 where $\delta_{ij}$ is the Kronecker symbol, $\sigma_i$ is the nominal uncertainty on measurement $i$. 
The priors on the parameters are presented in Table \ref{tab:priorhd69830}. As in \cite{hara2021a}, the posterior distributions are computed with \textsc{polychord} \citep[][]{handley2015,handley2015b}.
 
In Fig.~\ref{fig:hd69830} we represent the posterior distribution of $\tau$  (defined in Eq. \eqref{eq:taukep}) and $f$ (defined in Eq. \eqref{eq:f}) conditioned on the period $P$ being such that $|1/P -1/P_i|<1/T_{\mathrm{obs}}$ for $i=1,2,3$ and $P_1 =8.66$ days, $P_2 = 31.56$ days and $P_3 = 197$ days. $T_{\mathrm{obs}}$ is the total timespan of the observations. The $\tau$ distribution for the $P_3 = 197$ days planets peaks below $T_{\mathrm{obs}}$. In the $f$ distribution (right) it clearly appears that all distributions peak at 1. In conclusion, all the three known planets are indeed consistently detected. 
 
\begin{table}[]
\centering
\begin{tabular}{ccc}
\hline
Parameter & Units & Prior                     \\ \hline
$P$       & days  & log-Uniform: [0.7, 1000] \\
$K$       & m/s   & log-Uniform: [0.1, 20]    \\
$e$       &       & Beta: [0.867, 3.03]$^{\dagger}$       \\
$\omega$  & rad   & Uniform: [0, 2$\pi$]      \\
$M_0$     & rad   & Uniform: [0, 2$\pi$]      \\
$\tau_R$     & days   &  Uniform: [0, 10\,T$_{\rm obs}$]   \\
\hline
$k$     & - &   Uniform: [0, $k$]  \\ \hline
$\sigma_W$ & m/s & Uniform: [0, 20] \\ 
$\sigma_R$ & m/s  & Uniform: [0, 20] \\
$T$ & days  & log-Uniform: [0.1, 30] \\
\end{tabular}
\caption{Priors used for the computation of the FIP periodogram of HD 69830 and HD 13808. $^{\dagger}$ \cite{kipping2014}}
\label{tab:priorhd69830}
\end{table}

		\begin{figure*}
	    \centering
	    \includegraphics[width=0.9\linewidth]{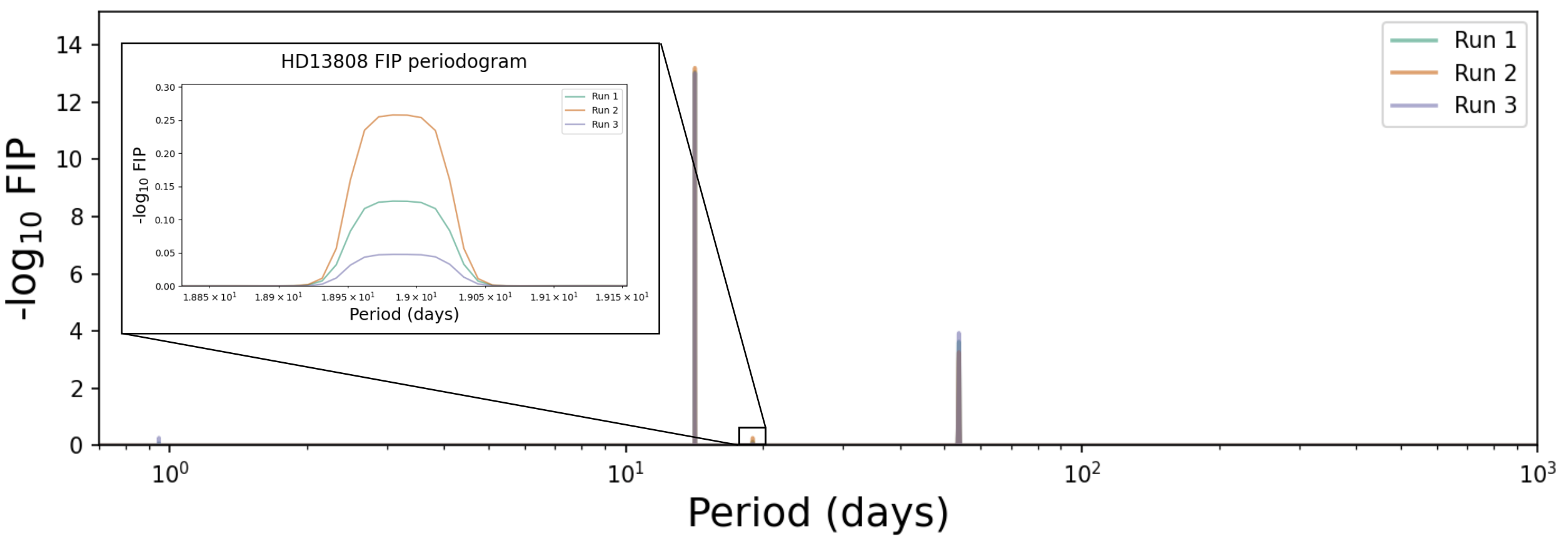}
	    \caption{FIP periodogram of HD 13808. $-\log_{10}$ FIP (in blue) and $\log_{10}$ TIP (in yellow) of the presence of a planet in a centered frequency interval $[\omega - \Delta \omega, \omega + \Delta \omega]$ as a function of $\omega$. $\Delta \omega = 2\pi/T_{\mathrm{obs}}$ where $T_{\mathrm{obs}}$ is the total observation time-span. Different colors correspond to independent computations with POLYCHORD. The plot inside black box represents a zoom on the region close to 18.9 days.  }
	    \label{fig:HD13808_FIP}
	\end{figure*}
	\begin{figure*}
\centering
	\includegraphics[width=0.49\linewidth]{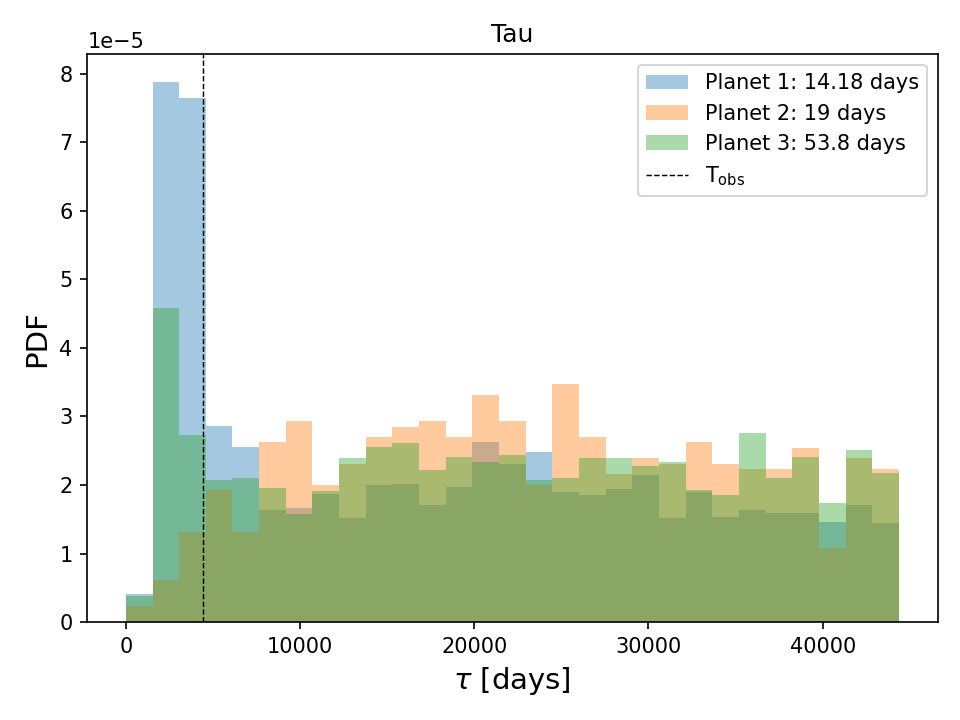}
	\includegraphics[width=0.49\linewidth]{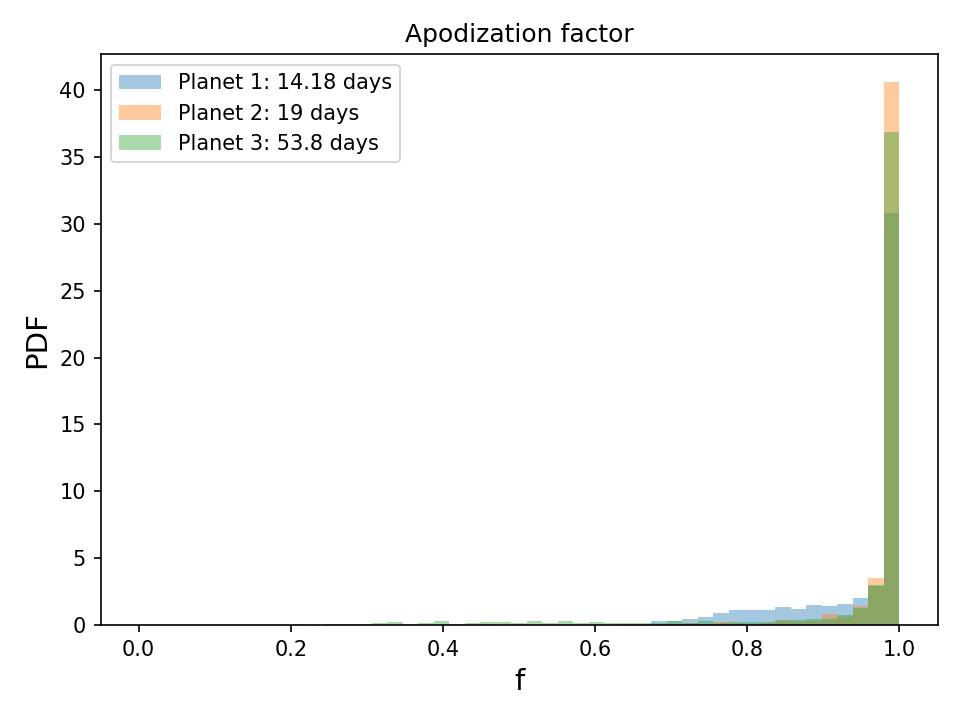}
	\caption{HD 13808 data set: left and right: posterior distribution of $\tau$  (defined in Eq. \eqref{eq:taukep}) and $f$ (defined in Eq. \eqref{eq:f}) conditioned on the period $P$ being such that $|1/P -1/P_i|<1/T_{\mathrm{obs}}$ for $i=1,2,3$ and $P_1 = 14.18$ days, $P_2 = 19$ days and $P_3 = 53.8$ days. $i=1,2,3$ correspond respectively to the blue, orange and green histograms. The black dotted line corresponds to $\tau = T_{\mathrm{obs}}$  }
	\label{fig:HD13808}
\end{figure*}

	\subsection{HD 13808 }
	
	The HD 13808 system has been observed with HARPS \citep{mayor2003} over a 10 year timespan. The dataset contains 246 radial velocity measurements, and presents three candidate signals at 14.1, 19 and 53.7 days.   In \cite{ahrer2021}, the HARPS is analysed with many different noise models based on the bisector span \citep{queloz2001}, and the $\log R'_{HK}$ \citep[][]{noyes1984}. The stellar activity models are built upon the framework of Gaussian processes \citep{rajpaul2015} and the $FF'$ method \citep[][]{aigrain2012}. The models with different numbers of planets and different stellar activity models are compared through their Bayesian evidence. The authors conclude that the  14.1 and 53.7 days planets are securely detected and the 19 days planet cannot be confidently claimed. 
	
	We here perform the same analysis as Section \ref{sec:hd69830}. Following the method of Section \ref{sec:bayesian}, we compute the joint posterior distribution of $\omega, t_0$, $\tau$, and the number of planets. We use the same noise model (see Eq. \eqref{eq:rnoise}), same priors (see Table \ref{tab:priorhd69830}) as for HD 69830 and the same numerical method, \textsc{polychord} \citep[][]{handley2015,handley2015b}.
	
	First, to check that the signals are significant, as in \cite{hara2021a}, we compute the FIP periodogram. This one is defined as follows. 	 We consider a grid of frequency intervals with a fixed length. The element $k$ of the grid, $I_k$, is defined as $[\omega_k - \Delta \omega/2, \omega_k + \Delta \omega/2]$ where $ \Delta \omega = 2\pi/T_\mathrm{obs}$, $T_\mathrm{obs}$ is the total observation timespan and  $ \omega_k = k\Delta \omega / N_\mathrm{oversampling}$. We take $N_\mathrm{oversampling} = 5$. We then compute the probability to have at least one planet in $I_k$ (the TIP), and the FIP = 1 - FIP. We then represent $-\log_{10} FIP$ as a function of $\omega_k$.  
	
	As suggested in \cite{hara2021a}, we compute the FIP periodogram for different maximum number of planets. We find that the FIP periodograms obtained for three and four planets are almost identical, and choose the three planets models. To assess the computational uncertainties, we compute the FIP periodogram with three dfferent runs of \textsc{polychord}.
	The resulting FIP periodogram is shown in Figure \ref{fig:HD13808_FIP}, each run corresponds to a FIP periodogram of a different color.  It appears that the probability of not having a planet at 14.1 and 53.7 days is below $10^{-10}$  and $10^{-2}$ respectively. The significance of the 18.9 days signal is much lower, with a FIP $\sim$ 70\%. We also did the same calculation with a white noise model, for which the 18.9 days signal is much more significant (FIP of 1\%). 
	
	We then compute the posterior distribution of \ch{the apodisation timescale} $\tau$ and \ch{the fraction of information captured by a window of given width}  $f$, as in Section \ref{sec:hd69830}. In Fig.~\ref{fig:HD13808} we represent the posterior distribution of $\tau$  (defined in Eq. \eqref{eq:taukep}) and $f$ (defined in Eq. \eqref{eq:f}) conditioned on the period $P$ being such that $|1/P -1/P_i|<1/T_{\mathrm{obs}}$ for $i=1,2,3$ and $P_1 = 14.18$ days, $P_2 = 19$ days and $P_3 = 53.8$ days. We find that all signals have an $f$ distribution peaking at 1.
	
	As a conclusion, in accordance with \cite{ahrer2021} we find that the 14.1 and 53.7 days signal are statistically significant. The significance of the 19 days signal is too low to claim a detection. However, it appears that the signal is consistent in time, so that a planetary origin cannot be excluded. However, this is not likely, since as shown in \cite{ahrer2021}, a 19 days signal is present in the ancillary indicators.  We have seen in the case of the Sun (see \ref{sec:sun}) that the first harmonic of the rotation period exhibits a constant period and phase, but not a constant amplitude. Since the stellar rotation period is $\sim$ 38 days, the 19 days signal could be another case of first, stable harmonic of the rotation period.

	\section{Discussion and conclusions}
	\label{sec:conclusion}
	
	In the present work, our goal is to assess whether an apparently periodic signal is strictly so, with a focus on the detection of planets in radial velocity data. This can be done with models intended to capture all variations due \ch{to} the planets and stellar activity~\citep{haywood2014, rajpaul2015, jones2017, hara2021a}. However, these models are not guaranteed to be trustworthy, which motivates the approach of the present work, whose methods offer a complementary outlook on the data.
	
	Following \citep{gregory2016}, we considered wavelet-like functions: periodic functions multiplied by an apodization term. 
	We suggested different representations and statistical tests to explore  whether a signal is truly periodic, by constraining its time-scale (see Section~\ref{sec:detec}), its variation of phase and amplitude (see Section~\ref{sec:amphase}) and its frequency variation (see Section~\ref{sec:period}). We considered tests based on periodograms (see Section~\ref{sec:def},~\ref{sec:grid},~\ref{sec:stat}) and on the Bayesian formalism (see Section~\ref{sec:bayesian}). Periodograms provide interpretable figures, and scan robustly and quickly all the periods, while their exploration might be difficult with random searches used for Bayesian inference. On the contrary, this one has the advantage of providing easily interpretable quantities. Depending on the objective and the size of the dataset, one of the two approaches might be more adapted. 
	
	The variability of the signal of interest can be assessed at different time-scale with a typical Heisenberg uncertainty principle: the shorter is the time-scale of variation probed, the higher is the uncertainty on the local estimates of phase, amplitude and period. We recommend to study  this variability on a grid of time-scales, chosen with an exponential decrease, as in wavelet decomposition~\citep[][]{mallat1989}.

	Our tests are defined to assess the consistency with a certain data model. If other signals are present (trends, other periodic or quasi periodic signals), or the shape of the signal examined is not sinusoidal -- for instance it corresponds to an eccentric planet -- our indicators might present significant variations, while the signals is truly periodic. As a consequence, for the results to be reliable, it is crucial to model simultaneously the signal of interest and other signals present. The approach we suggest is either to model the data as a sum of apodized periodic components, as in~\citep{gregory2016}, or to assume that all periodic signals are truly periodic, except the one under consideration.
	
	When assessing the phase and amplitude variability of a signal at frequency $\omega_0$ it is crucial to use all the data to estimate $\omega_0$, since the longest baseline gives the most precise estimate of $\omega_0$. An inaccurate estimate of $\omega_0$ might result in an apparent, \ch{spurious, linear phase shift with time}.

	We analysed the Solar radial velocity as measured by HARPS-N~\citep[][]{dumusque2020} as well as the HARPS data of HD 215152~\citep{delisle2018}, HD 69830 \citep{lovis2006} and HD 13808 \citep{ahrer2021}. The Solar radial velocity contains quasi-periodic signals at the first and second harmonic of the rotation period, \ch{the latter having the largest amplitude of the two}. The $\sim$25 - 27 days region shows a varying period, phase and amplitude. The second harmonic, appearing at 13.39 days, shows a period and phase constant within the uncertainties of statistical tests\ch{, showing that stellar activity signals can exhibit traits of purely periodic signals. Fortunately, the amplitude of the solar 13.39-day harmonic shows significant variations at the time-scale of 121 days (total observation time-span divided by nine). Such variability makes it less likely that such a stellar signal could be mistaken for a planet. } 
	These amplitude variations can be mapped to variations of the level of activity. This case shows that variations of amplitude, phase and frequencies due to stellar activity can be very subtle, and require a high signal to noise ratio to be statistically significant. \ch{These observations can speculatively be explained by the persistence of large active regions diametrically opposed, as suggested by \citep{borgniet2015}. Testing more rigorously this hypothesis is left for future work. }
	
	In HD 13808, there is a potential low amplitude, signal at 19 days, which also presents consistency in time. We therefore agree with the conclusion of \cite{ahrer2021} that two planets can be confidently claimed.  A planetary origin of the signal can although not be excluded, which might motivate further observations.

	 The detection of four and three planets in HD 215152 and HD 69830 were claimed in~\citep[][]{delisle2018} and \cite{lovis2011} respectively. We find in both cases that all the planets exhibit constant phases and amplitudes, which strengthens these detection claims.

	
	
In the present work, we have suggested several methods to assess whether a signal is strictly periodic or not, but many other can be considered. The techniques for simultaneous modelling of quasi periodic, periodic signals and noise, and the periodicity criteria could both be explored in further detail. 
We leave this exploration and the comparison of the different criteria for future work.

	\begin{acknowledgements}
	The authors thank the anonymous referee for their very insightful comments, which helped to improve the paper. 
	N. C. H thanks Michaël Cretigner for his input. 
	N. C. H. and J.-B. D. acknowledge the financial support of the National Centre for Competence in Research PlanetS of the Swiss National Science Foundation (SNSF).
\end{acknowledgements}
	
	\bibliography{biblio2.bib} 
	\bibliographystyle{aa}

	\appendix

	\section{Computations}
	\label{app:computations}

	\subsection{Distribution of the difference of two $\chi^2$ variables}
	\label{app:dz}

	Let us consider   two models $\K(\omega, t_0, \tau)$ and  $\K(\omega', t_0', \tau')$, which we denote by
	\begin{align}
	\K(\omega, t_0, \tau): y = Ax + \epsilon, \; \epsilon \sim G(0,V) \\
	\K(\omega', t_0', \tau'): y = Bx + \epsilon, \; \epsilon \sim G(0,V) \
	\end{align}
	where $G(0,V)$ is a multivariate Gaussian distribution of mean 0 and covariance $V$. We want to compute the  distribution of $z(\omega, t_{\tau,\omega}, \tau) - z(\omega, t_{\tau',\omega}, \tau') $ under the hypothesis that $\K(\omega, t_0, \tau)$ is correct. 
	We suppose the data is  $v_A+ \epsilon$ where $v_A  = A x_t$. 
	
	We further denote by $P_M = M (M^T V^{-1} M)^{-1} M^T V^{-1}$  and $H_M = V^{-1}P_M $for a given matrix $M$. The residuals of the fit of models $\mathcal{M}_A$ and $\mathcal{M}_B$ are
	\begin{align}
	r_A = &  (I- P_A)\epsilon  \\
	r_B  = &(I- P_B)(\epsilon +v_A)   \
	\end{align}
	We denote by $u_A:=(I- P_B)v_A$  and $V = LL^T$ where $L$ is the Cholesky decomposition.of $V$, and $\eta  := L^{-1} \epsilon$.   $\chi^2$ a
	\begin{align}
	\chi^2_A &=   \eta^T L^T( V^{-1} -  H_A) L \eta \\
	\chi^2_B & = \eta^T L^T( V^{-1} -  H_B) L \eta  +   2 u_AV^{-1}(I- P_B)L\eta    +u_A^T V^{-1}u_A
	\end{align}
	Then the difference of periodogram values is 
	\begin{align}
	D_z &= z(\omega, t_{\tau,\omega}, \tau) - z(\omega, t_{\tau',\omega}, \tau') =\chi^2_A  \ - \chi^2_B \\
	& = - (\eta^T D \eta + \beta ^T \eta  + \alpha)
	\end{align}
	where $D = H_A - H_B$, $\beta = 2 u_AV^{-1}(I- P_B)L$, $\alpha = u_A^T V^{-1}u_A$. Then the expectancy ad variance of $D_z$ are
	\begin{align}
	\mathbb{E} \{D_z\} = - \sum_i D_{ii} - \alpha \label{eq:dze}\\
	\mathbb{V} \{D_z\} =  \sum_{i,j} D_{ij}^2 + \sum_i \beta_i^2 \label{eq:dzv}
	\end{align}
	
	\subsection{Independence of semi-amplitude and phase estimate}
	\label{app:independence}
	Let us suppose the data is
	\begin{align}
	y = y_0 + \epsilon
	\end{align}
	where $y_0$ is the model and $\epsilon$ is a random variable following a Gaussian distribution with null mean and covariance $V$. We consider two alternative linear models: $y = Mx + \epsilon$ and $y = {M'}x + \epsilon$. Then their corresponding least square estimates are $x_M :=(M^TWM)^{-1} M^TW y$ and $x_{M'} :=({M'}^TW{M'})^{-1} {M'}^TW y$ where $W := V^{-1}$. Then the covariance of $x_M$ and $x_{M'}$ is 
	\begin{align}
	Cov(x_M, x_{M'}) =(M^TWM)^{-1} M^T W {M'} ({M'}^TW{M'})^{-1} 
	\label{eq:covarAB}
	\end{align}
	
	We now consider the case where $M = [m,  M_0 ]$ and ${M'} = [m',  M_0 ]$ where the brackets designate the concatenation of two matrices, and $m = [ w(t-t_0) \cos \omega t, w(t-t_0) \sin \omega t]$, $m' = [ w(t-t_0') \cos \omega t, w(t-t_0') \sin \omega t]$ where $w$ is a box-shaped time-window, zero everywhere except between $t_0 - \tau/2 $  and $t_0 + \tau/2$, and $|t_0 - t_0'|> \tau$. That is, fitting $M$ and ${M'}$ correspond to two model $\K$ (see eq.~\eqref{eq:mk}).
	
	If the noise is uncorrelated, $V$ is diagonal and so is $W$. Since $ w(t-t_0) w(t-t_0')=0$ The coefficients of columns $m$ and $m'$, from eq.~\eqref{eq:covarAB} are independent. 
	If  the noise is correlated $W$ is non diagonal and $w(t-t_0)^T W w(t-t_0')$ where $w(t)$ is the column vector $(w(t_i))_{i=1..N} $

	\section{FAP calculation}
	\label{app:fap}

	Here, we denote by the generic symbol  $Z(Y, \theta)$  a function of the data $Y$ which is either $z$, as defined in eq.~\eqref{eq:z}, or for a fixed $T$, $\omega, t_0 \rightarrow z(\omega, t_0, T)$.
	Denoting by $Z_{\mathrm{max}}$ the observed maximum of the periodogram on the data to be analysed, the false alarm probability (FAP) is defined as 
	\begin{align}
	\mathrm{FAP} = \mathrm{Pr} \{ \max _\theta Z(Y,\theta) \geqslant Z_{\mathrm{max}} | Y \sim \H\}  
	\end{align}
	where $\H$ is defined in eq.~\eqref{eq:mh}. The FAP can be computed by simulations generating data following model $\sim \H$ and computing the empirical distribution of $\max_\theta Z(Y,\theta)$, or generating $Y$ with random permutations of the input data. The value of $\theta_\H$ can be fixed to zero without loss of generality. This procedure can be very computationally heavy to reach low FAPs. The ASP is  special case of non linear ``Von Mises'' periodograms defined in~\cite{baluev2013_vonmises}, for which analytical approximations of the FAP can be derived. In that case, the specific form of the apodization function must be specified. We here adopt the approach of~\cite{suveges2014}, which consists in generating data like in the simulation approach, but then fitting a generalised extreme distribution (GEV) to the generated samples. The analytical form of a GEV distribution is
	\begin{align}
	p(z; \mu, \sigma, \xi) = \frac{1}{\sigma} t(z)^{\xi +1} \e^{-t(z)}
	\end{align}
	where 
	\begin{align}
	t(z) = \left\{
	\begin{array}{ll} \left(1 + \xi \frac{z - \mu}{\sigma} \right)^{-\frac{1}{\xi}} \text{ if } \xi \neq 0\\
	\e^{-\frac{z - \mu}{\sigma} } \text{ if } \xi = 0\\
	\end{array}
	\right.
	\end{align}
	The parameters $\mu, \sigma, \xi$ can be estimated with a maximum likelihood estimate, and the uncertainties obtained from the inverse of the Fisher information matrix can be propagated to  $p(z; \mu, \sigma, \xi)$ for a certain value of $z$, to ensure that the FAP computed is accurate enough.

	\begin{figure}
		\includegraphics[width=.94\linewidth]{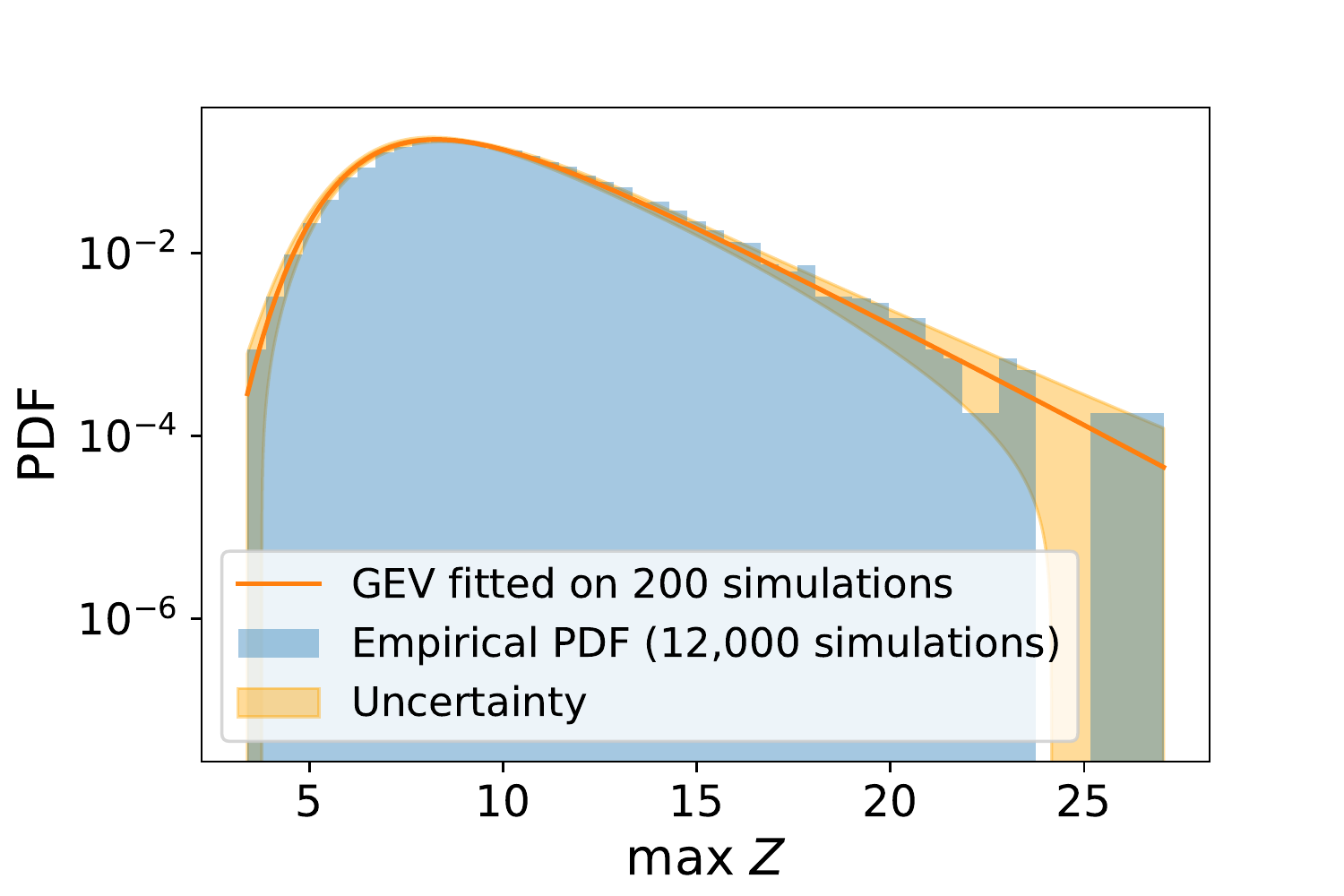}
		\caption{Empirical distribution of the maximum of periodograms generated }
		\label{fig:pdf}
	\end{figure}
	
	This approach allows to drastically reduce the number of simulations needed to estimate the low FAP levels. To illustrate this, we show in Fig.~\ref{fig:pdf} the histogram of 12,000 maxima of periodograms, and the distribution (blue) and the  fit of the 200 first simulations made with a Nelder-Mead algorithm (in orange). The simulation is made with the 70 first measurements of HD 85512~\citep{pepe2011}.

\end{document}